\documentclass[%
 aip,
 amsmath,amssymb,
 reprint,%
]{revtex4-1}
\usepackage{graphicx}
\usepackage{subfigure}
\usepackage{dcolumn}
\usepackage{bm}
\usepackage{color}
\usepackage[T1]{fontenc}
\usepackage{mathptmx}
\usepackage{multirow}
\usepackage{array}
\usepackage{float}
\usepackage{etoolbox}
\usepackage{booktabs}
\usepackage{url}
\usepackage{CJKutf8}
\usepackage{pifont}
\usepackage{epstopdf, epsfig}
\usepackage{graphicx}
\usepackage{longtable}
\usepackage[
    colorlinks=true,
    linkcolor=blue,
    urlcolor=blue,
    citecolor=blue
]{hyperref}

\usepackage{titlesec}
\titleclass{\subsubsubsection}{straight}[\subsection]
\newcounter{subsubsubsection}[subsubsection]
\renewcommand\thesubsubsubsection{\thesubsubsection.\alph{subsubsubsection}}
\titleformat{\subsubsubsection}
 {\normalfont\normalsize\bfseries}{\thesubsubsubsection}{1em}{}
\titlespacing*{\subsubsubsection}
{0pt}{3.25ex plus 1ex minus .2ex}{1.5ex plus .2ex}

\renewcommand\arraystretch{1.5}

\begin{document}
\preprint{AIP/123-QED}

\title{Stability analysis of discrete Boltzmann simulation for supersonic flows: Influencing factors, coupling mechanisms and optimization strategies}

\author{Yanhong Wu \begin{CJK*}{UTF8}{gbsn} (吴彦宏) \end{CJK*}}
\affiliation{Hebei Key Laboratory of Trans-Media Aerial Underwater Vehicle,
North China Institute of Aerospace Engineering, Langfang 065000, China}

\author{Yanbiao Gan \begin{CJK*}{UTF8}{gbsn} (甘延标) \end{CJK*}}
\thanks{Corresponding author: gan@nciae.edu.cn}
\affiliation{Hebei Key Laboratory of Trans-Media Aerial Underwater Vehicle,
North China Institute of Aerospace Engineering, Langfang 065000, China}

\author{Aiguo Xu \begin{CJK*}{UTF8}{gbsn} (许爱国) \end{CJK*}}
\affiliation{National Key Laboratory of Computational Physics, Institute of Applied Physics and Computational Mathematics, P. O. Box 8009-26, Beijing 100088, P.R.China}
\affiliation{National Key Laboratory of Shock Wave and Detonation Physics, Mianyang 621999, China}
\affiliation{HEDPS, Center for Applied Physics and Technology, and College of Engineering, Peking University, Beijing 100871, China}
\affiliation{State Key Laboratory of Explosion Science and Safety Protection, Beijing Institute of Technology, Beijing 100081, China}

\author{Bin Yang \begin{CJK*}{UTF8}{gbsn} (杨斌) \end{CJK*}}
\thanks{Corresponding author: binyang@tcu.edu.cn}
\affiliation{School of Energy and Safety Engineering, Tianjin Chengjian University, Tianjin 300384, China}

\date{\today}

\begin{abstract}
Supersonic flow simulations encounter challenges in threefold: trans-scale modeling, numerical stability and complex field analysis, which arise from inherent nonlinear, nonequilibrium and multiscale characteristics.
The discrete Boltzmann method (DBM) provides a multiscale kinetic modeling framework and analysis tool for capturing complex discrete/nonequilibrium states and effects.
Despite its fundamental role in DBM simulations, a comprehensive stability analysis is still lacking.
Similar to the lattice Boltzmann method, complexity in DBM lies mainly in the intrinsic coupling between velocity and spatiotemporal discretizations, which distinguishes it from traditional computational fluid dynamics.
This study conducts von Neumann stability analysis to examine factors influencing DBM simulation stability, including approaches for determining equilibrium distribution functions, thermodynamic nonequilibrium (TNE) levels, spatiotemporal discretization schemes, initial conditions, and model parameters.
Key findings include: (i) Among the equilibrium distribution discretization methods considered, the moment-matching approach outperforms the expansion- and weighting-coefficient-based methods in the test simulations. (ii) Increased TNE intensity/Knudsen number enhances system's nonlinear behavior and the intrinsic nonlinearity embedded in the matching model equation, thereby amplifying the instabilities in simulations.
(iii) Although additional viscous dissipation based on distribution functions improves stability, it distorts flow fields and alters constitutive relations, highlighting the need for careful trade-offs between stability and accuracy. (iv) Larger Courant-Friedrichs-Lewy numbers and relative time steps significantly degrade stability, necessitating appropriate time-stepping strategies.
To assess the stability regulation capability of DBMs across different TNE levels, stability-phase diagrams and stability probability curves are constructed within the moment-matching framework using morphological analysis.
These diagrams identify common stable parameter regions applicable across various TNE orders.
Finally, the effects of discrete velocity configurations on achieving both physical functionality and numerical stability are assessed through comparisons between numerical and analytical TNE solutions, as well as statistical properties of distribution functions.
This study reveals key factors and coupling mechanisms governing numerical stability in DBM simulation, and proposes general strategies for optimizing equilibrium distribution function discretization, discrete velocity design, and stability parameter selection across supersonic regimes.
\end{abstract}

\maketitle

\section{INTRODUCTION}\label{I}

Supersonic flow is a nonlinear, nonequilibrium, and multiscale phenomenon with wide applications in aerospace, defense, and astrophysics \cite{courant1999supersonic, bertin2006critical, candler2019rate}.
Understanding its evolution mechanisms and developing effective control strategies are critical for both scientific progress and engineering advancements.
In aerospace, hypersonic vehicles experience severe aerodynamic heating and complex shock-wave interactions such as shock–boundary-layer interactions, jet interference, turbulent mixing, and laminar–turbulent transition \cite{gaitonde2015progress, huh2018numerical, pan2010mixing}. Thus, high-fidelity numerical simulations are essential for optimizing aerodynamic configurations and enhancing the maneuverability, stability, and safety of these vehicles \cite{cao2024data}.
Similarly, during atmospheric re-entry, spacecraft are exposed to intense aerothermal environments \cite{gallais2007atmospheric}, where extreme temperatures generated by supersonic flows impose strict requirements on the design of thermal protection systems \cite{poloni2022carbon}. Accurate predictions of flow dynamics and heat loads are vital for ensuring vehicle stability and structural integrity in such harsh conditions \cite{du2023numerical}. Supersonic flows also play a fundamental role in the aerodynamics of combustion chambers and nozzles in advanced propulsion systems, including high-performance gas turbines and rocket engines \cite{jiang2021influence, emelyanov2019supersonic}. A comprehensive understanding of supersonic flow behavior is essential for optimizing fuel efficiency and optimizing propulsion performance \cite{tahsini2015investigating, huang2013multiobjective}.

A defining characteristic of supersonic flows is the coexistence of mesoscopic structures and kinetic modes evolving over disparate temporal and spatial scales \cite{gan2025supersonic}. The interplay between these mesoscopic structures and kinetic modes gives rise to complex, diverse, and widely distributed discrete and thermodynamic nonequilibrium (TNE) effects \cite{xu2012lattice, xu2022BSTP, xu2024advances}.
Among these structures, shock waves serve as a paradigmatic example in supersonic flows \cite{zapryagaev2018shock, gaitonde2023dynamics} . As the shock thickness approaches the mean free path of molecules, discrete effects become increasingly prominent \cite{andreopoulos2000shock}. Across the shock front, macroscopic quantities undergo abrupt variations, and the system lacks sufficient time to relax to thermodynamic equilibrium, resulting in pronounced nonequilibrium behavior.
A comprehensive understanding of complex supersonic flows requires the development of cross-scale models capable of accurately capturing the intensified TNE effects near mesoscopic structures, as well as the influence of nonequilibrium transport mechanisms, to enable reliable predictions of flow evolution and effective regulation of macroscopic behavior.

Compared with engineering experiments, numerical simulations offer distinct advantages, including controllable initial and boundary conditions, low cost, high reproducibility, and the ability to provide comprehensive and readily analyzable flow-field data. These benefits make numerical simulations indispensable tools for investigating supersonic flows \cite{xiwan2019survey, boccelli2024numerical}. Based on the underlying physical models, numerical simulations of supersonic flows can be broadly classified into macroscopic, microscopic, and mesoscopic approaches.
Macroscopic methods, founded on the continuum assumption and near-equilibrium approximation, primarily involve solving the Euler or Navier–Stokes (NS) equations \cite{ding2022numerical, bernardini2023streams}. However, quasi-continuum models fail to accurately capture discrete (rarefaction) effects, while near-equilibrium models are inadequate for representing strongly nonequilibrium behaviors. Consequently, macroscopic methods face intrinsic limitations in simulating supersonic flows.
Microscopic methods explicitly resolve the dynamics of individual particles, with molecular dynamics serving as a representative approach \cite{klima2018direct, li2021evaluation}. Despite their high accuracy, the extremely high computational cost of microscopic methods severely restricts their applicability to large-scale simulations and practical engineering scenarios.

Mesoscopic methods characterize the spatiotemporal evolution of the microscopic particle velocity distribution function by solving the Boltzmann equation, thereby bridging physical interactions across multiple scales \cite{tong2019review, qiu2020study}.
This approach is applicable to a wide range of flow regimes, including continuum, slip, transitional, and free molecular flows, making it a representative cross-scale modeling strategy.
However, the Boltzmann equation is inherently a complex integro-differential equation. Unlike conventional spatial and temporal discretization methods in traditional computational fluid dynamics (CFD), its numerical solution additionally requires discretization of velocity space.
Since the particle velocity distribution spans from negative to positive infinity, conventional discretization schemes are insufficient to capture its essential features.
Therefore, developing approximate and simplified kinetic models that retain the essential features of the Boltzmann equation has become an effective strategy. Representative approaches along this line  include the gas kinetic scheme \cite{xu2005multidimensional, sun2021numerical}, unified gas kinetic scheme \cite{zhu2016implicit, zhu2017unified}, discrete unified gas kinetic scheme \cite{guo2013discrete, guo2021progress}, discrete velocity method \cite{yang2019improved, yang2022efficient}, lattice Boltzmann method (LBM) \cite{succi2018lattice,watari2006supersonic, tran2022lattice, latt2020efficient,zhan2025thermodynamically,wang2022novel,fei2024pore,yang2025acoustic,hou2025lattice}, and discrete Boltzmann method (DBM) \cite{xu2022BSTP,xu2024advances,zhang2016kinetic,lin2019discrete,gan2018discrete, gan2022discrete, zhang2022discrete,guo2025POF,sun2024droplet,song2024plasma,song2025discrete,li2024kinetic,xu2025influence,he2025POF}, among others.

As a specialized discrete representation of the Boltzmann equation, the LBM has achieved substantial progress in simulating weakly compressible and even supersonic flows.
In 2003 and 2004, Watari and Kataoka \textit{et al.} developed NS-level LBMs, where temporal and spatial derivatives were discretized using finite-difference (FD) schemes \cite{watari2003two, kataoka2004lattice}.
The FDLBM decouples the velocity discretization from the spatial discretization inherent in traditional LBM formulations, significantly enhancing its flexibility and applicability.
Nevertheless, when the Mach number exceeds 1, these models often suffer from severe numerical instability.
To mitigate this issue, numerous stabilization strategies have been proposed, among which high-order numerical schemes has proven particularly effective.
Representative examples include the modified Lax–Wendroff (MLW) scheme \cite{gan2008two}, the non-oscillatory and non-free-parameter dissipation (NND) scheme \cite{qiu2017lattice, qiu2021mesoscopic}, the fifth-order weighted essentially non-oscillatory (WENO) scheme \cite{gan2011lattice, hejranfar2024computation}, the flux limiter method \cite{sofonea2004finite, gan2011flux}, the fully implicit LBM \cite{huang2024implicit}, and the implicit-explicit finite-difference scheme \cite{li2007coupled, gan2013lattice}.
Additionally, various alternative stabilization techniques have been developed, such as entropic LBM \cite{frapolli2015entropic, frapolli2020theory, yeoh2020lattice, hosseini2023entropic}, particles-on-demand formulation \cite{dorschner2018particles, kallikounis2022particles}, flux-solver-based approaches \cite{yang2019development, song2024double}, pressure-wave-dynamics convergence acceleration strategy \cite{yeoh2021utilization}, multi-relaxation-time (MRT) model \cite{chen2010multiple, li2015multiple}, shock-detection sensor techniques \cite{ghadyani2015use, esfahanian2015improvement}, and hybrid or coupled schemes.
These include hybrid LBM formulations \cite{li2023hybrid, zhao2020toward, guo2024hybrid}, hybrid recursive regularized LBM \cite{feng2019hybrid, feng2020grid}, coupled double-distribution-function LBM \cite{qiu2017lattice, qiu2020lattice, qiu2021mesoscopic}, semi-Lagrangian LBM \cite{wilde2020semi}, Lagrangian–Eulerian LBM \cite{saadat2020arbitrary}, and central-moment-based LBM \cite{lin2025central}, among others.

It is important to note that most existing models only account for the first-order deviation from the equilibrium distribution function, limiting their ability to capture significant discrete and thermodynamic nonequilibrium (TNE) effects in supersonic flows \cite{gan2018discrete}. To overcome this limitation, we adopt the Discrete Boltzmann Method (DBM), a multiscale modeling and complex physical fields analysis approach, developed from LBM framework by selectively discarding, retaining, and supplementing \cite{xu2012lattice, xu2024advances, xu2022BSTP}. In comparison with conventional approaches, DBM offers superior capabilities in capturing intricate nonequilibrium dynamics and multiscale phenomena in supersonic flows. This is achieved by employing higher-order non-conserved kinetic moments to quantify higher-order deviations from thermodynamic equilibrium \cite{gan2018discrete, gan2022discrete, zhang2022discrete, he2025POF, guo2025POF}.

The primary numerical distinction between mesoscopic methods and traditional computational fluid dynamics (CFD) approaches lies in the discretization of phase-space and its intrinsic coupling with spatial and temporal discretizations. A well-designed phase-space discretization is crucial to preserve the cross-scale descriptive capability of kinetic models \cite{he2025POF}. However, achieving a favorable balance among physical accuracy, numerical robustness, and computational efficiency in phase-space discretization—while maintaining consistent coupling with spatial and temporal schemes—remains a key challenge in the development of kinetic models. Therefore, stability analysis of kinetic models is essential for understanding the physical mechanisms governing their stability and for guiding the discretization of phase-space.

In the kinetic study of supersonic flows, low-order models (e.g., Euler and NS levels) differ significantly from high-order models (e.g., Burnett and super-Burnett levels) in their ability to capture TNE effects \cite{gan2018discrete,gan2022discrete}.
This distinction presents both advantages and limitations.
Low-order models are computationally efficient and numerically robust,  but are severely limited in applicability.
In contrast, high-order kinetic models incorporate higher-order deviations from equilibrium, which, at the macroscopic level, correspond to the adoption of extended constitutive relations to describe transport processes \cite{gan2018discrete, gan2022discrete, guo2025POF}.
However, high-order models face pronounced numerical stability challenges:
(i) The incorporation of multiscale effects enables high-order models to resolve finer mesoscopic structures. However, the competition and coupling between these delicate structures and fast-varying kinetic modes tend to trigger numerical instability.
(ii) The appearance of complex higher-order derivative terms imposes stringent requirements on spatial resolution and numerical precision, which may further exacerbate instability.
(iii) The construction of phase-space discretization in high-order models is inherently more sophisticated than in low-order models. Its stronger coupling with spatiotemporal discretizations further exacerbates numerical instability.

von Neumann stability analysis is a widely used approach for evaluating the stability characteristics of numerical schemes, particularly for solvers of partial differential equations (PDEs) \cite{kumari2021generalized, deriaz2020non}. Its key advantages can be summarized as follows:
\emph{(i) Simplified linearization framework:}
This method utilizes Fourier decomposition to represent the solution as a superposition of independent Fourier modes, reducing complex PDEs to a more tractable form. Numerical errors are treated as small perturbations, allowing nonlinear problems to be transformed into a linear framework, which significantly simplifies stability analysis.
\emph{(ii) Broad applicability:}
von Neumann analysis is compatible with various discretization schemes, including FD, finite volume, and finite element methods. It can also be extended to kinetic models, where phase-space discretization is strongly coupled with spatiotemporal discretizations.
\emph{(iii) Explicit stability criteria:}
By analyzing the amplification factor associated with each Fourier mode, this method provides clear and quantitative conditions for assessing numerical stability. These conditions offer a rigorous theoretical foundation for the design, optimization, and evaluation of numerical algorithms.
\emph{(iv) Intuitive frequency-domain insights:} By analyzing Fourier modes across a range of wavelengths, this approach provides clear insights into error propagation and amplification mechanisms.

Owing to its effectiveness in quantifying the propagation of numerical errors, von Neumann stability analysis has become a fundamental tool for stability assessment of kinetic models.
Sterling \textit{et al.} conducted one of the earliest von Neumann analyses on standard LBM lattices—including seven-velocity hexagonal, nine-velocity square, and fifteen-velocity cubic forms—and examined the influence of mean velocity, relaxation time, and perturbation wavenumber on stability \cite{sterling1996stability}.
Seta \textit{et al.} extended von Neumann analysis to FDLBMs, evaluating the impact of Courant–Friedrichs–Lewy (CFL) number, relaxation time, flow velocity, and FD schemes—including central and upwind methods—on stability. They also analyzed thirteen-velocity lattices and the destabilizing role of virtual forcing terms \cite{seta2001numerical, seta2002numerical, seta2001stability}.
Niu \textit{et al.} systematically analyzed local stability and dissipation characteristics of several LBM variants \cite{niu2004investigation}.
Gan \textit{et al.} demonstrated that the MLW scheme with artificial viscosity improves LBM stability for high-Mach flows, and this framework was later extended to 3D using the distributed weighting-coefficient method \cite{gan2008two, chen2009highly, chen2010three}.
Servan-Camas \textit{et al.} demonstrated that non-negativity of the equilibrium distribution ensures linear stability, and derived stability regions based on Peclet number, Courant number, relaxation time, and flow direction \cite{servan2009non}.
Kuzmin \textit{et al.} analyzed the two-relaxation-time (TRT) LBM, proposed an optimal formulation, and developed tuning strategies to enhance stability \cite{kuzmin2011role}.
Chen \textit{et al.} introduced an unconditionally stable LBM, performed the corresponding stability analysis, and later extended it to three-dimensional flows \cite{chen2017truly, chen2017three}.
Krivovichev \textit{et al.} studied the impact of spatiotemporal discretization schemes and model parameters on FDLBM stability \cite{krivovichev2019stability, krivovichev2020parametric},
incorporating dispersion and dissipation characteristics into FD schemes and proposing an optimization strategy to minimize numerical errors during the convection stage \cite{krivovichev2019stability, krivovichev2020parametric, krivovichev2020approach}.
Masset \textit{et al.} demonstrated that the stability of MRT models is governed by the ratio of relaxation frequencies \cite{masset2020linear}.
Wissocq \textit{et al.} systematically analyzed various collision models, including the Bhatnagar–Gross–Krook model and pre-collision regularization and recursive regularization strategies, and demonstrated that modal filtering, particularly recursive regularization, significantly enhances stability, especially at low viscosities \cite{wissocq2020linear}.

Beyond von Neumann analysis, weighted ${L^2}$-stability, matrix-based stability, and convergence analyses have also been employed to study numerical instability in kinetic and macroscopic models.
Worthing \textit{et al.} analyzed how background shear affects nine-velocity LBM stability \cite{worthing1997stability}.
Servan-Camas \textit{et al.} established TRT-LBM stability regions and showed that careful parameter tuning (relaxation parameters, Peclet numbers, and Courant numbers ) improves accuracy and stability \cite{servan2008lattice}.
Siebert \textit{et al.} found that higher-order LB approximations and suitable macroscopic parameters enhance stability \cite{siebert2008lattice}.
Perez \textit{et al.} used linearized equilibrium functions for global linear instability analysis, validating their solver against spectral element results \cite{perez2017lattice}.
Yang \textit{et al.} introduced an automated framework for weighted ${L^2}$-stability analysis of MRT models, which streamlines matrix decomposition and stability verification, enabling rapid derivation of stability conditions for a wide range of MRT formulations \cite{yang2024automatic}.
Ren \textit{et al.} introduced a matrix-based stability analysis method within the finite volume framework to address shock instabilities commonly encountered in Godunov-type schemes for hypersonic flows \cite{ren2024numerical}. They systematically explored the effects of spatial accuracy and Riemann solver choices on shock-induced instabilities, and extended the framework to fifth-order WENO finite volume schemes that enhances both shock stability and numerical accuracy \cite{ren2025numerical}.
Yeoh \textit{et al.} proposed an acceleration scheme based on pressure wave-dynamics, which utilizes pressure propagation information to reconstruct the flow field, thereby significantly accelerating the convergence of the LBM in both steady and unsteady flows \cite{yeoh2021utilization}.

Although the stability of kinetic models has been extensively studied, several limitations remain unaddressed:

(i) Most studies focus on incompressible or weakly compressible flows, while the stability of kinetic models in supersonic compressible regimes remains largely unexplored. The highly transient nature, strong coupling, pronounced nonlinearity, and multiscale characteristics of supersonic flows exacerbate model nonlinearity and significantly increase numerical complexity, making stability analysis particularly challenging.

(ii) While the stability of low-order kinetic models (e.g., Euler, NS level) has been thoroughly studied, high-order models are less understood and often exhibit reduced stability due to their stronger nonlinearity. Achieving reliable and robust stability for high-order models remains a major obstacle in simulating complex supersonic flows.

(iii) Despite notable differences among equilibrium distribution function discretization schemes, systematic comparative studies on their impact on numerical stability remain scarce. Accurate discretization plays a key role in DBM, ensuring the conservation of kinetic moments across phase-space transformation.
Several schemes—such as the moment-matching method\cite{song2024plasma,zhang2022discrete,li2024kinetic,he2025POF,guo2025POF}, the globally unified expansion-coefficient method\cite{watari2004possibility,watari2007finite,wissocq2020linear,perez2017lattice,yang2024automatic}, and the distributed weighting-coefficient method\cite{alexander1993lattice,chen1994thermal,kataoka2004lattice2}—have been proposed.
However, their stability characteristics differ significantly depending on the flow regime and numerical configuration.
A systematic investigation into their stability mechanisms, scope of applicability, and inherent trade-offs is essential for guiding the selection of appropriate schemes in practical modeling.

(iv) The coupling between phase-space and spatiotemporal discretization is not well understood, which distinguishes kinetic models from conventional CFD methods. Phase-space discretization requires selecting an appropriate discrete velocity set (DVS), which is critical for both model functionality and numerical stability. Although guidelines for phase-space discretization exist \cite{he2025POF}, a comprehensive theoretical framework is still lacking.

(v) The effects of high-order spatiotemporal discretization schemes, the relative time step ($\Delta t/\tau$), and the spatiotemporal step ratio ($\Delta t/\Delta r$) require further investigation. FDLBM improves numerical stability by eliminating the propagation–collision modes of the standard LBM and decoupling discrete velocities from the computational grid. However, existing stability analyses of FDLBM primarily focus on upwind and central difference schemes, with limited attention to total variation diminishing schemes, such as NND and WENO schemes, particularly in the context of high-order kinetic models. Moreover, key model parameters—such as the spatiotemporal step ratio (closely related to the CFL number) and the relative time step (governing the strength of nonequilibrium effects)—have not been systematically investigated in relation to high-order model stability.

To address the limitations outlined above, this study employs von Neumann stability analysis to systematically evaluate the stability of both the DBM and LBM in supersonic flow regimes. The objective is to identify the key physical mechanisms and model parameters that govern numerical stability and to provide theoretical guidance for improving the robustness of high-order kinetic models.

The structure of the paper is organized as follows:
Section \ref{II} introduces the fundamental principles of DBM.
Section \ref{III} presents three discretization schemes for the equilibrium distribution function.
Section \ref{IV} describes the FD schemes employed for spatial discretization.
Section \ref{V} outlines the von Neumann stability analysis and derives the general form of the amplification matrix.
Section \ref{VI} provides a comprehensive stability analysis of various DBMs and LBMs under different flow conditions. A stability control strategy is proposed based on stability-phase diagrams, and a morphological method is introduced to generate stability probability curves. Numerical simulations are performed to validate the effectiveness of the proposed strategy and to compare the numerical stability and accuracy and of different DVSs.
Finally, Section \ref{VII} concludes the paper with a summary of the main findings and a discussion of potential future research directions.

\section{DISCRETE BOLTZMANN METHOD}\label{II}

The evolution equations of DBM consist of the discrete Boltzmann equation (\ref{e1}) for the discrete distribution function ${f_i}$ and the constraint equation (\ref{e3}) for the discrete equilibrium distribution function (DEDF) $f_i^{eq}$.
In DBM, ``discrete'' refers to the discretization of phase-space, where continuous particle velocities are approximated by a finite set of discrete velocities:
\begin{equation}\label{e1}
\frac{\partial f_i}{\partial t} + {\mathbf{v}}_i \cdot \bm{\nabla} f_i = - \frac{1}{\tau} \left( f_i - f_i^{eq} \right),
\end{equation}where $i$ denotes the index of the discrete velocities.

The constraint equation for ${f_i^{eq}}$ is derived based on the principle that the kinetic moments of ${f_i}$, which characterize the system’s macroscopic state and behavior, must remain invariant before and after phase-space discretization:
\begin{equation}\label{e2}
\bm{\Phi}' = \int^\infty_{-\infty} \int^\infty_{-\infty} f \bm{\Psi}'(\mathbf{v}, \mathbf{\eta}) d \mathbf{v} d \mathbf{\eta} = \sum_i f_i \bm{\Psi}'(\mathbf{v}_i, \mathbf{\eta}_i),
\end{equation}
where $\bm{\Psi '}(\mathbf{v}, \mathbf{\eta}) = \left[ 1, \mathbf{v}, \frac{1}{2}(v^2 + \eta^2), \mathbf{v}\mathbf{v}, \frac{1}{2}(v^2 + \eta^2) \mathbf{v}, \mathbf{v}\mathbf{v}\mathbf{v}, \cdots \right]$. These kinetic moments serve as invariants in coarse-grained modeling, providing a unique perspective for analyzing kinetic behavior, and are continuously refined to accommodate the growing demands of nonequilibrium studies.

Chapman–Enskog (CE) multiscale analysis forms the theoretical foundation of DBM. It demonstrates that ${f_i}$ can be expressed as a nonlinear combination of the spatiotemporal derivatives of ${f_i^{eq}}$.
Consequently, the constraints on ${f_i}$ can be reformulated as constraints on ${f_i^{eq}}$: \begin{equation}\label{e3}
\bm{\Phi} = \int^\infty_{-\infty} \int^\infty_{-\infty} f \bm{\Psi}(\mathbf{v}, \mathbf{\eta}) d \mathbf{v} d \mathbf{\eta}= \sum_i f_i^{eq} \bm{\Psi}(\mathbf{v}_i, \mathbf{\eta}_i).
\end{equation}
This requires $\bm{\Psi}$ to include more components and higher-order terms of $(\mathbf{v}, \mathbf{\eta})$ than $\bm{\Psi '}$.

Specifically, when ${f_i^{eq}}$ satisfies the following seven kinetic moment relations:
$\bm{\Phi}=(\mathbf{M}_0, \mathbf{M}_1, \mathbf{M}_{2,0}, \mathbf{M}_2, \mathbf{M}_{3,1}, \mathbf{M}_{3}, \mathbf{M}_{4,2})$,
DBM considers the first-order TNE effects in viscous stress and heat flux;
When ${f_i^{eq}}$ additionally satisfies $\bm{\Phi}= (\mathbf{M}_{4}, \mathbf{M}_{5,3})$, it captures second-order TNE effects in viscous stress and heat flux \cite{gan2018discrete};
Similarly, when ${f_i^{eq}}$ additionally satisfies $\bm{\Phi}= (\mathbf{M}_{m}, \mathbf{M}_{m+1,m-1})$, DBM accounts for $(m-2)$th-order TNE effects\cite{guo2025POF}.
In other words, for every two additional independent moment relations satisfied by ${f_i^{eq}}$, DBM achieves one higher order of accuracy in describing viscous stress and heat flux.
Here, ${{\bf{M}}_{m,n}} = \sum\limits_i {{{\left( {\frac{1}{2}} \right)}^{1 - {\delta _{m,n}}}}} f_i^{eq}{\bf{v}}_i^n{\left( {{\bf{v}}_i^2 + {\eta ^2}} \right)^{\frac{{m - n}}{2}}}$.

As the degree of discrete/nonequilibrium increases, the DBM—being a direct kinetic modeling tool—does not experience a significant increase in complexity compared with macroscopic models, which rely on ``\emph{first deriving extended hydrodynamic equations and then solving them numerically}''.
In contrast, these macroscopic equations become increasingly complicated, involving numerous high-order derivatives and nonlinear couplings, making the numerical solution process exceptionally complex.

In the analysis of complex physical fields, DBM employs the non-conserved moments of (${f_i} - f_i^{eq}$) to characterize the system’s deviation from thermodynamic equilibrium \cite{xu2022BSTP,xu2024advances}, enabling multilevel exploration of physical phenomena that were previously inaccessible.
Specifically, DBM defines two types of non-conserved moments: $\bm{\Delta}_{m,n}^*$, which characterizes pure TNE effects, and $\bm{\Delta}_{m,n}$, which accounts for the combined influence of hydrodynamic nonequilibrium (HNE) and TNE effects.

\section{DISCRETIZATION METHODS FOR THE EQUILIBRIUM DISTRIBUTION FUNCTION}\label{III}

\begin{figure*}[htbp]
	\centering
	\includegraphics[width=0.8\textwidth]{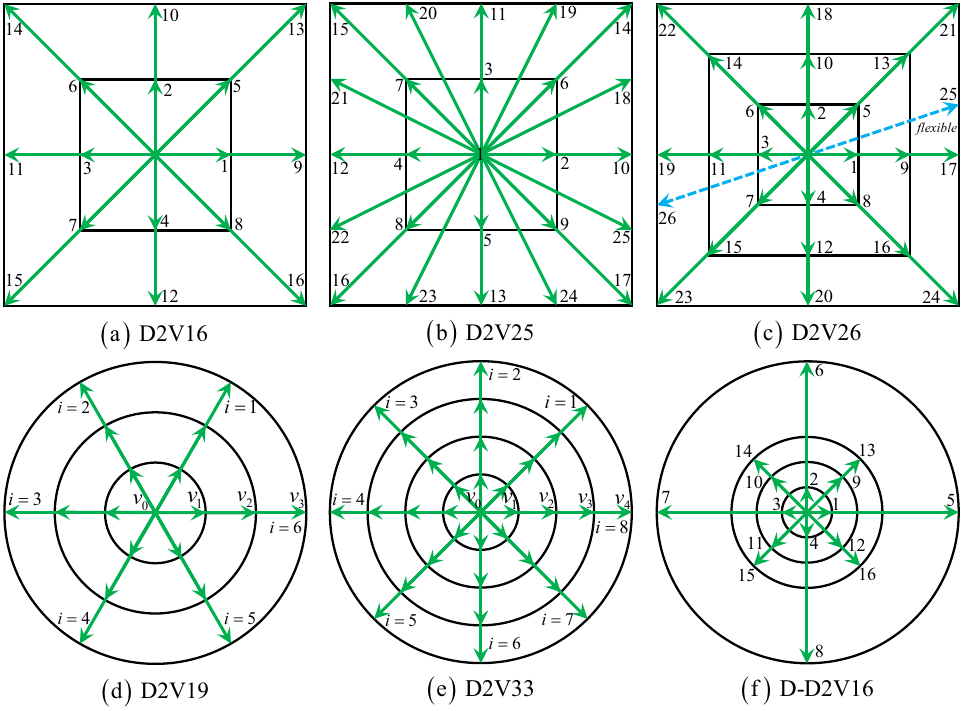}
	\caption{\centering{Schematic representation of six discrete velocity sets.}}
	\label{Fig01}
\end{figure*}

The formulation of the DEDF determines the construction of the DVS, thereby influencing both the physical fidelity and numerical stability of the model.
This section briefly introduces three discrete methods for constructing the DEDF:
(i) the moment-matching method;
(ii) the globally unified expansion-coefficient method;
(iii) the distributed weighting-coefficient method.

\subsection{Moment-matching method}\label{III_A}

The essence of the moment-matching method lies in expressing the kinetic moment relations satisfied by the DEDF in matrix form:
\begin{equation}\label{e4}
\bm{\Phi} = {\mathbf{C}} \cdot {{\mathbf{f}}^{eq}},
\end{equation}
where ${{\mathbf{f}}^{eq}} = {(f_1^{eq},f_2^{eq},f_3^{eq}, \cdots)^T}$ is the vector of DEDFs, $\bm{\Phi} = {\left( {{M_0},{M_{1x}},{M_{1y}}, \cdots} \right)^T}$ denotes the prescribed kinetic moments, and the matrix ${\mathbf{C}} = \left( {{{\mathbf{c}}_1},{{\mathbf{c}}_2},{{\mathbf{c}}_3}, \cdots} \right)$ is the transformation matrix that links the DEDFs  to the corresponding moments. Each column ${{\mathbf{c}}_i}$ of ${\mathbf{C}}$ is defined as ${{\mathbf{c}}_i} = {\left( {1,{v_{ix}},{v_{iy}},\frac{1}{2}\left( {v_i^2 + \eta _i^2} \right), \cdots} \right)^T}$.
Accordingly, the DEDFs can be obtained by:
\begin{equation}\label{e5}
{{\mathbf{f}}^{eq}} = {{\mathbf{C}}^{-1}} \cdot {\mathbf{M}},
\end{equation}
 where ${{\mathbf{C}}^{-1}}$ denotes the inverse of ${\mathbf{C}}$.

This approach offers several key advantages:

(i) Simplicity and directness: This method directly enforces the preservation of kinetic moments. If the DVS guarantees the existence of ${{\mathbf{C}}^{-1}}$, the DEDF can be explicitly obtained in a closed form.

(ii) Flexibility and generality: The approach is applicable to DBM construction in arbitrary spatial dimensions, at any desired order of nonequilibrium effects, and for any prescribed set of moment constraints.

(iii) High computational efficiency:  The moment vector $\bm{\Phi}$ includes only the minimal set of kinetic moments required to describe the targeted nonequilibrium effects. Each direction in the DVS corresponds to an independent moment constraint, making the formulation both physically compact and computationally efficient.

The moment-matching method has been widely used in DBM development for a range of complex systems, including high-speed compressible flows \cite{guo2025POF,zhang2022discrete}, multiphase flows \cite{sun2024droplet}, reactive flows \cite{zhang2016kinetic}, hydrodynamic instabilities \cite{he2025POF}, and plasma kinetics \cite{song2025discrete}, etc.
Figures~\ref{Fig01}(a)–(c) show representative DVS configurations: D2V16 corresponds to a DBM capable of capturing first-order TNE effects, while D2V25 and D2V26 are designed to incorporate second-order TNE effects.

\subsection{Globally unified expansion-coefficient method}\label{III_B}

In 2003, Watari \textit{et al.} expanded the DEDF using a Taylor series truncated at the fourth-order velocity terms and applied a unified weighting-coefficient throughout the formulation  \cite{watari2003two}.
This resulted in a NS-level LBM with a D2V33 DVS.
The DEDF is given by:
\begin{equation}\label{e6}
	f_{ki}^{eq} = \rho {F_k}\left[ \begin{array}{l}
		\left( {1 - \frac{{{u^2}}}{{2T}} + \frac{{{u^4}}}{{8{T^2}}}} \right) + \frac{{\left( {{{\bf{v}}_{ki}} \cdot {{\bf{u}}}} \right)}}{T}\left( {1 - \frac{{{u^2}}}{{2T}}} \right) \\
		+ \frac{{{{\left( {{{\bf{v}}_{ki}} \cdot {{\bf{u}}}} \right)}^2}}}{{2{T^2}}}\left( {1 - \frac{{{u^2}}}{{2T}}} \right)+ \frac{{{{\left( {{{\bf{v}}_{ki}} \cdot {{\bf{u}}}} \right)}^3}}}{{6{T^3}}} + \frac{{{{\left( {{{\bf{v}}_{ki}} \cdot {{\bf{u}}}} \right)}^4}}}{{24{T^4}}}
	\end{array} \right].
\end{equation}

The procedure for determining the expansion coefficients $F_k$ includes:
(i) substituting $f_{ki}^{eq}$ into the moment constraints and matching terms of equal velocity order;
(ii) simplifying the resulting equations to derive constraints on $F_k$;
(iii) imposing isotropy conditions of DVS to minimize the number of required constraints on ${F_k}$;
(iv) selecting discrete particle speeds $v_k$ and solving for $F_k$ analytically.

This method exhibits several limitations:

(i) Low computational efficiency:
All moment constraints are imposed on the global coefficients $F_k$, resulting in strong nonlinearity in $F_k$, which is related to the eighth power of the particle velocity. This leads to two issues: first, the model’s numerical stability becomes excessively sensitive to changes in macroscopic flow, and second, it necessitates highly isotropic DVS configurations.
For instance, recovering the NS equations with fixed specific-heat ratio requires 33 discrete velocities in two dimensions, whereas the moment-matching method needs only 13.
For models with adjustable specific-heat ratio, this number increases to 65 versus 16 for the moment-matching method.
As computational cost scales with the number of discrete velocities, this method is considerably less efficient and more complex in terms of both coefficient determination and DVS design than the moment-matching method.

(ii) Excessive number of free parameters:
To adjust the specific-heat ratio, this approach introduces $3k$ free parameters\cite{watari2007finite}, namely $v_1, v_2, \dots, v_{2k}$ and $\eta_1, \eta_2, \dots, \eta_k$. While this increases the model's flexibility, it is possible to find parameters that enhance numerical stability under certain conditions. However, the parameter space is vast, and with such strong nonlinearity in the model, effective parameter selection remains a practical challenge.

(iii) Poor numerical stability:
The Taylor expansion truncation neglects higher-order terms in DEDF, making the method highly susceptible to instability and inadequate for accurately capturing high-speed compressible flows.

Using the same framework, Gan \textit{et al.} later developed an Euler-level D2V19 LBM to investigate the effects of density and velocity gradients on the Kelvin–Helmholtz instability\cite{gan2011lattice}.
Figures~\ref{Fig01}(d)–(e) display the DVS configurations for the D2V19 and D2V33 models.

\subsection{Distributed weighting-coefficient method}\label{III_C}

To decouple the weighting coefficients $F_k$ from the DVS and reduce the number of discrete velocities, Kataoka \textit{et al.}  proposed a distributed weighting-coefficient strategy, in which individual coefficients are assigned to each term in the Taylor expansion of the DEDF \cite{kataoka2004lattice}.
This resulted in a NS-level D2V16 model\cite{kataoka2004lattice}. The DEDF is expressed as:
\begin{equation}\label{e7}
	f_i^{{{eq}}} = \rho \left[ \begin{array}{l}
		{a_{0i}} + {a_{1i}}T + {a_{2i}}{T^2} + \left( {{a_{3i}} + {a_{4i}}T} \right)u^2\\
		+ {a_{5i}}u^4 + \left( {{b_{0i}} + {b_{1i}}T + {b_{2i}}u^2} \right)(\mathbf{v}_i \cdot \mathbf{u})\\
		+ \left( {{d_{0i}} + {d_{1i}}T + {d_{2i}}u^2} \right)(\mathbf{v}_i \cdot \mathbf{u})^2
		+ {e_i}(\mathbf{v}_i \cdot \mathbf{u})^3
	\end{array} \right].
\end{equation}
While the coefficients can be found in Ref. \onlinecite{kataoka2004lattice}, the method generates a large and complex set of fixed coefficients.
The underlying principle for constructing these coefficients is unclear,  and the fixed model parameters limit its flexibility in simulating high-Mach-number flows.
To distinguish this approach from the moment-matching D2V16, and to emphasize its distributed characteristic, we refer to it as the D-D2V16 DVS [see Fig.~\ref{Fig01}(f)].

\section{FINITE DIFFERENCE SCHEMES}\label{IV}

The choice of FD schemes directly affects the stability, accuracy, and computational efficiency of numerical solutions. Commonly used FD schemes include central difference (CD), upwind difference, Lax-Wendroff (LW), non-oscillatory and non-free-parameter dissipation (NND)\cite{zhang1988non}, and weighted essentially non-oscillatory (WENO)\cite{jiang1996efficient}. The DBM does not prescribe a fixed temporal or spatial discretization scheme. Instead, the discretization scheme should be chosen based on the problem's specific characteristics, ensuring that the physical consistency of the DBM is maintained.

In this study, the time derivative of the discrete Boltzmann equation is discretized using the first-order forward Euler scheme. For spatial derivatives, five schemes are considered: CD, CD with third-order dispersion correction, LW, modified LW (MLW)\cite{gan2008two}, NND, and WENO.

\subsection{MLW scheme}\label{IV_A}

In 2008, Gan \textit{et al.} developed the MLW scheme by adding a third-order dispersion term and a second-order artificial viscosity term \cite{gan2008two}:
\begin{equation}\label{e8}
\begin{aligned}
	 \frac{{\partial {f_i}}}{{\partial t}} + {v_{i\alpha }}\frac{{\partial {f_i}}}{{\partial {r_\alpha }}} =  &- \frac{1}{\tau }\left( {{f_i} - f_i^{eq}} \right) + \frac{{{v_{i\alpha }}\left( {1 - c_{i\alpha }^2} \right)\Delta r_\alpha ^2}}{6}\frac{{{\partial ^3}{f_i}}}{{\partial r_\alpha ^3}} \\
	 & + {\theta _{\alpha I}}\left| {{\kappa _\alpha }} \right|\left( {1 - \left| {{\kappa _\alpha }} \right|} \right)\frac{{\Delta r_\alpha ^2}}{{2\Delta t}}\frac{{{\partial ^2}{f_i}}}{{\partial r_\alpha ^2}},
\end{aligned}
\end{equation}
where ${c_{i\alpha }} = {v_{i\alpha }}\frac{{\Delta t}}{{\Delta {r_\alpha }}}$, ${\kappa _\alpha } = {u_\alpha }\frac{{\Delta t}}{{\Delta {r_\alpha }}}$, ${\theta _{\alpha I}} = \lambda \left| {\frac{{{P_{\alpha I + 1}} - 2{P_{\alpha I}} + {P_{\alpha I - 1}}}}{{{P_{\alpha I + 1}} + 2{P_{\alpha I}} + {P_{\alpha I - 1}}}}} \right|$, with $\lambda$ as the artificial viscosity coefficient.

In the numerical implementation, the convection term  is discretized using the LW scheme, while the third-order dispersion and second-order artificial viscosity terms are discretized with the CD scheme.
The resulting discrete evolution equation of the distribution function is:
\begin{equation}\label{e9}
	\begin{aligned}
		f_{i,I}^{{\rm{new }}} = &{f_{i,I}} - \frac{{{c_{i\alpha }}}}{2}\left( {{f_{i,I + 1}} - {f_{i,I - 1}}} \right) - \frac{{\Delta t}}{\tau }\left( {{f_{i,I}} - f_{i,I}^{eq}} \right)\\
		&+ \frac{{c_{i\alpha }^2}}{2}\left( {{f_{i,I + 1}} - 2{f_{i,I}} + {f_{i,I - 1}}} \right)\\
		&+ \frac{{{c_{i\alpha }}\left( {1 - c_{i\alpha }^2} \right)}}{{12}}\left( {{f_{i,I + 2}} - 2{f_{i,I + 1}} + 2{f_{i,I - 1}} - {f_{i,I - 2}}} \right)\\
		&+ \frac{{{\theta _{\alpha I}}\left| {{\kappa _\alpha }} \right|\left( {1 - \left| {{\kappa _\alpha }} \right|} \right)}}{2}\left( {{f_{i,I + 1}} - 2{f_{i,I}} + {f_{i,I - 1}}} \right),
\end{aligned}
\end{equation}
where $I$ denotes the grid node in the $x$ or $y$ direction.
Retaining only the first row of Eq. (\ref{e9}) gives the CD scheme; while adding the second row results in the standard LW scheme.
The third and fourth rows correspond to the dispersion and artificial viscosity terms, respectively.
Numerical simulations show that the dispersion term effectively suppresses oscillations near discontinuities, while the artificial viscosity term enhances numerical stability \cite{gan2008two}.

\subsection{NND scheme}\label{IV_B}

The NND scheme is designed to suppress spurious oscillations in flow fields with steep gradients and strong discontinuities \cite{zhang1988non}.
After applying the NND scheme to discretize spatial derivatives, the evolution equation for the discrete distribution function becomes:
\begin{equation}\label{e10}
f_{i,I}^{{\rm{new }}} = {f_{i,I}} - {c_{i\alpha }}\left( {{h_{i,I + 1/2}} - {h_{i,I - 1/2}}} \right) - \frac{{\Delta t}}{\tau }\left( {{f_{i,I}} - f_{i,I}^{eq}} \right),
\end{equation}
where $h$ denotes the numerical flux, defined by:
\begin{equation}\label{e11}
{h_{i + 1/2}} = \left\{ {\begin{array}{*{20}{l}}
		{{f_{i,I}} + \frac{1}{2} \rm{minmod} \,\left[ \begin{array}{l}
				\left( {{f_{i,I + 1}} - {f_{i,I}}} \right),\\
				\left( {{f_{i,I}} - {f_{i,I - 1}}} \right)
			\end{array} \right]},&{{v_{i\alpha }} \ge  0},\\
		{{f_{i,I + 1}} - \frac{1}{2} \rm{minmod} \,\left[ \begin{array}{l}
				\left( {{f_{i,I + 2}} - {f_{i,I + 1}}} \right),\\
				\left( {{f_{i,I + 1}} - {f_{i,I}}} \right)
			\end{array} \right]},&{{v_{i\alpha }} < 0},
\end{array}} \right.
\end{equation}
with the limiter function:  $\rm{minmod} \left( {X,Y} \right) = \frac{1}{2}\min (\left| X \right|,\left| Y \right|)\left[ {Sign\left( X \right) + Sign\left( Y \right)} \right]$.

\subsection{WENO scheme}\label{IV_C}

The WENO scheme achieves high-order accuracy by performing a nonlinear weighted average over multiple discrete stencils \cite{jiang1996efficient}.
It effectively reduces numerical dissipation while suppressing spurious oscillations and maintaining high accuracy in smooth regions.

After spatial discretization with the WENO scheme, the evolution equation becomes:
\begin{equation}\label{e12}
f_{i,I}^{{\rm{new }}} = {f_{i,I}} - {c_{i\alpha }}\left( {{h_{i,I + 1/2}} - {h_{i,I - 1/2}}} \right) - \frac{{\Delta t}}{\tau }\left( {{f_{i,I}} - f_{i,I}^{eq}} \right).
\end{equation}

For ${v_{i\alpha }} \ge 0$, the numerical flux is expressed as:
\begin{equation}\label{e13}
{h_{i,I+1/2}}={\varpi _{1}}h_{i,I+1/2}^{1}+{\varpi _{2}}h_{i,I+1/2}^{2}+{%
\varpi _{3}}h_{i,I+1/2}^{3},
\end{equation}
with $h_{i,I+1/2}^{1}=\frac{1}{3}{F_{i,I-2}}-\frac{7}{6}{F_{i,I-1}}+\frac{{11}}{6}{%
F_{i,I}}$, $h_{i,I+1/2}^{2}=-\frac{1}{6}{F_{i,I-1}}+\frac{5}{6}{F_{i,I}}+\frac{1}{3}{%
F_{i,I+1}}$, and $h_{i,I+1/2}^{3}=\frac{1}{3}{F_{i,I}}+\frac{5}{6}{F_{i,I+1}}-\frac{1}{6}{%
F_{i,I+2}}$, ${\varpi _q} = {\tilde \omega _q}/\left( {{{\tilde \omega }_1} + {{\tilde \omega }_2} + {{\tilde \omega }_3}} \right)$, ${\tilde \varpi _q} = {\delta _q}/{\left( {{{10}^{ - 6}} + {\sigma _q}} \right)^2}$, ${\delta _1} = 0.1,{\delta _2} = 0.6,{\delta _3} = 0.3$. The smoothness indicators are: ${\sigma _1} = {\left( {{F_{i,I - 2}} - 2{F_{i,I - 1}} + {F_{i,I}}} \right)^2}
	+ {\left( {{F_{i,I - 2}} - 4{F_{i,I - 1}} + 3{F_{i,I}}} \right)^2}$,
${\sigma _2} = {\left( {{F_{i,I - 1}} - 2{F_{i,I}} + {F_{i,I + 1}}} \right)^2} + {\left( {{F_{i,I - 1}} - {F_{i,I + 1}}} \right)^2}$, and
${\sigma _3} = {\left( {{F_{i,I}} - 2{F_{i,I + 1}} + {F_{i,I + 2}}} \right)^2}
	+ {\left( {3{F_{i,I}} - 4{F_{i,I + 1}} + {F_{i,I + 2}}} \right)^2}$.
For ${v_{i\alpha }} < 0$, the above expressions are applied with mirrored indices: $\left( {I + a} \right) \to \left( {I - a} \right)$.

\section{VON NEUMANN STABILITY ANALYSIS}\label{V}

von Neumann stability analysis is a classical framework for assessing the stability of numerical methods.
In this framework, the solution of the FD equation is expressed as a Fourier series, from which the amplification matrix is derived.
A numerical method is stable if the absolute values of all eigenvalues of the amplification matrix do not exceed $1$.

The procedure for von Neumann stability analysis of kinetic models is as follows:
(i) Define the small perturbation as $\Delta {f_i}({r_\alpha },t) = {f_i}({r_\alpha },t) - \bar {f}_i^{eq} $, where $\bar {f}_i^{eq}$ is the global equilibrium which depends only on the mean density, velocity, and temperature.
(ii) Express the small perturbation as a series of complex exponents: $\Delta {f_i}\left( {{r_\alpha },t} \right) = F_i^t\exp \left( {{\bf{i}}{k_\alpha }{r_\alpha }} \right)$, where $F_i^t$ denotes the amplitude, ${\bf{i}}$ is the imaginary unit, and ${k_\alpha }$ is the wavenumber.
(iii) Substitute the small perturbation expression into the evolution equation of the discrete distribution function to obtain the amplification matrix: ${G_{ij}} = {{F_i^{t + \Delta t}}}/{{F_j^t}}$.

The general formulations of the amplification matrix ${G_{ij}}$ for the MLW, NND, and WENO schemes are presented below:
\onecolumngrid
\begin{equation}\label{e14}
	\begin{aligned}
			{G_{ij-\rm{MLW}}} = &\left( {1 - \frac{{\Delta t}}{\tau }} \right){\delta _{ij}} + \frac{{\Delta t}}{\tau }\frac{{\partial f_i^{eq}}}{{\partial {f_j}}} - \frac{{{c_{i\alpha }}}}{2}\left( {{{\rm{e}}^{{\bf{i}}{k_\alpha }\Delta {r_\alpha }}} - {{\rm{e}}^{ - {\bf{i}}{k_\alpha }\Delta {r_\alpha }}}} \right){\delta _{ij}} + \frac{{c_{i\alpha }^2}}{2}\left( {{{\rm{e}}^{{\bf{i}}{k_\alpha }\Delta {r_\alpha }}} - 2} \right.\left. { + {{\rm{e}}^{ - {\bf{i}}{k_\alpha }\Delta {r_\alpha }}}} \right){\delta _{ij}}\\
			&+ \frac{{{c_{i\alpha }}\left( {1 - c_{i\alpha }^2} \right)}}{{12}}\left( {{{\rm{e}}^{{\bf{i}}2{k_\alpha }\Delta {r_\alpha }}} - 2{{\rm{e}}^{{\bf{i}}{k_\alpha }\Delta {r_\alpha }}} + 2{{\rm{e}}^{ - {\bf{i}}{k_\alpha }\Delta {r_\alpha }}} - {{\rm{e}}^{ - {\bf{i}}2{k_\alpha }\Delta {r_\alpha }}}} \right){\delta _{ij}}\\
			&+ \frac{{{\theta _{\alpha I}}\left| {{\kappa _\alpha }} \right|\left( {1 - \left| {{\kappa _\alpha }} \right|} \right)}}{2}\left( {{{\rm{e}}^{{\bf{i}}{k_\alpha }\Delta {r_\alpha }}} - 2 + {{\rm{e}}^{ - {\bf{i}}{k_\alpha }\Delta {r_\alpha }}}} \right){\delta _{ij}},
	\end{aligned}
\end{equation}
\begin{equation}\label{e15}
	{G_{ij-\rm{NND}}} = \left( {1 - \frac{{\Delta t}}{\tau }} \right){\delta _{ij}} + \frac{{\Delta t}}{\tau }\frac{{\partial f_i^{eq}}}{{\partial {f_j}}} - {c_{i\alpha }}{\delta _{ij}}\left\{ {\begin{array}{*{20}{l}}
			{\left\{ \begin{array}{l}
					\left[ {1 + \frac{1}{2} \rm{minmod}\left( {{e^{{\bf{i}}{k_\alpha }\Delta {r_\alpha }}} - 1,{\kern 1pt} 1 - {e^{ - {\bf{i}}{k_\alpha }\Delta {r_\alpha }}}} \right)} \right]\\
					- \left[ {{e^{ - {\bf{i}}{k_\alpha }\Delta {r_\alpha }}} + \frac{1}{2} \rm{minmod} \left( {1 - {e^{ - {\bf{i}}{k_\alpha }\Delta {r_\alpha }}},{\kern 1pt} {e^{ - {\bf{i}}{k_\alpha }\Delta {r_\alpha }}} - {e^{ - 2{\bf{i}}{k_\alpha }\Delta {r_\alpha }}}} \right)} \right]
				\end{array} \right\},}&{{v_{i\alpha }} \ge 0,}\\
			{\left\{ \begin{array}{l}
					\left[ {{e^{{\bf{i}}{k_\alpha }\Delta {r_\alpha }}} - \frac{1}{2} \rm{minmod}\left( {{e^{2{\bf{i}}{k_\alpha }\Delta {r_\alpha }}} - {e^{{\bf{i}}{k_\alpha }\Delta {r_\alpha }}},{\kern 1pt} {e^{{\bf{i}}{k_\alpha }\Delta {r_\alpha }}} - 1} \right)} \right]\\
					- \left[ {1 - \frac{1}{2} \rm{minmod}\left( {{e^{{\bf{i}}{k_\alpha }\Delta {r_\alpha }}} - 1,{\kern 1pt} 1 - {e^{ - {\bf{i}}{k_\alpha }\Delta {r_\alpha }}}} \right)} \right]
				\end{array} \right\},}&{{v_{i\alpha }} < 0,}
	\end{array}} \right.
\end{equation}
\begin{equation}\label{e16}
{G_{ij - {\rm{WENO}}}} = \left( {1 - \frac{{\Delta t}}{\tau }} \right){\delta _{ij}} + \frac{{\Delta t}}{\tau }\frac{{\partial f_i^{eq}}}{{\partial {f_j}}} - {c_{i\alpha }}{\delta _{ij}}\left\{ {\begin{array}{*{20}{l}}
		{\left\{ {\begin{array}{*{20}{l}}
					{\left[ {\begin{array}{*{20}{l}}
								{0.1\left( {\frac{1}{3}{e^{ - 2{\bf{i}}{k_\alpha }\Delta {r_\alpha }}} - \frac{7}{6}{e^{ - {\bf{i}}{k_\alpha }\Delta {r_\alpha }}} + \frac{{11}}{6}} \right)}\\
								{ + 0.6\left( { - \frac{1}{6}{e^{ - {\bf{i}}{k_\alpha }\Delta {r_\alpha }}} + \frac{5}{6} + \frac{1}{3}{e^{{\bf{i}}{k_\alpha }\Delta {r_\alpha }}}} \right)}\\
								{ + 0.3\left( {\frac{1}{3} + \frac{5}{6}{e^{{\bf{i}}{k_\alpha }\Delta {r_\alpha }}} - \frac{1}{6}{e^{2{\bf{i}}{k_\alpha }\Delta {r_\alpha }}}} \right)}
						\end{array}} \right]}\\
					{ - \left[ {\begin{array}{*{20}{l}}
								{0.1\left( {\frac{1}{3}{e^{ - 3{\bf{i}}{k_\alpha }\Delta {r_\alpha }}} - \frac{7}{6}{e^{ - 2{\bf{i}}{k_\alpha }\Delta {r_\alpha }}} + \frac{{11}}{6}{e^{ - {\bf{i}}{k_\alpha }\Delta {r_\alpha }}}} \right)}\\
								{ + 0.6\left( { - \frac{1}{6}{e^{ - 2{\bf{i}}{k_\alpha }\Delta {r_\alpha }}} + \frac{5}{6}{e^{ - {\bf{i}}{k_\alpha }\Delta {r_\alpha }}} + \frac{1}{3}} \right)}\\
								{ + 0.3\left( {\frac{1}{3}{e^{ - {\bf{i}}{k_\alpha }\Delta {r_\alpha }}} + \frac{5}{6} - \frac{1}{6}{e^{{\bf{i}}{k_\alpha }\Delta {r_\alpha }}}} \right)}
						\end{array}} \right]}
			\end{array}} \right\},}&{{v_{i\alpha }} \ge 0,}\\
		{\left\{ {\begin{array}{*{20}{l}}
					{\left[ {\begin{array}{*{20}{l}}
								{0.1\left( {\frac{1}{3}{e^{3{\bf{i}}{k_\alpha }\Delta {r_\alpha }}} - \frac{7}{6}{e^{2{\bf{i}}{k_\alpha }\Delta {r_\alpha }}} + \frac{{11}}{6}{e^{{\bf{i}}{k_\alpha }\Delta {r_\alpha }}}} \right)}\\
								{ + 0.6\left( { - \frac{1}{6}{e^{2{\bf{i}}{k_\alpha }\Delta {r_\alpha }}} + \frac{5}{6}{e^{{\bf{i}}{k_\alpha }\Delta {r_\alpha }}} + \frac{1}{3}} \right)}\\
								{ + 0.3\left( {\frac{1}{3}{e^{{\bf{i}}{k_\alpha }\Delta {r_\alpha }}} + \frac{5}{6} - \frac{1}{6}{e^{ - {\bf{i}}{k_\alpha }\Delta {r_\alpha }}}} \right)}
						\end{array}} \right]}\\
					{ - \left[ {\begin{array}{*{20}{l}}
								{0.1\left( {\frac{1}{3}{e^{2{\bf{i}}{k_\alpha }\Delta {r_\alpha }}} - \frac{7}{6}{e^{{\bf{i}}{k_\alpha }\Delta {r_\alpha }}} + \frac{{11}}{6}} \right)}\\
								{ + 0.6\left( { - \frac{1}{6}{e^{{\bf{i}}{k_\alpha }\Delta {r_\alpha }}} + \frac{5}{6} + \frac{1}{3}{e^{ - {\bf{i}}{k_\alpha }\Delta {r_\alpha }}}} \right)}\\
								{ + 0.3\left( {\frac{1}{3} + \frac{5}{6}{e^{ - {\bf{i}}{k_\alpha }\Delta {r_\alpha }}} - \frac{1}{6}{e^{ - 2{\bf{i}}{k_\alpha }\Delta {r_\alpha }}}} \right)}
						\end{array}} \right]}
			\end{array}} \right\},}&{{v_{i\alpha }} < 0.}
\end{array}} \right.
\end{equation}

\twocolumngrid

In these equations, the derivative $\frac{\partial f_i^{eq}}{\partial f_j}=\frac{\partial f_i^{eq}}{\partial \rho} \frac{\partial \rho}{\partial f_j} + \frac{\partial f_i^{eq}}{\partial T} \frac{\partial T}{\partial f_j} + \frac{\partial f_i^{eq}}{\partial u_\alpha} \frac{\partial u_\alpha}{\partial f_j}$, and $\delta_{ij}$ denotes the Kronecker delta.
The stability criterion requires that the maximum modulus of all eigenvalues ${|\omega|_{\max}}$ of the amplification matrix satisfies ${\left| \omega  \right|_{\max }} \le 1$.
If this condition holds for all wavenumbers, the numerical method is stable.

\section{MODEL STABILITY ANALYSIS AND VERIFICATION}\label{VI}

\onecolumngrid

\begin{table}[htbp]
	\centering
	\caption{Influencing factors on DBM simulation stability. The first three columns are ranked by priority (red for highest, followed by blue and green). The remaining columns reflect each factor's relative influence on stability, with red indicating the strongest effect, followed by blue and green.}
	\label{tableI}\centering
	\renewcommand{\arraystretch}{1.4}
	
	\begin{tabular}{|>{\centering\arraybackslash}m{5.3cm}
			|>{\centering\arraybackslash}m{1.3cm}
			|>{\centering\arraybackslash}m{4.1cm}
			|>{\centering\arraybackslash}m{2.8cm}
			|>{\centering\arraybackslash}m{3.4cm}|}
		\hline
		\textbf{Phase-space discretization approaches} & \textbf{DVS} & \textbf{Spatial discretization schemes} & \textbf{Initial macroscopic quantities} & \textbf{Model parameters} \\
		\hline
		
		\multirow{3}{*}{\textcolor[rgb]{1,0,0}{Moment-matching method}}
		& \textcolor[rgb]{0.06, 0.68, 0.34}{D2V16} & \textcolor[rgb]{0.06, 0.68, 0.34}{Central difference} & \multirow{2}{*}{\textcolor[rgb]{1,0,0}{Velocity}} & \multirow{3}{*}{\textcolor[rgb]{0,0,1}{Relative time step}} \\
		\cline{2-3}
		& \textcolor[rgb]{0,0,1}{D2V25} & \textcolor[rgb]{0.06, 0.68, 0.34}{Lax-Wendroff} & & \\
		\cline{2-4}
		& \textcolor[rgb]{1,0,0}{D2V26} & \textcolor[rgb]{0.06, 0.68, 0.34}{Central difference + dispersion} & \multirow{2}{*}{\textcolor[rgb]{0,0,1}{Temperature}} & \\
		\cline{1-3} \cline{5-5}
		
		\multirow{2}{*}{\centering\shortstack{\textcolor[rgb]{0.06, 0.68, 0.34}{Globally unified} \\ \textcolor[rgb]{0.06, 0.68, 0.34}{expansion-coefficient method}}}
		& \textcolor[rgb]{0.06, 0.68, 0.34}{D2V19} & \textcolor[rgb]{0.06, 0.68, 0.34}{Modified Lax-Wendroff} & & \multirow{3}{*}{\textcolor[rgb]{0,0,1}{Time-space step ratio}} \\
		\cline{2-4}
		& \textcolor[rgb]{0.06, 0.68, 0.34}{D2V33} & \textcolor[rgb]{1,0,0}{NND} & \multirow{2}{*}{\textcolor[rgb]{0.06, 0.68, 0.34}{Density}} & \\
		\cline{1-3}
		
		\textcolor[rgb]{0,0,1}{Distributed weighting-coefficient method}
		& \textcolor[rgb]{0.06, 0.68, 0.34}{D-D2V16} & \textcolor[rgb]{0,0,1}{WENO} & & \\
		\hline
	\end{tabular}
\end{table}

\twocolumngrid

This section presents a comprehensive analysis of the numerical stability of the DBM and related kinetic models. The main objectives and corresponding methodologies are outlined as follows.

\textbf{(i) Stability analysis based on ${\left| \omega \right|_{\max }}$:}
Using von Neumann stability analysis, we systematically examine the impact of several key factors on the numerical stability of DBMs, including spatial discretization schemes, initial macroscopic quantities, relative time step, and the ratio of temporal to spatial discretization steps, as summarized in Table~\ref{tableI}.
The different colors represent the recommended priority and relative significance of each influencing factor. These priorities are determined through the analysis in Sec.~\ref{VI} and further discussed in Sec.~\ref{VII}.
\emph{It is worth noting that the effect of phase-space discretization—specifically, the construction of the equilibrium distribution function—is inherently coupled with all of these factors. Moreover, aside from the discretization scheme itself, each factor contributes to varying degrees of TNE intensity. Consequently, the impacts of phase-space discretization and TNE intensity on model stability are intrinsically embedded throughout the entire analysis.}

\textbf{(ii) Identification of stability-controlling factors and construction of stability-phase diagrams:}
We identify the dominant factors governing DBM stability and construct stability-phase diagrams that delineate the stability control capacities of various DBMs across different modeling levels and flow conditions.

\textbf{(iii) Morphological quantification of stability-controlling performance:}
Morphological analysis is employed to quantify the geometric features of the stability-phase diagrams over the full wavenumber space. Furthermore, model stability probability curves are extracted to provide a statistical characterization of the stability performance.

\textbf{(iv) Numerical verification and evaluation of the stability-phase diagrams:}
The reliability of the proposed stability-phase diagrams is validated through numerical simulations. In addition, the effects of DVS on model stability and accuracy are examined by evaluating the consistency between numerical and analytical TNE solutions, as well as the continuity of the distribution function.

\subsection{Model stability analysis from the perspective of ${\left| \omega  \right|_{\max }}$}\label{VI_A}

\subsubsection{Influence of spatial discretization schemes}\label{VI_A_1}

\begin{figure*}[htbp]
	\centering
	\includegraphics[width=0.8\textwidth]{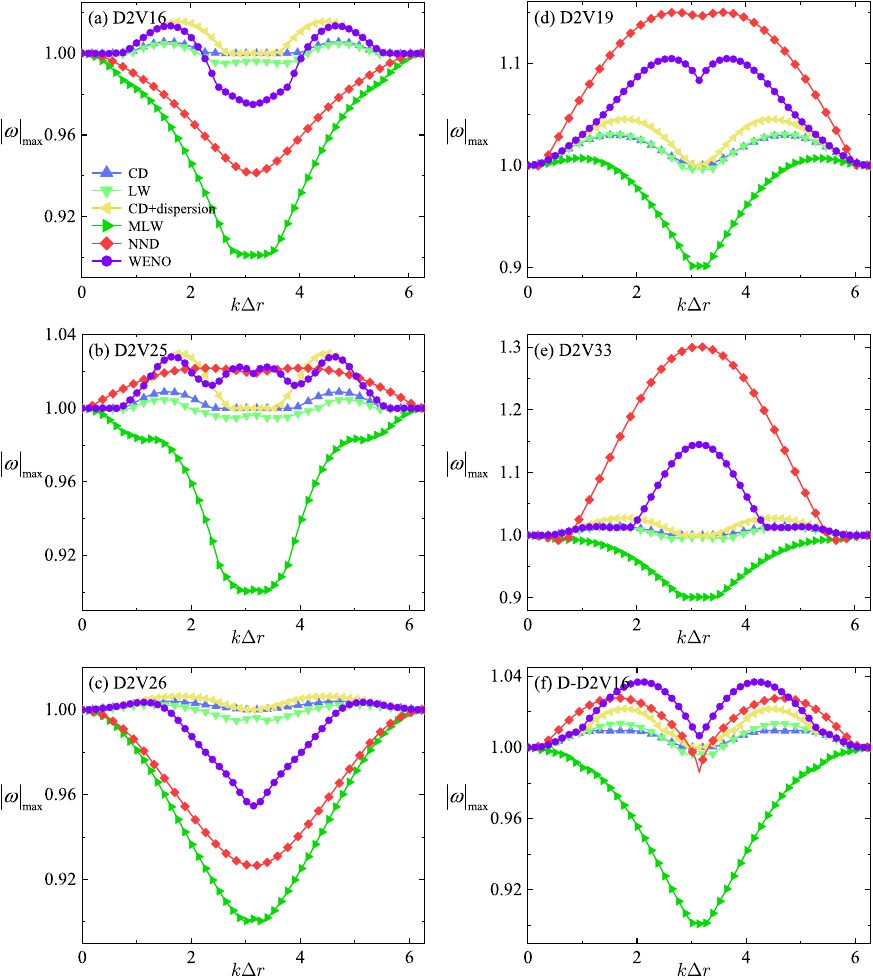}
	\caption{\centering{Influence of spatial discretization schemes on the stability of various kinetic models.}}
	\label{Fig02}
\end{figure*}

For the problem considered, the macroscopic quantities are set as ${(\rho ,T,{u_x},{u_y}) = (1,1,5,0)}$. The common parameters for all models are ${\Delta r = 10^{-3}}$, ${\Delta t = 10^{-5}}$, ${\tau = 10^{-4}}$, ${\gamma = 1.67}$, and ${\lambda = 1}$. The specific adjustable parameters for each model are as follows: for the moment-matching method, ${c = 2.628}$ and ${\eta_0 = 1.546}$.
Here $c$ represents the magnitude of the fundamental velocity in the DVS, as shown in Fig.~\ref{Fig01}(a) where $|\bm{v}_1| = c$. The vector $\bm{\eta}_i$ is introduced to adjust the specific-heat ratio, and $\eta_0$ denotes the magnitude of its base velocity \cite{gan2013lattice, gan2018discrete};
for the globally unified-coefficient method, ${v_1 = 1}$, ${v_2 = 2}$, ${v_3 = 3}$, and ${v_4 = 4}$; and for the distributed weighting-coefficient method, ${v_1 = 1}$, ${v_2 = 6}$, ${v_3 = 2}$, and ${v_4 = 3}$. Figure~\ref{Fig02} illustrates the effect of different spatial discretization schemes on the stability of various kinetic models. The key findings are summarized as follows:

(i) For all kinetic models, the CD scheme is highly unstable due to insufficient numerical dissipation, which fails to suppress numerical oscillations.
Although the dispersion term can mitigate oscillations near discontinuities \cite{gan2008two}, it substantially amplifies the maximum eigenvalue ${\left| \omega  \right|_{\max }}$ of the amplification matrix.
In contrast, the LW scheme improves stability by introducing a viscous term, yet this effect is insufficient to overcome the inherent instability of the CD scheme.

(ii) The MLW scheme, incorporating artificial viscosity, substantially reduces ${\left| \omega \right|_{\max }}$, particularly at $k = \pi/\Delta r$, and thus enhances the stability of all models, demonstrating broad applicability. The Laplacian term ${{\bm{\nabla}^2}{f_i}}$ in the artificial viscosity effectively smooths the spatial gradients of ${f_i}$ and suppresses numerical oscillations, while the shock-detection coefficient $\theta_{\alpha}$ ensures that additional dissipation is introduced only near strong discontinuities. Previous studies have shown that this scheme enables stable LBM simulations for shock waves with ${Ma} > 30$ \cite{gan2008two}, confirming its robustness. Nevertheless, since the introduction of artificial viscosity is based on the distribution function, it has a more fundamental impact compared to the macroscopic-based artificial viscosity in traditional CFD. Not only does it inevitably distort the flow field, but it also alters the constitutive relations, thereby compromising the accuracy of physical solutions, particularly near mesoscopic structures where nonequilibrium effects are significant. Therefore, artificial viscosity should be minimized as much as possible in practical simulations, while ensuring numerical stability is maintained.

(iii) When employing the NND scheme, the D2V16 and D2V26 DBMs based on the moment-matching method exhibit superior numerical stability compared with other phase-space discretization schemes or equilibrium distribution function construction methods. This improvement stems from the fact that the moment-matching method computes $f_i^{eq}$ via matrix inversion, ensuring the strict preservation of kinetic moment relations. Such physical consistency markedly enhances numerical stability. In contrast, both the globally unified expansion-coefficient method and the distributed weighting-coefficient method approximate $f_i^{eq}$ through a finite-term Taylor expansion of ${\left( {\mathbf{v}_{ki}} \cdot \mathbf{u} \right)^n} / (n!{T^n})$, where truncating higher-order terms deteriorates stability at high flow velocities. These results highlight the significant advantage of the moment-matching method in terms of stability. However, stability also depends on the specific configuration of DVS; for example, the D2V25 model exhibits inferior stability compared with the D2V16 and D2V26 models. A detailed analysis of the effects of DVS on both numerical stability and the achievement of the model's physical accuracy is presented in Sec.~\ref{VI_D}.

(iv) The WENO scheme exhibits relatively poor stability across all models. This is attributed to the fact that WENO utilizes more stencils (spatial nodes) to compute spatial derivatives, which enhances numerical accuracy but reduces numerical dissipation, thereby compromising stability. However, for the D2V26 model, ${\left| \omega \right|_{\max }}$ obtained using the WENO scheme approaches 1, suggesting that appropriate adjustment of other model parameters may stabilize the model further. This observation once again highlights the superior stability performance of the moment-matching method.

\subsubsection{Influence of flow velocity}\label{VI_A_2}

\begin{figure*}[htbp]
	\centering
	\includegraphics[width=0.8\textwidth]{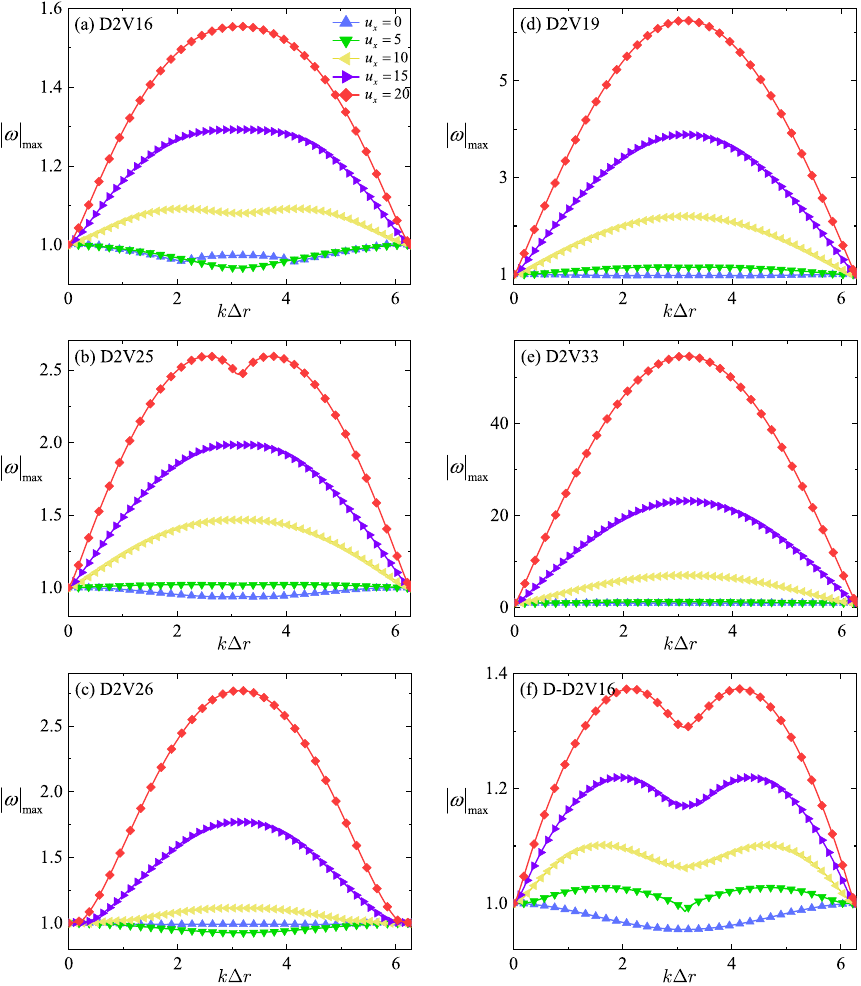}
	\caption{\centering{Influence of flow velocity on the stability of various kinetic models.}}
	\label{Fig03}
\end{figure*}

The influence of flow velocity on the numerical stability of various kinetic models is further examined.
As shown in Fig.~\ref{Fig02}, the MLW scheme compromises the model's accuracy, while the WENO scheme suffers from poor numerical stability.
Therefore, the NND scheme is adopted for spatial discretization scheme, while all other parameters remain the same as in Fig.~\ref{Fig02}.
Figure~\ref{Fig03} illustrates the effects of initial flow velocity on the stability of various kinetic models, from which the following conclusions are drawn:

(i) As the flow velocity increases, ${\left| \omega \right|_{\max }}$ rises rapidly for all models, leading to a significant decline in numerical stability. The increase in flow velocity raises the Mach number, which in turn enhances system compressibility, nonequilibrium driving forces, nonequilibrium intensity, and nonlinear effects.
Initial numerical instabilities accumulate during iteration, potentially leading to oscillations or even divergence in the numerical solution.  This suggests that improving model stability becomes increasingly important as discrete and nonequilibrium effects intensify.

(ii) The average growth rate of ${\left| \omega \right|_{\max }}$ with flow velocity highlights significant differences in the sensitivity of different DVSs to numerical instabilities as the Mach number increases: ${S_{{\rm{D2V16}}}} < {S_{{\rm{D2V26}}}} < {S_{{\rm{D2V25}}}} < {S_{{\rm{D2V19}}}} < {S_{{\rm{D2V33}}}}$. The D2V19 and D2V33 models, based on the globally unified expansion-coefficient method, exhibit rapid deterioration in stability as the Mach number increases. These models rely on finite-term Taylor expansions of velocity. Specifically, the D2V19 model retains third-order velocity terms, whereas the D2V33 model includes velocity terms up to the fourth order. At higher flow velocities, higher-order velocity terms become increasingly important for accurately capturing system behavior. In contrast, models based on the moment-matching method, which directly solve $f_i^{eq}$, avoid this limitation and thus exhibit greater resistance to stability degradation at higher flow velocities.

(iii) The comparisons of ${S_{{\rm{D2V19}}}} < {S_{{\rm{D2V33}}}}$ and ${S_{{\rm{D2V16}}}} < {S_{{\rm{D2V25}}}}({S_{{\rm{D2V26}}}})$ indicate that higher-order models are more sensitive to numerical instabilities. This sensitivity arises from their stronger nonlinearity, inclusion of additional multiscale effects, higher-order derivative terms, more complex phase-space discretization schemes, and stronger coupling between phase-space and spatiotemporal discretizations. Therefore, developing stability-enhancing strategies for higher-order models at high Mach numbers is essential.

(iv) \emph{Multiscale effects act as a double-edged sword for simulation stability.
High-order DBMs enhance the ability to capture multiscale effects, by retaining higher-order kinetic moment relations during coarse-grained modeling \cite{gan2018discrete}. However, this also introduces additional nonlinearity into the DEDF due to the inclusion of higher-order velocity terms. As the flow velocity and Mach number increase, the system becomes more compressible and the nonequilibrium effects intensify, which in turn amplifies nonlinearities and increases the risk of numerical instability. In contrast, low-order models exhibit less nonlinearity and a limited capacity to capture multiscale effects, making them less sensitive to velocity-induced instabilities compared to high-order models. While low-order models provide greater numerical stability, their limited physical accuracy makes them unsuitable for high-accuracy simulations and they should be avoided in such cases.}

\subsubsection{Influence of temperature}\label{VI_A_3}

\begin{figure*}[htbp]
	\centering
	\includegraphics[width=0.8\textwidth]{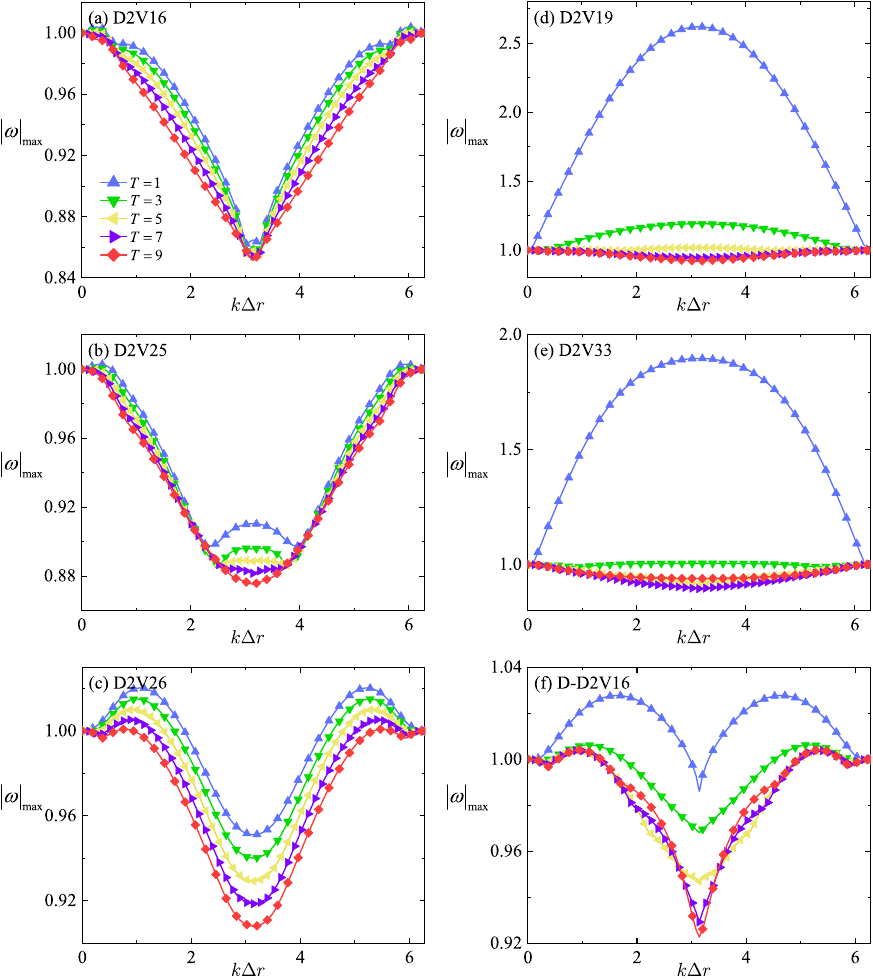}
	\caption{\centering{Influence of temperature on the stability of various kinetic models.}}
	\label{Fig04}
\end{figure*}

The influence of temperature on the model's numerical stability is examined similarly.
The adjustable parameters are updated as follows: for the moment-matching method, ${c = 5}$ and ${\eta_0 = 10}$; for the globally unified expansion-coefficient method, ${v_1 = 4}$, ${v_2 = 5}$, ${v_3 = 6}$, and ${v_4 = 7}$.
All other parameters are consistent with those in Fig.~\ref{Fig02}, and the NND scheme is also used.
Figure~\ref{Fig04} presents the influence of temperature on the stability of different kinetic models. The key findings are summarized below:

(i) Increasing temperature leads to a gradual decrease in ${\left| \omega \right|_{\max }}$, indicating enhanced numerical stability. This is attributed to the elevated speed of sound at higher temperatures, which in turn reduces the Mach number and TNE intensity.

(ii) A comparison of the average variation rates of ${\left| \omega \right|_{\max }}$ in Figs.~\ref{Fig03} and \ref{Fig04} reveals that flow velocity has a stronger impact on model stability than temperature, and this difference becomes more pronounced with increasing Mach number.
This difference arises from two main factors:
At the macroscopic level, both velocity and temperature influence Mach number, $Ma = u/\sqrt{\gamma R T}$, where $Ma$ is proportional to flow velocity but inversely proportional to the square root of temperature. For equal variation amplitudes, changes in velocity cause a larger shift in $Ma$, thereby exerting a stronger influence on numerical stability.
At the mesoscopic level, moment relations contain more velocity-dependent terms than temperature-dependent ones, making velocity the dominant contributor to ${f_i^{eq}}$ and, consequently, to numerical stability.

(iii) It is further observed that kinetic models constructed using the moment-matching method exhibit superior numerical stability compared with those developed by the other two approaches. At low temperatures (\(T \leq 1\)), the D2V19 and D2V33 models become highly unstable, with \(|\omega|\) peaking at \(k = \pi /\Delta r \). This behavior is in stark contrast not only to
the other models but also to the high-temperature ($T > 5$) performance of D2V19 and D2V33 models, which display minimum \(|\omega|\) at the same wavenumber.

\subsubsection{Influence of density}\label{VI_A_4}

\begin{figure*}[htbp]
	\centering
	\includegraphics[width=0.8\textwidth]{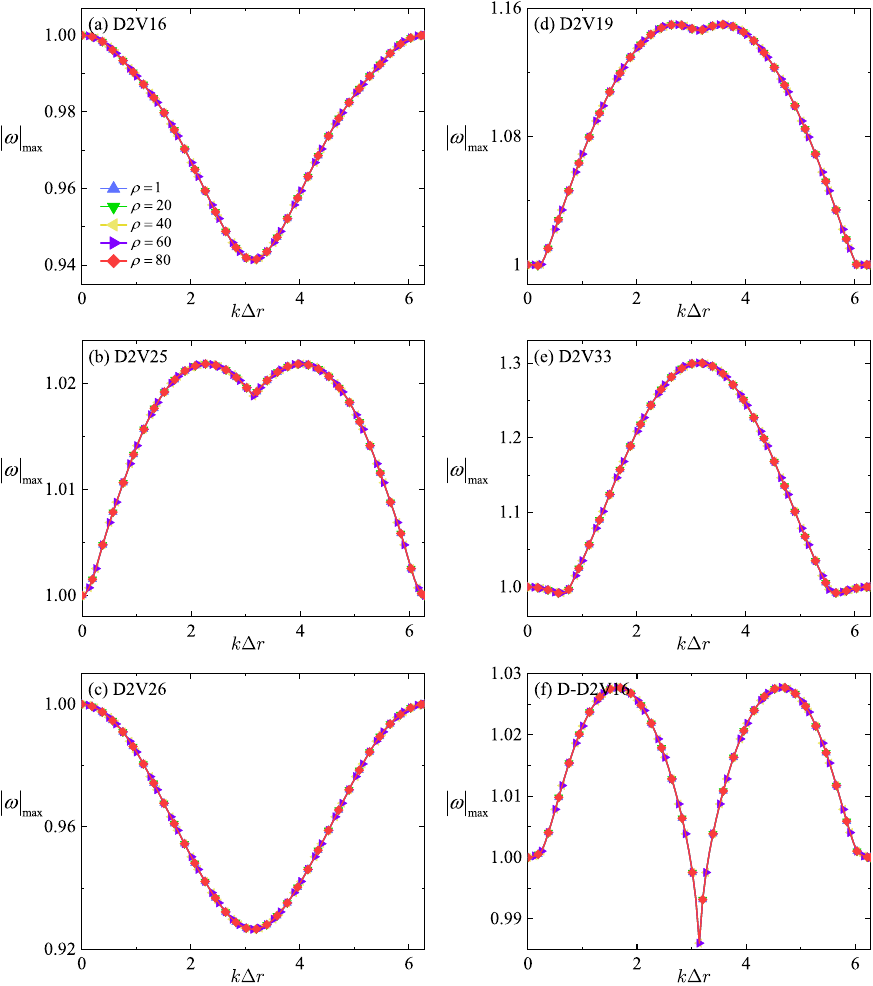}
	\caption{\centering{Influence of density on the stability of various kinetic models.}}
	\label{Fig05}
\end{figure*}

Figure~\ref{Fig05} illustrates the influence of density on the stability of various kinetic models with all other parameters consistent with those in Fig.~\ref{Fig02}. The following conclusions can be drawn:

(i) For all kinetic models, \(|\omega|_{\max}\) remains nearly constant as density increases. Numerically, this invariance stems from the fact that density can be normalized in the amplification matrix \(G_{ij}\). Physically, density does not directly affect the Mach number, and it enters the equilibrium distribution function \(f_i^{eq}\) only as a linear factor.

(ii) Comparing the average variation rates of \(|\omega|_{\max}\) across Figs. \ref{Fig03}–\ref{Fig05} shows that the influence of macroscopic quantities on model stability follows the order: \(\rho \ll T < u\).

\subsubsection{Influence of relative time step}\label{VI_A_5}

\begin{figure*}[htbp]
	\centering
	\includegraphics[width=0.8\textwidth]{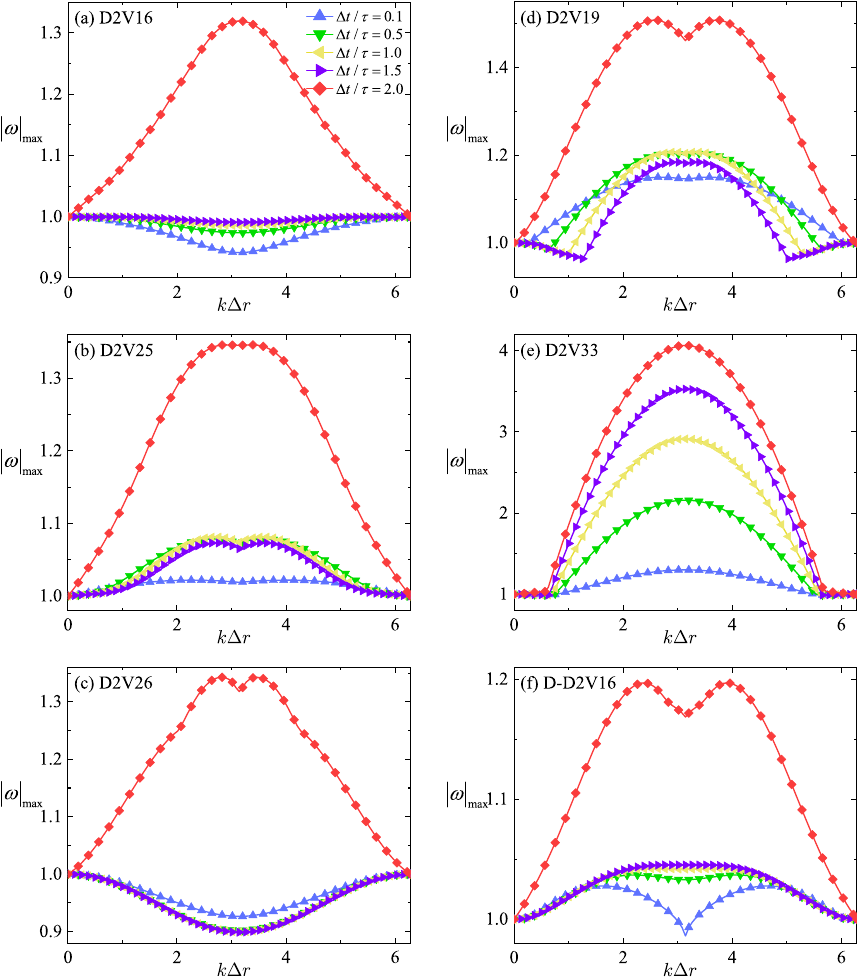}
	\caption{\centering{Influence of relative time step on the stability of various kinetic models.}}
	\label{Fig06}
\end{figure*}

von  Neumann stability analysis reveals that the amplification matrix \(G_{ij}\) depends on the ratio \(\Delta t/\tau\) of the time step \(\Delta t\) to the relaxation time \(\tau\). Physically, the relative time step \(\Delta t/\tau\) characterizes the competition between flow and collision-induced nonequilibrium. Figure~\ref{Fig06} shows how \(\Delta t/\tau\) affects the stability of different kinetic models. All other parameters follow Fig.~\ref{Fig02}, and the NND scheme is employed. The main conclusions are as follows:

(i) For all kinetic models, an increase in \(\Delta t/\tau\) gradually raises \(|\omega|_{\max}\), reducing numerical stability. This mismatch occurs because the temporal resolutions of flow evolution and the relaxation process become inconsistent. Consequently, the numerical scheme fails to accurately capture nonequilibrium effects within one time step. The reduced accuracy weakens the model's capability to resolve nonlinear and cross-scale dynamics, ultimately causing numerical oscillations or divergence.

(ii) At \(\Delta t/\tau = 1.5\), the D2V16 and D2V26 models remain stable, but the D2V19, D2V33, and D-D2V16 models lose stability even at \(\Delta t/\tau = 0.1\). This result indicates that the first two models have significantly higher stability thresholds \((\Delta t/\tau)_{\mathrm{th}}\), primarily because of the superior moment-preserving capability provided by the moment-matching method. For a fixed relaxation time, a lower stability threshold \((\Delta t/\tau)_{\mathrm{th}}\) demands a smaller \(\Delta t\), thus decreasing computational efficiency. Notably, the D2V25 model also exhibits a low stability threshold, whereas the D2V26 model retains a higher threshold. This suggests that selecting an appropriate DVS can substantially enhance stability, even among kinetic models of the same order.

\subsubsection{Influence of time-space step ratio}\label{VI_A_6}

\begin{figure*}[htbp]
	\centering
	\includegraphics[width=0.8\textwidth]{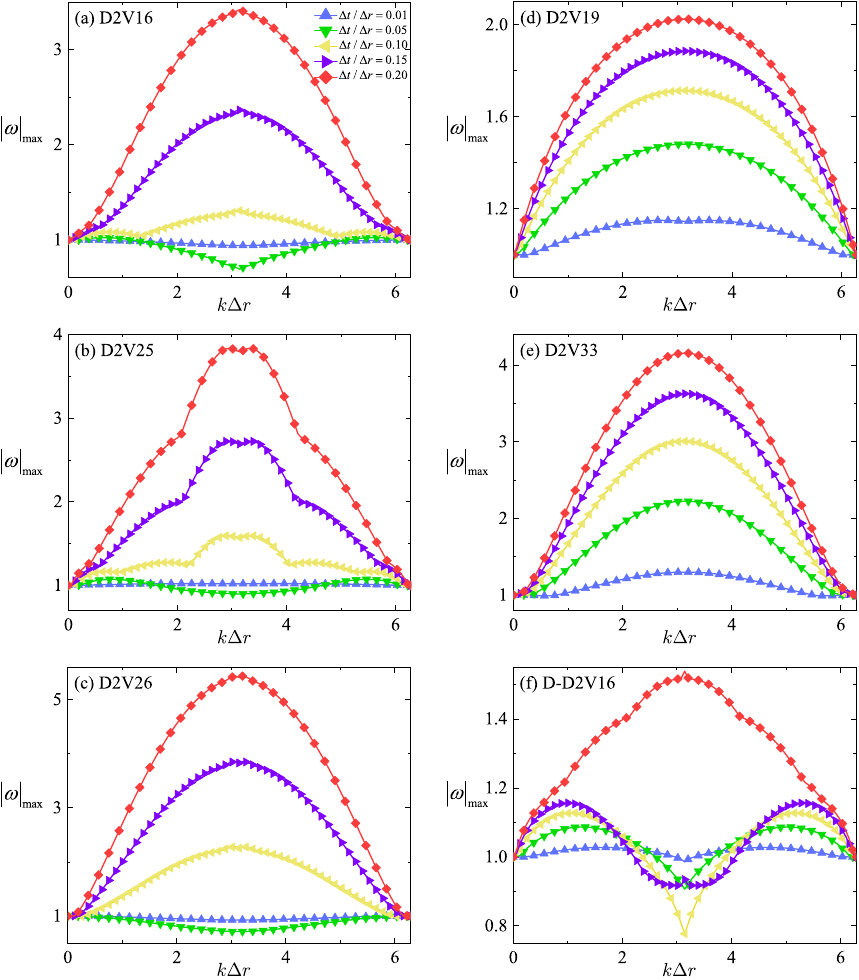}
	\caption{\centering{Influence of time-space step ratio on the stability of various kinetic models.}}
	\label{Fig07}
\end{figure*}

Similarly, the amplification matrix \(G_{ij}\) depends only on the ratio of the time step \(\Delta t\) to the spatial step \(\Delta r\). Physically, the ratio \(\Delta t/\Delta r\) is proportional to the CFL number, defined as \( v\Delta t/\Delta r \), which measures the distance traveled by a particle within one time step relative to the spatial grid size. Figure~\ref{Fig07} illustrates how the time-space step ratio \(\Delta t/\Delta r\) influences the stability of different kinetic models. The key findings are as follows:

(i) For all kinetic models, increasing \(\Delta t/\Delta r\) results in a higher CFL number and a corresponding increase in \(|\omega|_{\max}\). A larger CFL number leads numerical information to propagate faster than the spatial grid can accurately resolve, causing accumulated errors and eventual instability. As the CFL number increases, numerical oscillations become more pronounced, especially at high Mach numbers where stronger nonequilibrium and nonlinear effects intensify the destabilizing effects.

(ii) At \(\Delta t/\Delta r = 0.01\), the three moment-matching-based models remain stable, whereas the other three models become unstable. This indicates that moment-matching-based models possess a higher stability threshold \((\Delta t/\Delta r)_{\mathrm{th}}\), attributed to their superior preservation of moments and more accurate phase-space discretization. For a fixed spatial step, a lower stability threshold necessitates a smaller time step, substantially decreasing computational efficiency.

\subsection{Analysis of model stability regulation capability under fixed wavenumber }\label{VI_B}

von Neumann stability analysis indicates that \(\left| \omega \right|_{\max}\) is a multivariable function dependent on phase-space discretization, spatiotemporal discretization, initial conditions, and model parameters. Due to this complexity, a coarse-graining approach is employed in Sec.~\ref{VI_A}, where specific parameters are held constant to identify the primary stability factors.

For the six kinetic models examined, the discretization of the equilibrium distribution function introduces adjustable parameters: specifically, \(c\) and \(\eta_0\) for models D2V16, D2V25, and D2V26, and discrete velocities \(v_i\) for D2V19, D2V33, and D-D2V16. Although the physical selection criterion for parameters \(c\) and \(\eta_0\) merely requires invertibility of matrix \(\mathbf{C}\), numerical experiments demonstrate their significant influence on both
numerical stability and accuracy.
Appropriately chosen \(c\) and \(\eta_0\) enhance the continuity and smoothness of \(f_i^{eq}\), thereby improving numerical stability by reducing spatial oscillations in \(\bm{\nabla} f_i\).

This subsection explores the effects of parameters \((c, \eta_0)\) on model stability across an extensive parameter space. The objective is to define a stability domain—termed the \((c, \eta_0)\) stability domain—within which the model remains stable. The area of this domain is positively correlated with the model's stability regulation capability.

Based on findings in Sec.~\ref{VI_A}, two key insights guide the present analysis:

(i) The maximum eigenvalue modulus \(\left|\omega\right|_{\max}\) typically peaks at a particular wavenumber for each model. Thus, stability at the most unstable wavenumber provides a sufficient condition to define the stable parameter domain.

(ii) The moment-matching discretization significantly enhances stability. Furthermore, using moment-matching alongside the NND spatial discretization scheme provides superior stability compared with other approaches.

Consequently, the stability analysis below exclusively considers the three moment-matching models (D2V16, D2V25, D2V26), employs the NND scheme, and fixes the wavenumber \(k\) at its most unstable value. Unless otherwise specified, additional model parameters and macroscopic quantities remain consistent across subsections: \((\rho, T, u_x, u_y) = (1,1,5,0)\), \(\Delta r = 10^{-3}\), \(\Delta t = 10^{-5}\), \(\tau = 10^{-4}\), \(k=\pi/\Delta r\), and \(\gamma = 1.67\).

In this study, $c$ and ${\eta _0}$ are free parameters. By introducing additional moment relations satisfied by the DEDF, analytical expressions for $c$ and ${\eta _0}$ can be derived. This allows for the establishment of their relationships with macroscopic quantities, which can be used to determine the values of $c$ and ${\eta _0}$ based on macroscopic values. Moreover, the empirical guideline \cite{gan2013lattice,gan2018discrete} proposed by Gan \textit{et al.} also indicates that when $c$ and $\eta_0$ are set within $(0, 50)$, the corresponding Mach number falls within $(0, 38.69)$. Considering these findings and the simulation cases designed in this section, we adopt $c \in (0, 50)$ and $\eta_0 \in (0, 50)$ as the parameter ranges.

\subsubsection{Influence of flow velocity}\label{VI_B_1}

\begin{figure*}[htbp]
	\centering
	\includegraphics[width=0.75\textwidth]{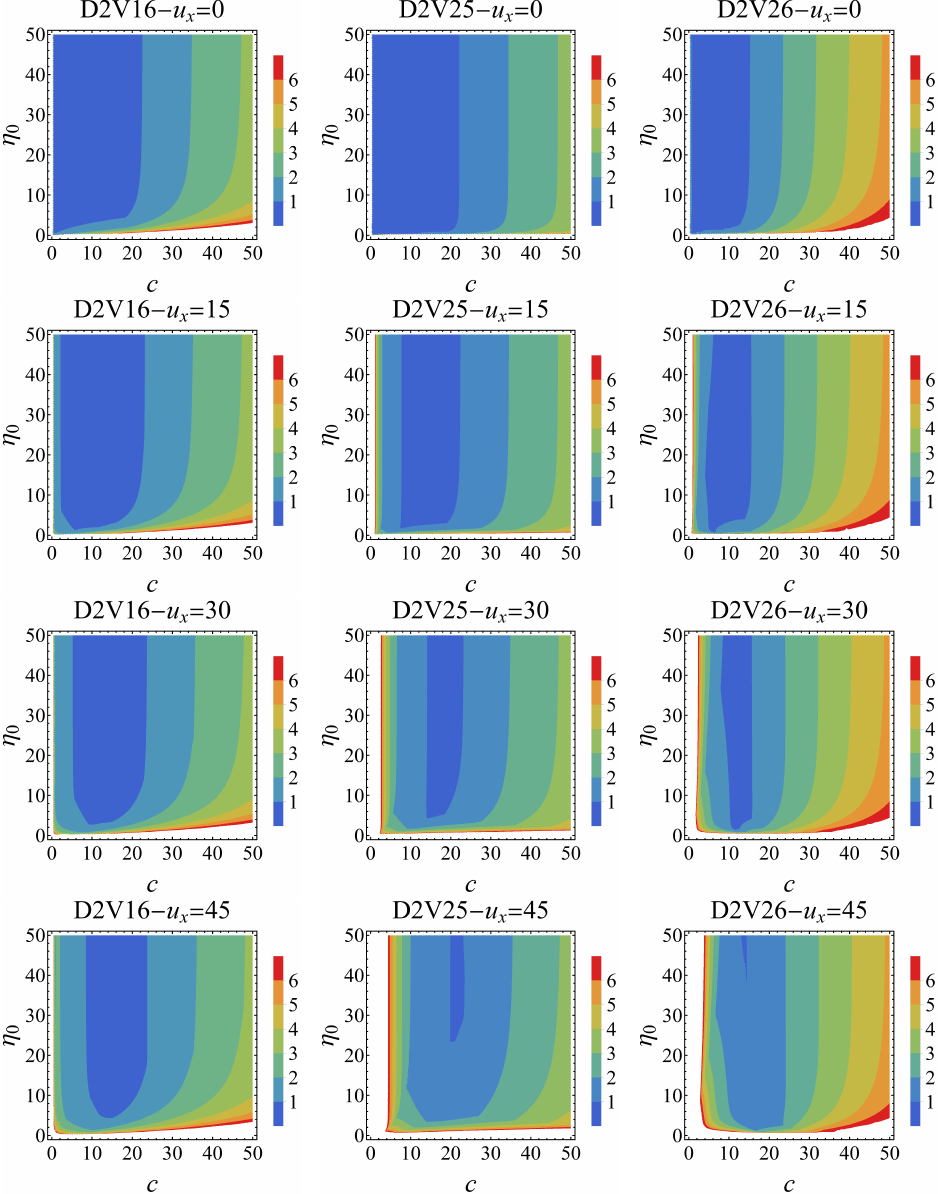}
	\caption{\centering{Influence of flow velocity on the stability-phase diagrams of various kinetic models at \(k=\pi/\Delta r\).}}
	\label{Fig08}
\end{figure*}

The initial flow velocities considered are $u_{x} = {0, 15, 30, 45}$. Figure~\ref{Fig08} presents the stability-phase diagrams, showing how initial flow velocity influences the stability of various DBMs. The key findings are as follows:

(i) For all kinetic models, increasing flow velocity reduces the (\(c, \eta_0\)) stability domain area, weakening the model’s stability regulation capability. This result aligns with the analysis in Sec.~\ref{VI_A_2}, where higher Mach numbers intensify compressibility, nonequilibrium, and nonlinear effects, thus impairing stability regulation.

(ii) Increasing flow velocity shifts the (\(c, \eta_0\)) stability domain rightward, indicating that particle velocities must closely match the macroscopic flow velocity to maintain model stability. Microscopically, insufficient particle velocities relative to the flow velocity result in incomplete momentum transfer and collisions, causing numerical instability. Mesoscopically, mismatched particle velocities impose kinetic moment constraints primarily on \(f_i^{eq}\), leading to large local spatial gradients in \(f_i\) and \(f_i^{eq}\), thus inducing oscillations or even divergence.

(iii) Increasing the flow velocity rapidly narrows the stability range of \(c\), whereas the stability range of \(\eta_0\) remains nearly unchanged. Moreover, the stability range of \(\eta_0\) is substantially larger than that of \(c\), forming a strip-like \((c, \eta_0)\) stability domain. This arises because the particle velocity \(c\) is directly coupled to the macroscopic flow velocity \(u\), while \(\eta_0\) is only associated with internal energy from additional degrees of freedom, primarily related to temperature \(T\). As a result, \(c\) is considerably more sensitive to variations in flow velocity than \(\eta_0\). In addition, the moment relations impose weaker constraints on \(\eta_0\), which involve only contracted moments with higher isotropy. Compared with non-contracted moments, the higher isotropy of contracted moments reduces the numerical stability requirements, thereby keeping the stability range of \(\eta_0\) essentially unchanged across varying flow conditions.

(iv) At the same flow velocity, the ($c$, ${\eta_0}$) stability domains of the higher-order models D2V25 and D2V26 are smaller than that of the lower-order model D2V16, indicating reduced stability regulation capacity. This reduction arises because higher-order models incorporate multiscale effects, include higher-order derivative terms, and exhibit stronger coupling between phase-space and spatiotemporal grids.

(v) In the flow velocity stability analysis in Sec.~\ref{VI_A_2}, all three models become unstable at ${u_x} = 10$.
In contrast, as shown in Fig.~\ref{Fig08} , all three models retain a ($c,{\eta _0}$) stability domain even at ${u_x} = 45$.
This demonstrates the significant potential of the ($c,{\eta _0}$) parameter in stability regulation. As the ($c,{\eta _0}$) parameter space expands, the model's stability regulation capability improves accordingly.

\subsubsection{Influence of temperature}\label{VI_B_2}

\begin{figure*}[htbp]
	\centering
	\includegraphics[width=0.75\textwidth]{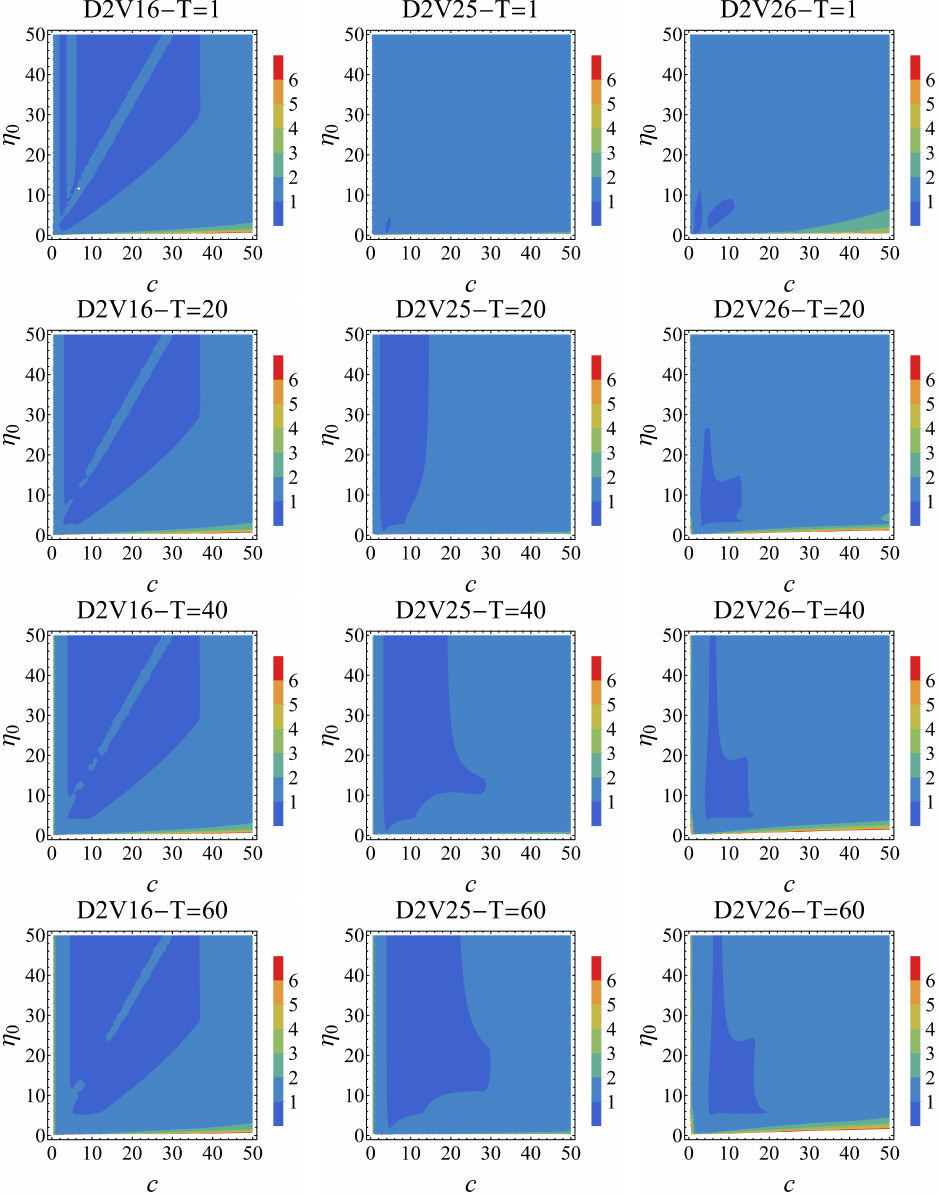}
	\caption{\centering{Influence of temperature on the stability-phase diagrams of various kinetic models at \(k=\pi/(12 \Delta r)\).}}
	\label{Fig09}
\end{figure*}

The initial temperature is set to $T = {1, 20, 40, 60}$, and the most unstable wavenumber corresponds to $k = \pi / (12 \Delta r)$, as shown in Fig.~\ref{Fig04}. Figure~\ref{Fig09} presents stability-phase diagrams showing the effects of initial temperature on various kinetic models.

(i) For all kinetic models, increasing temperature expands the ($c$, ${\eta_0}$) stability domain and enhances stability regulation capability, consistent with the analysis in Sec.~\ref{VI_A_3}. Moreover, the ($c$, ${\eta_0}$) stability domain shifts upward and to the right as temperature increases, indicating that $c$ and ${\eta_0}$ should scale with temperature to maintain numerical stability.

(ii) At a fixed temperature, the ($c$, ${\eta_0}$) stability domains of the higher-order models D2V25 and D2V26 are significantly smaller than that of the lower-order model D2V16, indicating reduced stability regulation capability. Additionally, D2V25 exhibits a much larger stability region than D2V26 at high temperatures, while the opposite trend is observed at low temperatures. This suggests that D2V25 DBM is more suitable for simulating nonequilibrium flows with low compressibility.

\subsubsection{Influence of density}\label{VI_B_3}

\begin{figure*}[htbp]
	\centering
	\includegraphics[width=0.75\textwidth]{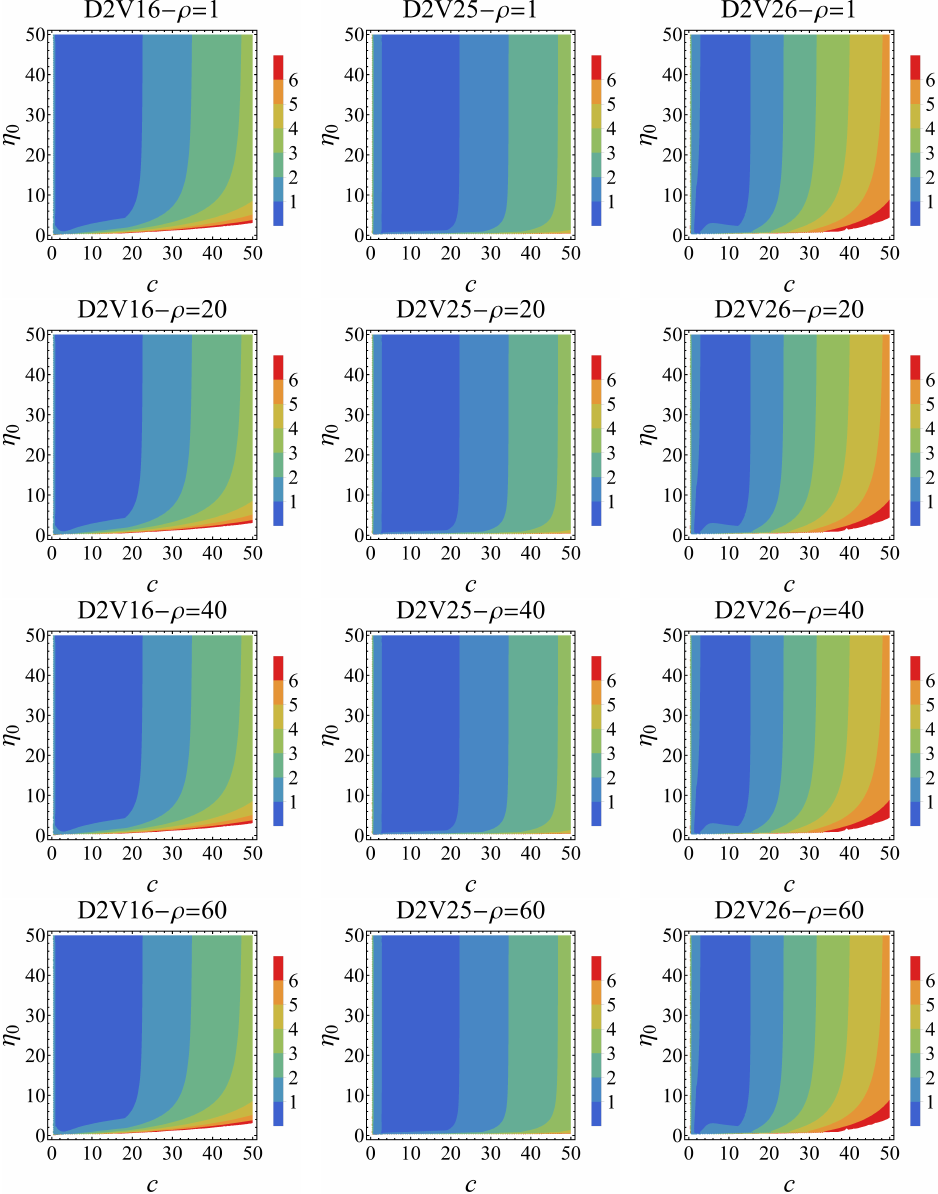}
	\caption{\centering{Influence of density on the stability-phase diagrams of various kinetic models at \(k=\pi/\Delta r\).}}
	\label{Fig10}
\end{figure*}

The initial density is set to ${\rho} = {1, 20, 40, 60}$. Figure~\ref{Fig10} presents stability-phase diagrams showing the effects of initial density on different kinetic models.

(i) For all kinetic models, increasing the density leaves the ($c$, ${\eta_0}$) stability domain nearly unchanged, indicating that the model's stability regulation is largely insensitive to density. This finding is consistent with the analysis in Sec.~\ref{VI_A_4}.

(ii) At fixed density, the ($c$, ${\eta_0}$) stability domain of the higher-order model D2V26 is significantly smaller than that of the lower-order model D2V16, indicating weaker stability regulation capability. In contrast, D2V25 exhibits a stability domain comparable to D2V16.

\subsubsection{Influence of relative time step}\label{VI_B_4}

\begin{figure*}[htbp]
	\centering
	\includegraphics[width=0.75\textwidth]{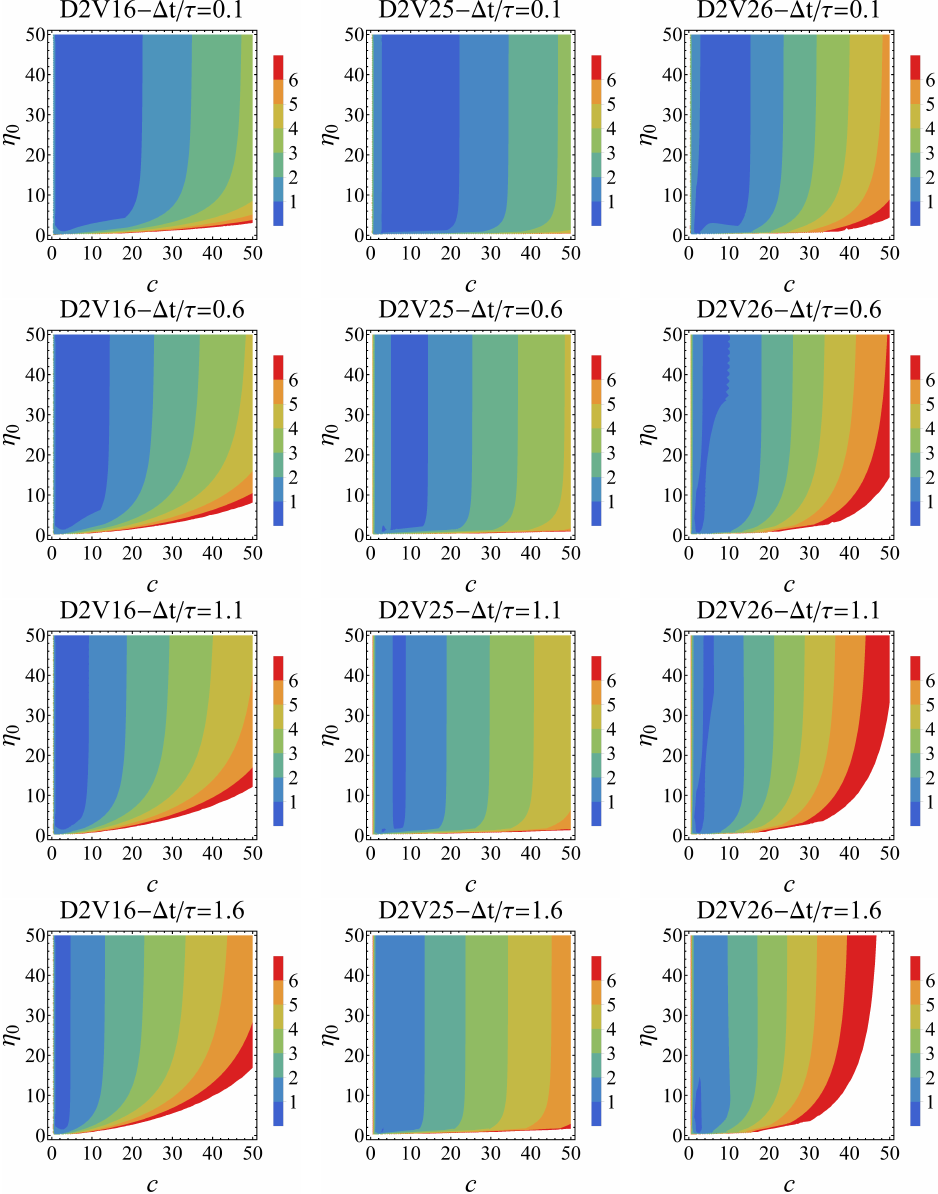}
	\caption{\centering{Influence of relative time step on the stability-phase diagrams of various kinetic models at \(k=\pi/\Delta r\).}}
	\label{Fig11}
\end{figure*}

The relative time step is adjusted by varying the relaxation time $\tau$, with values of $ \Delta t / \tau  = {0.1, 0.6, 1.1, 1.6}$.
Figure~\ref{Fig11} shows the stability-phase diagrams, illustrating the influence of relative time step on various kinetic models. The following conclusions can be drawn:

(i) For all kinetic models, increasing $\Delta t/\tau$ reduces the ($c$, ${\eta_0}$) stability domain, indicating a decline in stability regulation capability. This finding is consistent with the analysis in Sec.~\ref{VI_A_5}. A small $\Delta t/\tau$ corresponds to a large relaxation time $\tau$ and thus a stronger dissipation mechanism, such as enhanced viscosity and heat conduction. These effects smooth the interface, reduce $\bm{\nabla} f_i$, and improve model stability.

(ii) At fixed $\Delta t/\tau$, the ($c$, ${\eta_0}$) stability domains of the higher-order models D2V25 and D2V26 are significantly smaller than that of the lower-order model D2V16, suggesting that higher-order models involve higher-order deviations of the distribution function, which depend nonlinearly on $\tau$ and thus exhibit greater sensitivity to its variation. In comparison, at $\Delta t/\tau = 1.6$, the stability domain of the D2V25 model nearly vanishes, indicating poorer numerical stability than the D2V26 model.


\subsubsection{Influence of time-space step ratio}\label{VI_B_5}

\begin{figure*}[htbp]
	\centering
	\includegraphics[width=0.75\textwidth]{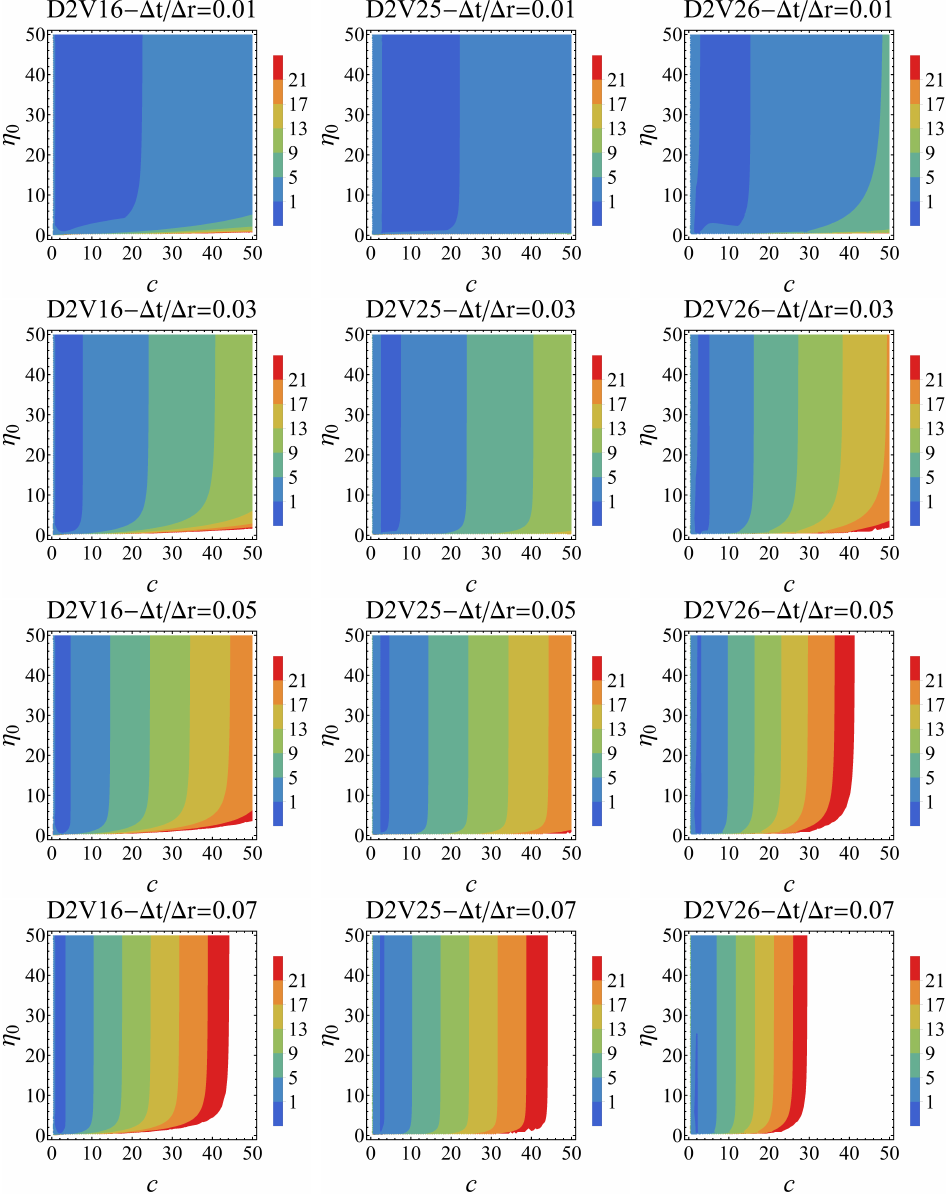}
	\caption{\centering{Influence of time-space step ratio on the stability-phase diagrams of various kinetic models at \(k=\pi/\Delta r\).}}
	\label{Fig12}
\end{figure*}

Here, the time step $\Delta t$ is fixed, and the spatial step $\Delta r$ is varied to yield a time-space step ratio of $\Delta t/\Delta r = {0.01, 0.03, 0.05, 0.07}$. As shown in Fig.~\ref{Fig12}, increasing $\Delta t/\Delta r$ reduces the ($c$, ${\eta_0}$) stability domain for all kinetic models, indicating a weakening of stability regulation capability. A larger $\Delta t/\Delta r$ corresponds to a smaller $\Delta r$, which significantly amplifies $\bm{\nabla} f_i$ and rapidly degrades stability. At fixed $\Delta t/\Delta r$, the ($c$, ${\eta_0}$) stability domains of the higher-order models D2V25 and D2V26 remain smaller than that of the lower-order model D2V16. This is because higher-order models involve higher-order spatial derivatives and are therefore more sensitive to variations in $\Delta r$.


\subsection{Analysis of model stability regulation capability under full wavenumber }\label{VI_C}

\begin{figure*}[htbp]
	\centering
	\includegraphics[width=0.75\textwidth]{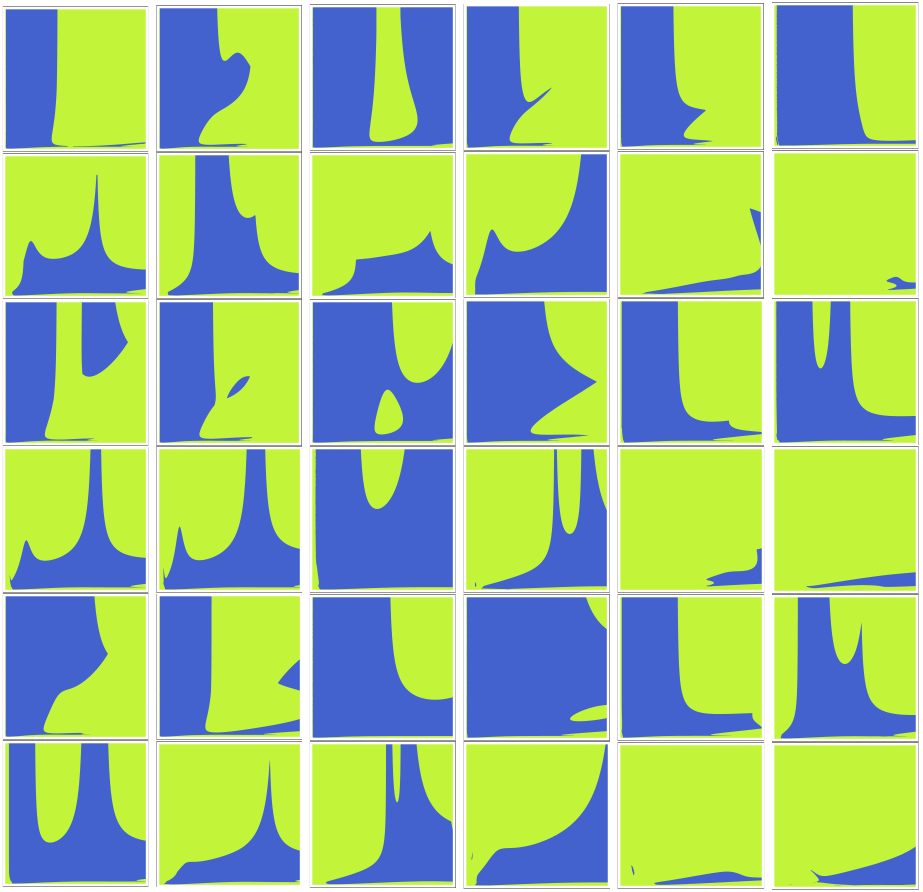}
	\caption{\centering{Complex stability configurations (blue regions) of the D2V26 DBM under different flow velocities and wavenumbers using the MLW scheme, with $c$ and $\eta_0$ representing the horizontal and vertical axes, respectively, in each panel.}}
	\label{Fig13}
\end{figure*}

The stability analysis in Sec.~\ref{VI_B} is performed under a fixed wavenumber, which corresponds to the most unstable mode under the given macroscopic conditions and model parameters. According to von Neumann stability theory, a model is stable when the largest modulus of the eigenvalues of the amplification matrix satisfies ${\left| \omega \right|_{\max}} \le 1$ for all wavenumbers.
As macroscopic quantities and model parameters vary, the most unstable wavenumber also changes, leading to highly complex configurations of the ($c$, ${\eta_0}$) stability domain, see Fig.~\ref{Fig13}. To identify a universally stable region across varying conditions, it is necessary to scan across the full wavenumber space.

Since ${\left| \omega \right|_{\max}}$ is periodic, a discrete set of wavenumbers is selected for analysis: ${k_i} = 2\pi i \times 0.005/\Delta r$, where $i \in [1, 200]$. Each wavenumber ${k_i}$ defines a corresponding stability domain in ($c$, ${\eta_0}$), denoted by ${A_i}$. The intersection of all ${A_i}$ defines the common stability domain ${A_{\mathrm{com}}}$, which strictly satisfies the von Neumann stability criterion. The full parameter space is defined as ${A_{\mathrm{tot}}}: c \in (0, 50), \eta_0 \in (0, 50)$.
Using the Minkowski functional $A = {A_{\mathrm{com}}}/{A_{\mathrm{tot}}}$\cite{sofonea1999morphological,gan2011phase}, a morphological analysis is performed to extract the stability area from the complex structure of the $\omega(c, \eta_0)$  across wavenumbers. This ratio quantifies the proportion of the common stability domain within the total parameter space.
This approach enables the construction of stability regulation phase diagrams for various DBMs over the full parameter space and allows the derivation of model stability probability curves under full wavenumber. All model parameters, except for the wavenumber, are consistent with those used in the corresponding subsections of Sec.~\ref{VI_B}.

\subsubsection{Influence of spatial discretization schemes and flow velocity}\label{VI_C_1}

\begin{figure*}[htbp]
	\centering
	\includegraphics[width=0.99\textwidth]{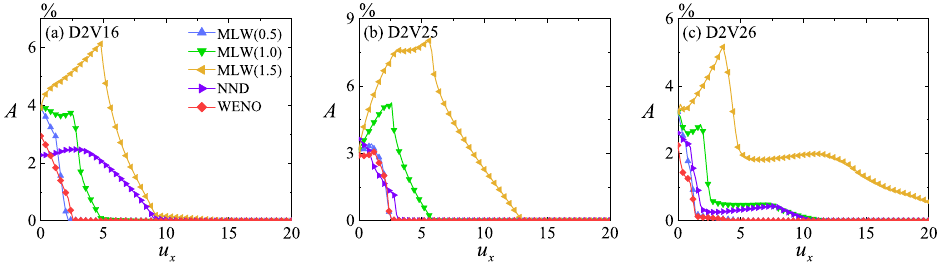}
	\caption{\centering{Influence of spatial discretization schemes and flow velocity on the model stability probability.}}
	\label{Fig14}
\end{figure*}

Figure~\ref{Fig14} presents the probability curves of stability for DBMs employing different spatial discretization schemes under various flow velocities. The following conclusions can be drawn:

(i) A larger $A$ indicates a higher probability that the model will achieve stability. The area $S_{\rm{stab}}$ enclosed by the $A$ curve and the coordinate axes quantifies the model's overall stability probability under the given parameter variation. In Fig.~\ref{Fig14}, a larger $S_{\rm{stab}}$ signifies the model's enhanced ability to maintain stability across varying flow velocities through stability adjustment.

(ii) When the MLW scheme is used, the relation ${A_{\mathrm{MLW}(0.5)}} < {A_{\mathrm{MLW}(1.0)}} < {A_{\mathrm{MLW}(1.5)}}$ holds for all flow velocities and all models, where the numbers in parentheses represent the artificial viscosity coefficients shown in Eq. (\ref{e8}). This trend indicates that increasing the artificial viscosity  enhances the model’s stability regulation capability. The improvement arises from the increased diffusion, which suppresses the spatial gradients of ${f_i}$ and promotes numerical stability. However, excessive diffusion may over-smooth multiscale structures and distort the physical field. Therefore, selecting an appropriate artificial viscosity coefficient requires balancing numerical stability and physical fidelity.

(iii) For most DBMs, there exists a critical flow velocity, ${u_c}$. When ${u_x} < {u_c}$, the stability measure $A$ increases with ${u_x}$, while for ${u_x} > {u_c}$, $A$ decreases with ${u_x}$. This behavior results from the interplay between two competing mechanisms: (a) On one hand, increasing flow velocity intensifies compressibility, nonequilibrium, and nonlinear effects, which reduce the stability domain and the value of $A$; (b) On the other hand, a higher flow velocity requires expanding the considered ($c$, ${\eta_0}$) parameter space. Higher velocities can match new stable regions with increasing $c$ and $\eta_0$, converting previously unstable regions into stable ones, thus enlarging the stability domain. The dominant mechanism dictates the overall trend of $A$. If compressibility effects dominate, $A$ decreases; if the benefit from parameter space expansion prevails, $A$ increases. Specifically, when ${u_x} > {u_c}$, expanding the parameter space becomes insufficient to counterbalance the detrimental effects of increased compressibility, leading to reduced stability. \emph{These findings do not contradict those in Sec.~\ref{VI_B_1}, but rather extend them, as the difference arises from changes in (a) wavenumber: from fixed to full ones; (b) the number of sampling points for macroscopic quantities: from 4 ($u_x=0,15,30,45$) to 101 ($0 \leq u_x \leq 20$). This broader analysis improves the accuracy, generality, and credibility of the results.}

(iv) When the MLW(1.5) scheme is adopted, ${u_x} = 10$ yields $A = 0.16\%$ for the D2V16 model, $A = 2.28\%$ and $1.96\%$ for the D2V25 and D2V26 models, respectively. These results indicate that higher-order models exhibit higher stability probabilities than lower-order models. Moreover, when ${u_x} > 15$, the D2V26 model exhibits significantly higher stability than the other two models. Even when ${u_x}$ increases to 20, the D2V16 and D2V25 models become completely unstable, while the D2V26 model remains stable.
The discrepancy with the conclusion in Sec.\ref{VI_B} arises from the inclusion of the full wavenumber spectrum in the analysis. As shown in Fig.~\ref{Fig13}, the shape of the stability domain varies complexly with wavenumber, even for the same DVS. In fact, different DVSs produce distinct stability domains across wavenumbers for different physical problems, thereby inevitably influencing the common stability domain. \emph{Essentially, the degree of alignment among the DVS, wavenumber, and macroscopic quantities is the key factor governing model stability. }Under the current configuration, the D2V26 DVS aligns more effectively with both the wavenumber and macroscopic quantities. This alignment enhances the model’s resistance to stability degradation caused by variations in macroscopic physical quantities.

When the NND scheme is applied, similar but slightly different conclusions can be drawn: the D2V16 and D2V26 models maintain $A > 0$ at ${u_x} = 10$, while the D2V25 model reaches the cutoff stability speed at ${u_x} = 4.4$. This conclusion differs from that obtained with the MLW(1.5) scheme, highlighting the substantial impact of the coupling between phase-space discretization and spatiotemporal discretization on model stability. A well-constructed phase-space discretization can compensate for symmetry breaking, thereby improving cross-scale representation and numerical stability

(v) When the WENO scheme is used, the cutoff stability speed is significantly lower, and the stability area $S_{\rm{stab}}$ is generally smaller compared with other spatial discretization schemes. This indicates that although the WENO scheme significantly improves numerical accuracy, it is more challenging to identify the stable region.
In the parameters corresponding to Fig.~\ref{Fig02}(c), when $u_x = 4$, the D2V26 DBM using the WENO scheme is close to stable but actually unstable. However, in Fig.~\ref{Fig14}(c), by expanding the parameter range of ($c, \eta_0$), when $u_x = 4$, $A = 0.00193\%$, which, although small, indicates the presence of a stability domain. This demonstrates the necessity of in-depth analysis of numerical stability.

\subsubsection{Influence of temperature, density, relative time step, and time-space step ratio}\label{VI_C_2}

\begin{figure*}[htbp]
	\centering
	\includegraphics[width=0.8\textwidth]{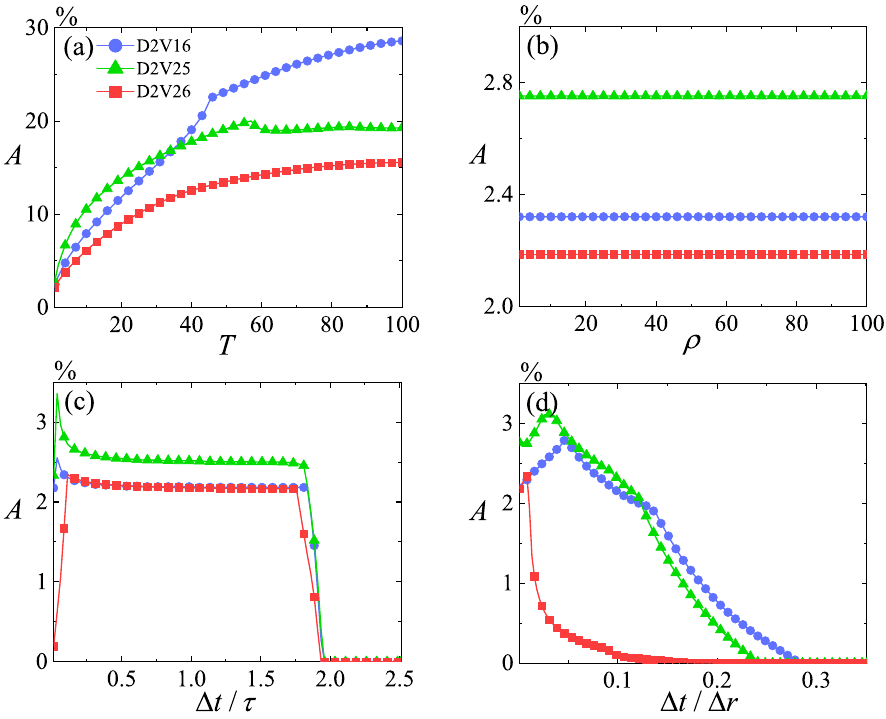}
	\caption{\centering{Influence of temperature, density, relative time step size and time-space step ratio on the model stability probability.}}
	\label{Fig15}
\end{figure*}

Figure~\ref{Fig15} shows the effects of temperature, density, relative time step $\Delta t/\tau$, and time-space step ratio $\Delta t/\Delta r$ on the model’s stability probability curves.
The following conclusions can be drawn:

(i) As temperature increases, the stability probability measure $A$ increases gradually,  indicating enhanced numerical stability across all three models.

(ii) In contrast, as density increases, $A$ remains nearly constant, and the stability probability shows little variation.

(iii) For the relative time step, as $\Delta t/\tau$ increases, $A$ eventually approaches zero, indicating a complete loss of stability.
During this process, all three models exhibit a plateau where $A$ remains constant while $\Delta t/\tau$ increases.
This suggests that, with fixed relaxation time $\tau$, increasing the time step $\Delta t$ to improve computational efficiency is permissible without compromising stability—up to a certain threshold.

(iv) As $\Delta t/\Delta r$ increases, $A$ also approaches zero.
Unlike $\Delta t/\tau$, $A$ decreases sharply with increasing $\Delta t/\Delta r$ and does not exhibit a plateau.
Therefore, with a fixed spatial step, minimizing the time step is essential for ensuring model stability.

(v) No single DVS consistently achieves the highest stability probability across all conditions, indicating that no DVS is universally optimal.

\emph{(vi) The slope of each stability probability curve reflects the model's sensitivity to the corresponding parameter. A steeper slope indicates greater sensitivity, implying that the parameter requires higher weighting in numerical simulations.
The intersection point between the probability curve and the horizontal axis represents the critical threshold of the parameter beyond which the model becomes unstable. A smaller threshold implies weaker stability regulation capability.
Although $S_{\rm{stab}}$ quantifies the overall stability probability of the model under the given parameter variation, we prefer to focus on the extension of the $A$ curve along the $x$-axis. That is, the goal is not to maximize the area of $S_{\rm{stab}}$, but to increase the physical range covered by the $A$ curve.}

\subsection{Numerical verification of the effectiveness of stability-phase diagrams}\label{IX}\label{VI_D}

Based on the above analysis, the numerical stability of the kinetic model is closely related to the intensity of nonequilibrium effects.
To assess the effectiveness of the stability-phase diagrams under varying nonequilibrium intensities, and to identify differences in both numerical stability and the accuracy of nonequilibrium representation among different DVS configurations, three test cases are performed using the following initial conditions:
\begin{equation}\label{e24}
	\rho(x, y) = \frac{{\rho_L + \rho_R}}{2} - \frac{{\rho_L - \rho_R}}{2} \tanh \left( \frac{{x - N_x \Delta r / 2}}{{L_\rho}} \right),
\end{equation}
\begin{equation}\label{e25}
	u_x(x, y) = \frac{{u_{xL} + u_{xR}}}{2} - \frac{{u_{xL} - u_{xR}}}{2} \tanh \left( \frac{{x - N_x \Delta r / 2}}{{L_u}} \right).
\end{equation}
Table~\ref{tableII} summarizes the parameters for the three test cases. The temperature field is defined by $T(x, y) = P(x, y)/\rho(x, y)$, and the transverse velocity  is set to $u_y = 0$. A uniform grid of $1000 \times 4$ cells is employed with $\Delta x = \Delta y = 1.5 \times 10^{-3}$, $\Delta t = 10^{-5}$, and $\gamma = 1.5$. The values of $c$ and $\eta_0$ are uniformly set to $c = 0.8$ and $\eta_0 = 6$ for all cases, based on the intersection of the stability domains for the three models. Spatial discretization is carried out using the 5th-WENO scheme.

\onecolumngrid

\begin{table}[htbp]
	\caption{Initial value configurations for viscous stress calculation under different scenarios.}
	\label{tableII}\centering
	\begin{tabular}{cccccccc}
		\toprule \textbf{Case} & ${\bm{\Delta }}_{2xx}^*$ & \textbf{Density} &
		\textbf{Velocity} & \textbf{Pressure} & \textbf{Width} & \textbf{Relaxtion Time} & \textbf{Computational time}
		\\
		\midrule I & weak & $2{\rho _L} = {\rho _R} = 2$ & ${u_{xL}} = \;{u_{xR}} = 0$ & $P = 2$ & ${L_\rho } = {L_u} = 20$ & ${10^{ - 3}}$ & 0.02
		\\
		II & strong & $2{\rho _L} = {\rho _R} = 2$ & ${u_{xL}} = {-u_{xR}} =   0.5$ & $P = 2$ & ${L_\rho } = {L_u} = 30$ & $2 \times {10^{ - 3}}$ & 0.025
		\\
		III & strong & $2{\rho _L} = {\rho _R} = 2$ & ${u_{xL}} = {-u_{xR}} =  0.5$ & $P = 2$ & ${L_\rho } = {L_u} = 30$ & $2 \times {10^{ - 3}}$ & 0.031
		\\
		\bottomrule
	\end{tabular}
\end{table}

\twocolumngrid

\subsubsection{Weak TNE case}\label{VI_D_1}

\begin{figure*}[htbp]
	\centering
	\includegraphics[width=0.98\textwidth]{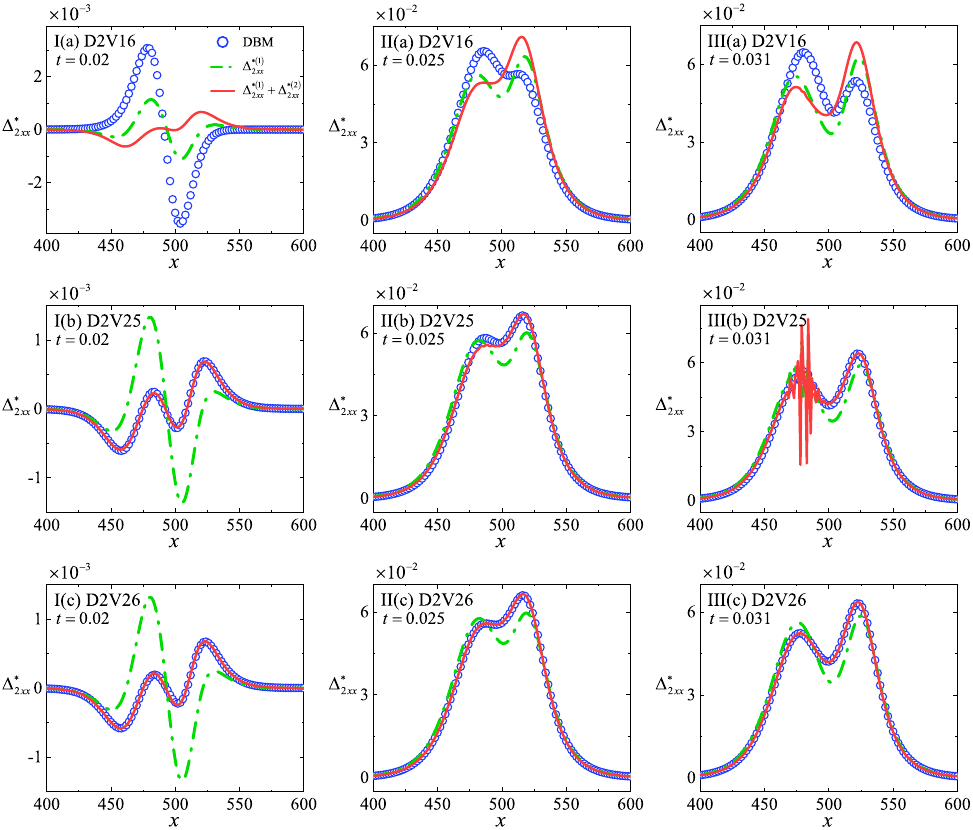}
	\caption{\centering{Comparisons of numerical and first/second-order analytical solutions of $\Delta _{2xx}^ * $ obtained from D2V16 (first row), D2V25 (second row) and D2V26 (third row) models under different nonequilibrium intensities. The left, middle and right columns correspond to the weak TNE case, and the early and later stages of the strong TNE case, respectively.
}}
 \label{Fig16}
\end{figure*}

\begin{figure*}[htbp]
	\centering
	\includegraphics[width=0.8\textwidth]{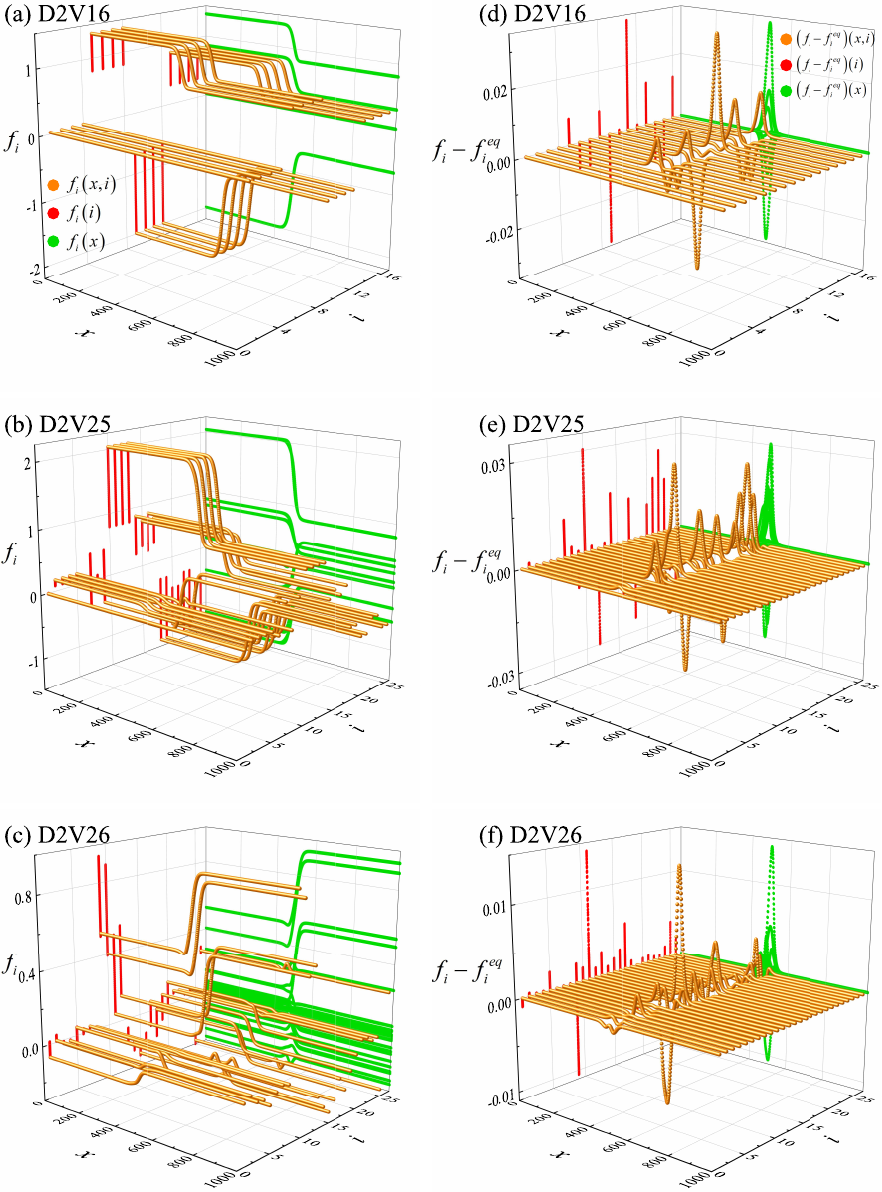}
	\caption{\centering{Distributions of ${f_i}$ and (${f_i} - f_i^{eq}$) along the centerline in the weak TNE case at $t = 0.02$. The first, second, and third rows correspond to the results from the D2V16, D2V25, and D2V26 DBMs, respectively.}}
	\label{Fig17}
\end{figure*}

The first column of Fig.~\ref{Fig16} compares the numerical results and analytical solutions for the $xx$ component of viscous stress $\Delta _{2xx}^*$ at $t = 0.02$. The blue circles represent the numerical solution from DBM simulations, while the green dashed and red solid lines correspond to the first- and second-order analytical solutions of $\Delta _{2xx}^*$\cite{gan2018discrete}, respectively. The following conclusions are drawn:

(i) This case initially does not include a velocity gradient; therefore, the first-order $\Delta _{2xx}^{*\left( 1 \right)} \sim 0$, and the second-order $\Delta _{2xx}^{*\left( 2 \right)}$, induced by density and temperature gradients, dominates at this point. The absence of first-order TNE effects and the limited density and temperature gradients result in a relatively weak overall TNE intensity.

(ii) At $t = 0.02$, both the numerical and analytical solutions for all DBMs are smooth, showing no oscillations or divergence. This confirms that, under weak nonequilibrium intensities, the parameter set ($c, {\eta_0}$), obtained by intersecting the stability domains of different models, effectively ensures numerical stability.

(iii) For all DBMs, the numerical solutions do not align with the first-order analytical solutions, indicating that an accurate characterization of TNE effects requires considering the second-order deviation ${f^{(2)}}$, despite the weak TNE intensity. The numerical solution of the D2V16 model does not align with the second-order analytical solution because it considers only the contribution of ${f^{(1)}}$. The insufficient selection of invariants in physical modeling, along with the inadequate preservation of moment relations in constructing the DVS, leads to its inability to accurately capture TNE effects, despite its numerical stability. In contrast, the numerical solutions of the D2V25 and D2V26 models align closely with the second-order analytical solutions, due to the consideration of ${f^{(2)}}$'s contribution.

Since DBM characterizes mesoscale features through the kinetic moments of ${f_i}$ and (${f_i} - {f_i}^{eq}$), the continuity and smoothness of these quantities are essential for ensuring numerical stability and realizing the model's physical functions. Figure~\ref{Fig17} presents the distributions of ${f_i}$ and (${f_i} - {f_i}^{eq}$) along the centerline at the same moment as in Fig.~\ref{Fig16}. To enable a quantitative comparison, Fig.~\ref{Fig20}(a) reports the descriptive statistics of ${f_i}$ and (${f_i} - {f_i}^{eq}$), including the range $R$, standard deviation $\sigma$, and average spatial gradient $\overline {|\bm{\nabla}|}$ . The vertical axis is plotted on a logarithmic scale.

(i) As shown in Figs.~\ref{Fig17}(a)–(c), the D2V16 and D2V25 models exhibit greater discreteness in ${f_i}$ compared with the D2V26 model. For D2V16 and D2V25, the ranges of (${f_{i\min}}, {f_{i\max}}$) are ($-1.99$, $1.37$) and ($-1.38$, $2.04$), with $R(f_i) = 3.36$ and $3.42$, $\sigma(f_i) = 0.97$ and $0.86$, and $\overline{|\bm{\nabla} f_i|} = 6.44$ and $10.71$, respectively. In contrast, the distribution of ${f_i}$ in the D2V26 model is considerably smoother, with a range of ($-0.30$, $0.94$), a span of only $1.24$, a standard deviation of $0.25$, and a mean spatial gradient of $2.27$. The smaller span, standard deviation, and spatial gradient indicate that the D2V26 model achieves improved continuity in ${f_i}$.

(ii) In the D2V16 model, the first four components ${f_i}$ ($i = 1$–$4$) are zero, resulting in a limited number of active ${f_i}$ during simulation. In contrast, in the D2V25 model only ${f_1}$ and ${f_6}$ are zero, whereas in the D2V26 model all ${f_i}$ remain nonzero. Moreover, the distribution patterns of nonzero ${f_i}$ differ significantly across models. In D2V16, only three distinct groups are present: $f_{5}=f_{6}=f_{7}=f_{8}$, $f_{9}=f_{10}=f_{11}=f_{12}$, and $f_{13}=f_{14}=f_{15}=f_{16}$. The D2V25 model exhibits 8 distinct ${f_i}$ values, while the D2V26 model includes all 26 components as independently active. A greater number and diversity of active ${f_i}$ enhance the model's capacity to capture high-order nonequilibrium effects, as illustrated in Figs.~\ref{Fig16}I(b)–I(c). \emph{Therefore, an effective DVS should activate a broader set of ${f_i}$ to partially share the burden of moment constraints and ensure better continuity of the distribution function.}

(iii) Similarly, as shown in Figs.~\ref{Fig17}(d)–(f) and Fig.~\ref{Fig20}(a), the D2V26 model also exhibits superior continuity and smoothness in (${f_i} - f_i^{eq}$) compared to the other two models. Its span, standard deviation, and mean spatial gradient are significantly lower—only about $30\%$ of the corresponding values in the D2V16 model. This effectively supports the continuity and smoothness of nonequilibrium effects.

(iv) As shown in Figs.~\ref{Fig17}(d)–(f), the minimum values of (${f_i} - f_i^{eq}$) appear at $i = 9$, $i = 12$, and $i = 9$ for the D2V16, D2V25, and D2V26 models, respectively.
The corresponding maximum values occur at $i = 11$, $i = 10$, and $i = 11$.
At these six positions, (${f_i} - f_i^{eq}$) exhibits the strongest discontinuities and the most pronounced deviations from equilibrium, forming sharp spike-like structures.
As shown in Figs.~\ref{Fig01}(a)–(c), these discrete velocity directions correspond to the positive or negative $x$-axis.
This is primarily due to the initial macroscopic gradients being applied along the $x$-direction, which significantly enhances the TNE intensity and nonlinearity in that direction.

\subsubsection{Strong TNE case: early stage}\label{VI_D_2}

\begin{figure*}[htbp]
	\centering
	\includegraphics[width=0.8\textwidth]{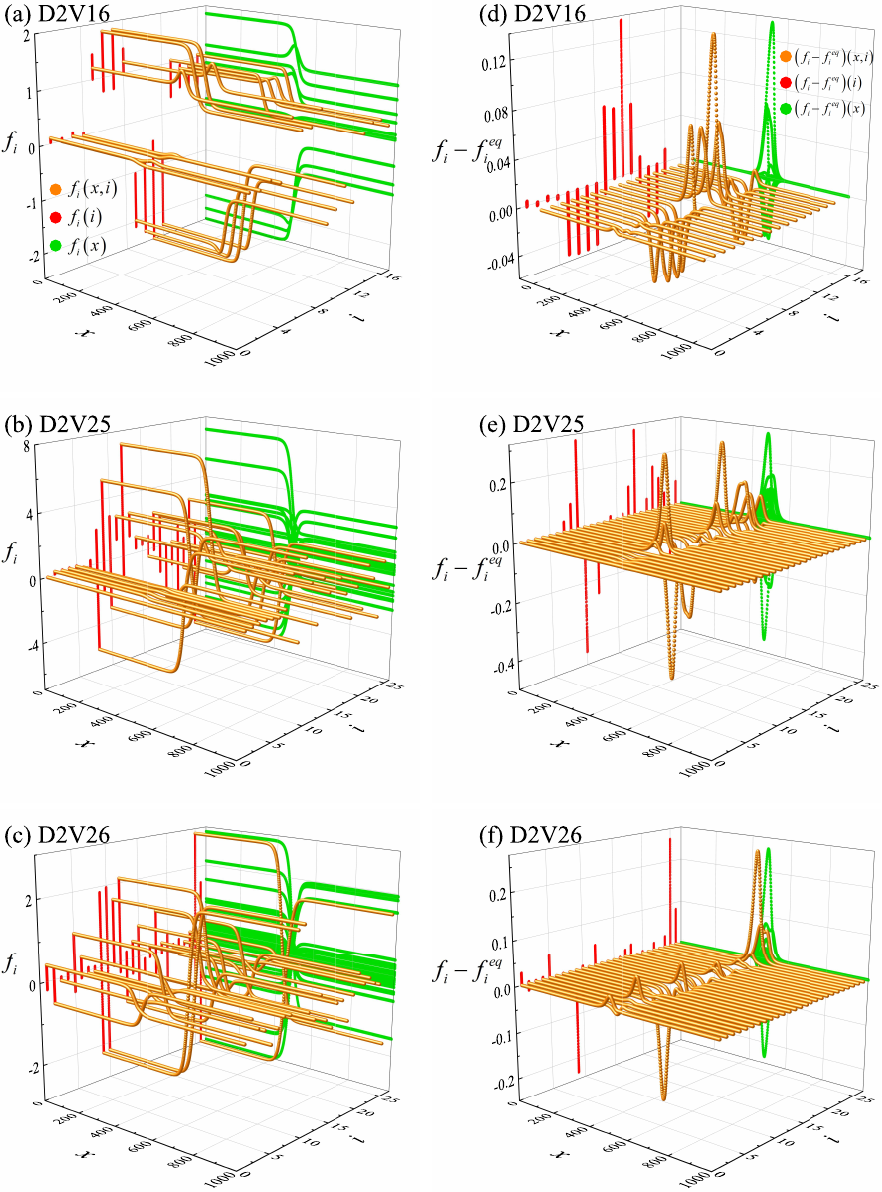}
	\caption{\centering{Distributions of ${f_i}$ and (${f_i} - f_i^{eq}$) along the centerline in the strong TNE case at $t = 0.025$. The first, second, and third rows correspond to the results from the D2V16, D2V25, and D2V26 DBMs, respectively.}}
	\label{Fig18}
\end{figure*}

The second column of Fig.~\ref{Fig16} compares the numerical results with the analytical solutions for $\Delta_{2xx}^*$ at $t = 0.025$, from which the following conclusions are drawn:

(i) Compared with the weak TNE case, the introduction of velocity gradients and an increase in relaxation time intensify $\Delta_{2xx}^*$ from $10^{-3}$ to $6 \times 10^{-2}$.

(ii) At $t = 0.025$, all three models remain numerically stable, confirming that the stability control strategy is effective even under intensified TNE conditions.

(iii) As the TNE intensity increases, the numerical solution of the D2V25 model deviates from the second-order analytical solution near the left peak of $\Delta_{2xx}^*$. In contrast, the D2V26 model maintains excellent agreement with the analytical solution. This suggests that the D2V26 DVS is better suited to realizing the physical functionality of the model.

Similarly, Fig.~\ref{Fig18} and Fig.~\ref{Fig20}(b) present the distributions and statistical characteristics of ${f_i}$ and (${f_i} - f_i^{eq}$), respectively.

(i) Compared with the weak TNE case, the discreteness of ${f_i}$ in the D2V25 model increases significantly. The span, the standard deviation, and the mean spatial gradient of $f_i$ are increased by a factor of 3 to 5. In contrast, as shown in Fig.~\ref{Fig20}(b), the corresponding statistical quantities in the D2V26 model are only about $50\%$ of those in the D2V25 model.

(ii) An increasing nonequilibrium intensity requires the DVS to activate a broader range of distribution function patterns. As shown in Fig.~\ref{Fig18}, the number of nonzero ${f_i}$ components in the D2V25 model increases from 8 to 21 relative to the weak TNE case, while the D2V26 model maintains 25 nonzero components. The limited diversity of nonzero patterns in the D2V25 model constrains its ability to capture high-order TNE effects, particularly under strong TNE conditions, as illustrated in Fig.~\ref{Fig16}II(b). An effective DVS should activate a wider variety of nonzero ${f_i}$ patterns with diverse magnitudes to ensure accurate representation of TNE effects.

(iii) As shown in Figs.~\ref{Fig17}(d)–(f), the minimum values of (${f_i} - f_i^{eq}$) occur at $i = 6, 7$, $i = 10$, and $i = 9$ for the D2V16, D2V25, and D2V26 models, respectively. The corresponding maximum values appear at $i = 11$, $i = 9, 18$, and $i = 25$. Unlike in the weak TNE case, the directions associated with $i = 6, 7$ in D2V16, $i = 9, 18$ in D2V25, and $i = 25$ in D2V26 no longer align with the positive or negative $x$-axis. \emph{This suggests that nonequilibrium effects exhibit stronger anisotropy as the degree of discreteness and nonequilibrium increases. }The constraints imposed on ${f_i}$ are significantly intensified, especially along directions with weak spatial symmetry—such as direction $i = 25$ in the D2V26 model.

(iv) As the degree of nonequilibrium increases, the nonlinearity of both ${f_i}$ and ${f_i^{eq}}$ becomes markedly stronger due to the involvement of more high-order and strongly coupled macroscopic derivatives (i.e., nonlinear nonequilibrium driving forces) in $f_i$. As a result, the continuity and smoothness of both ${f_i}$ and (${f_i} - f_i^{eq}$) are disrupted, leading to significant increases in all related statistical quantities, as shown in Figs.~\ref{Fig20}(a)–(b). This, in turn, degrades the model’s ability to accurately characterize TNE effects, as demonstrated in Figs.~\ref{Fig16}I–II(b).

\subsubsection{Strong TNE case: later stage}\label{VI_D_3}

\begin{figure*}[htbp]
	\centering
	\includegraphics[width=0.8\textwidth]{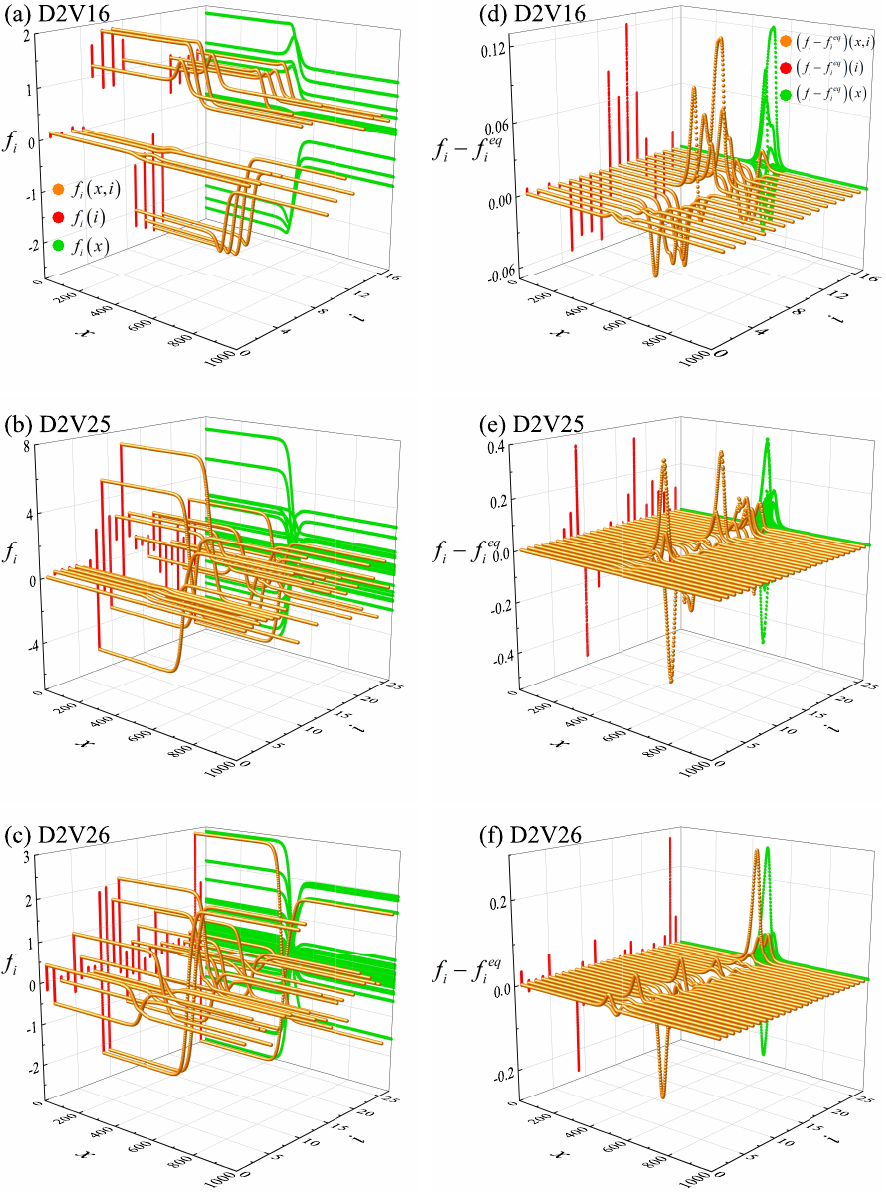}
	\caption{\centering{Distributions of ${f_i}$ and (${f_i} - f_i^{eq}$) along the centerline in the strong TNE case at $t = 0.031$. The first, second, and third rows correspond to the results from the D2V16, D2V25, and D2V26 DBMs, respectively.}}
	\label{Fig19}
\end{figure*}

The third column of Fig.~\ref{Fig16} shows results at $t = 0.031$.

(i) Compared with the early stage, the D2V25 model exhibits severe numerical oscillations near the left peak of $\Delta_{2xx}^*$. This is because the von Neumann stability analysis is performed based on the initial conditions. As the TNE intensity increases and numerical errors accumulate, the initially stable parameters ($c$, $\eta_0$) may become invalid. Therefore, it is necessary to reconstruct the stability parameter space using updated stability analysis and to dynamically adjust the model parameters. This is precisely the fundamental idea behind the particle-on-demand strategy\cite{dorschner2018particles, kallikounis2022particles}.

(ii) A comparison between Fig.~\ref{Fig16}II(b) and Fig.~\ref{Fig16}III(b) reveals that the region where numerical oscillations occur in the D2V25 model at the later stage corresponds exactly to the region where the numerical and analytical solutions of $\Delta_{2xx}^*$ begin to diverge at the early stage. This indicates that the loss of accuracy caused by an inadequate DVS can be amplified by nonlinear system evolution, eventually leading to numerical instability.

(iii) At this stage, the D2V25 model exhibits nonlinear instability, despite meeting the criteria for linear stability, revealing a key limitation of von Neumann analysis [see Fig. 16III(b)]. This highlights the importance of nonlinear stability analysis \cite{lakshmikantham1989stability,sastry2013nonlinear}, which assesses the boundedness and physical consistency of the full system evolution over finite time intervals.

\begin{figure*}[htbp]
	\centering
	\includegraphics[width=0.98\textwidth]{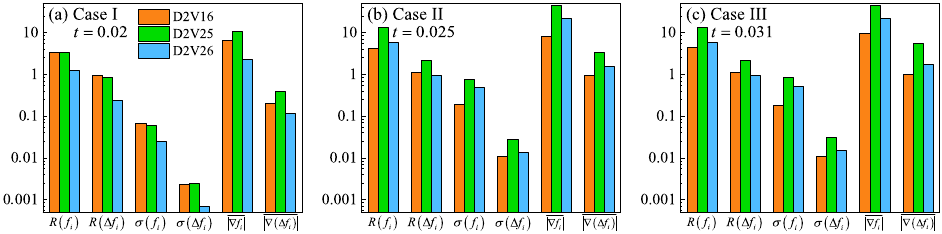}
	\caption{\centering{Statistical characteristics of ${f_i}$ and $\Delta f_i = {f_i} - f_i^{eq}$ along the centerline for each model under three degrees of nonequilibrium. $R$: range; $\sigma $: standard deviation; $\overline {\left| \bm{\nabla}  \right|} $: mean spatial gradient. The vertical axis is displayed on a logarithmic scale.}}
	\label{Fig20}
\end{figure*}

As shown in Fig.~\ref{Fig19}(e), the discreteness of (${f_i} - f_i^{eq}$) in the D2V25 model increases significantly compared with the early stage, with the mean spatial gradient increasing by $63.04\%$.
Compared with Fig.~\ref{Fig16}II(B), this significantly amplifies both the frequency and amplitude of numerical oscillations in $\Delta_{2xx}^*$.
In contrast, all statistical quantities in the D2V26 model increase by only about $10\%$, enabling accurate characterization of TNE effects throughout all stages of the strong TNE case.

In summary, the DVS plays a fundamental role in maintaining numerical stability and ensuring the physical functionality of the model.
\emph{An effective DVS should preserve spatial symmetry while incorporating a greater number and wider range of discrete velocities, thereby activating a larger variety of ${f_i}$ patterns, ensuring continuity and smoothness of ${f_i}$, and enabling accurate characterization of multiscale TNE effects.}
The D2V26 model surpasses the D2V16 model in terms of physical functionality, and outperforms the D2V25 model in both numerical stability and nonequilibrium characterization.
However, it should be noted that the D2V26 model merely represents one feasible design that satisfies the current requirements, and is not necessarily the optimal solution.
In practical applications, the DVS should be designed according to the specific nonequilibrium characteristics of the problem, following the principles outlined above.

\section{CONCLUSIONS, DISCUSSIONS, AND FUTURE PERSPECTIVES}\label{VII}

Numerical simulation of supersonic flows faces three major challenges: cross-scale modeling, numerical stability and complex field analysis. The DBM, as a kinetic approach designed for multiscale modeling and complex physical field analysis, offers a powerful framework for addressing these issues. However, the inherent nonlinear, nonequilibrium, and multiscale nature of supersonic flows poses significant challenges to DBM stability.
This challenge is further exacerbated by the intrinsic coupling between phase-space discretization and spatiotemporal discretization—a fundamental characteristic that distinguishes kinetic methods from conventional CFD.

To address this challenge, we perform von Neumann stability analysis to systematically investigate the influence of phase-space discretization approaches—including discretization of equilibrium distribution and discrete velocity construction—as well as TNE intensity, spatiotemporal discretization schemes, initial conditions, and model parameters on the numerical stability of DBM. The main findings are as follows:
(i) Among approaches for determining equilibrium distribution functions considered,
the moment-matching approach, which strictly preserves kinetic moment relations from coarse-grained modeling and reconstructs $f_i^{eq}$ accurately, offers significantly higher stability than the globally unified expansion-coefficient and distributed weighting-coefficient methods, both of which rely on truncated Taylor expansions.
(ii) As discreteness and nonequilibrium intensify, higher-order DBMs generally exhibit lower stability compared to lower-order models. The underlying reasons are as follows.
As the Knudsen number increases, the system becomes more nonlinear, and this nonlinear enhancement is inherently embedded in the governing equations of the corresponding model.
In numerical simulations, instabilities caused by rapidly varying features—such as shock waves—are often amplified by nonlinear terms in the model equations. Consequently, the higher the nonlinear order of the model, the weaker its capability to handle such sharp variations, and the more stringent the requirements on the numerical scheme. This reflects a common phenomenon in multiscale modeling.
(iii) Among the spatial discretization schemes, MLW scheme with additional viscosity based on the distribution function, most effectively enhances stability but alters the constitutive relations, requiring a trade-off between accuracy and robustness.
WENO scheme provides superior accuracy but poor stability due to insufficient numerical dissipation. While the NND scheme achieves a favorable balance, when coupled with moment-matching method, owing to appropriate coupling between phase-space and spatial discretizations.
(iv) Initial macroscopic conditions affect model numerical stability primarily through Mach number variations, with influence ranking  $\rho \ll T < u$.
(v) Larger relative time steps weaken the coupling between flow- and collision-induced nonequilibrium, and higher time-space step ratios allow particle information to propagate beyond the resolution scale, both of which degrade numerical stability.

To propose a stability control strategy, we construct stability-phase diagrams for various DBMs under different TNE levels within the moment-matching framework by systematically varying the discretization parameters ($c$, ${\eta_0}$).
The fundamental particle speed $c$ has a more pronounced impact on stability than the energy control parameter ${\eta_0}$, as the former is closely associated with flow velocity, while the latter primarily relates to temperature. Moreover, in both the moment relations and the distribution function, the power of flow velocity typically appears at twice the order of that of temperature, amplifying its influence on numerical stability.
Since stability domains vary complexly with wavenumber, DVS, macroscopic variables, model parameters, and spatiotemporal schemes, morphological analysis is performed to evaluate stability probabilities across the full wavenumber spectrum and under diverse flow conditions.
The resulting probability curves demonstrate that the degree of matching among DVS, wavenumber, and macroscopic variables is a key determinant of stability.

We validate the proposed strategy through numerical simulations across a range of TNE conditions and DBM orders. Evaluation metrics include consistency between numerical and analytical TNE results, as well as the range, standard deviation, and spatial gradient of the distribution function. Results show the strategy remains effective across multiscale scenarios. Moreover, we clarify the impact of DVS on achieving physical accuracy and numerical stability, establishing a guideline: an effective DVS should preserve spatial symmetry while incorporating a diverse set of discrete velocities with sufficiently distinct directions and magnitudes, enabling smooth, continuous activation of rich $f_i$ patterns and accurate realization of multiscale physics.
Our results demonstrate that no single DVS achieves uniformly optimal performance. Instead, DVS selection should be flow-specific, reflecting the trade-offs among stability, physical accuracy, and computational efficiency.
Comparing algorithmic features is meaningful only when the models own the same physical functionality. Some of the observations in this study may help strike a balance between physical functions requirements and numerical algorithmic demands.

Despite this progress, several challenges remain.
(i) This study considers up to second-order TNE; yet, higher-order TNE interactions grow significantly under strong discreteness and nonlinearity, requiring further investigation into the stability of higher-order DBMs.
(ii) All current models are two-dimensional; in 3D cases, increased spatial dimensionality and nonequilibrium drivers lead to more complex coupling, demanding advanced DVS optimization and stability frameworks.
(iii) von Neumann analysis is inherently initial-state-based. As flows evolve nonlinearly, the stability domain of the initial ($c$, ${\eta_0}$) pair may become invalid. In practical flows, the spatiotemporal evolution of macroscopic fields necessitates time-adaptive alignment of the DVS with local flow dynamics.
(iv) While von Neumann analysis offers efficient linear stability diagnostics, key nonlinear phenomena—including modal interactions, subcritical instabilities, and energy feedback mechanisms \cite{tripathi2024global,dubois2024beyond}—lie beyond the scope of linear criteria.

To further advance the stability analysis of DBM simulation in supersonic flows, future work will focus on multidimensional models that incorporate higher-order TNE effects. To ensure stability throughout the entire evolution process, we propose a three-step strategy: (a) assess the stability-sustaining capacity of various initial configurations, (b) identify optimal parameter sets, and (c) dynamically update parameters during flow evolution, following von Neumann criteria.
In future work, we also plan to introduce tools such as energy norm analysis, Lyapunov-based methods, and local modal spectrum analysis to systematically investigate the nonlinear stability of the DBM \cite{lakshmikantham1989stability,sastry2013nonlinear}.
These efforts are expected to offer theoretical insights and practical methodologies for stability analysis and control in higher-order kinetic models, expanding DBM applications in aerospace, defense engineering, and advanced energy systems.

\begin{acknowledgments}
We acknowledge support from the National Natural Science Foundation of China
(Grant Nos. 52278119, 11875001, and 12172061), Hebei Outstanding Youth Science Foundation (Grant No. A2023409003), Central Guidance on Local Science and Technology Development Fund of Hebei Province (Grant No. 226Z7601G), and Science Foundation of NCIAE (Grant Nos. ZD-2025-06 and YKY-2024-77),
the Foundation of National Key Laboratory of
Shock Wave and Detonation Physics (Grant no. JCKYS2023212003), the Opening Project of State Key Laboratory of Explosion Science and Safety Protection (Beijing Institute of Technology) (Grant No. KFJJ25-02M).
\end{acknowledgments}

\section*{Data Availability}
The data that support the findings of this study are available from the corresponding author upon reasonable request.

\section*{References}
\bibliography{Stability_Analysis}

\begin{thebibliography}{130}%
\makeatletter
\providecommand \@ifxundefined [1]{%
 \@ifx{#1\undefined}
}%
\providecommand \@ifnum [1]{%
 \ifnum #1\expandafter \@firstoftwo
 \else \expandafter \@secondoftwo
 \fi
}%
\providecommand \@ifx [1]{%
 \ifx #1\expandafter \@firstoftwo
 \else \expandafter \@secondoftwo
 \fi
}%
\providecommand \natexlab [1]{#1}%
\providecommand \enquote  [1]{``#1''}%
\providecommand \bibnamefont  [1]{#1}%
\providecommand \bibfnamefont [1]{#1}%
\providecommand \citenamefont [1]{#1}%
\providecommand \href@noop [0]{\@secondoftwo}%
\providecommand \href [0]{\begingroup \@sanitize@url \@href}%
\providecommand \@href[1]{\@@startlink{#1}\@@href}%
\providecommand \@@href[1]{\endgroup#1\@@endlink}%
\providecommand \@sanitize@url [0]{\catcode `\\12\catcode `\$12\catcode
  `\&12\catcode `\#12\catcode `\^12\catcode `\_12\catcode `\%12\relax}%
\providecommand \@@startlink[1]{}%
\providecommand \@@endlink[0]{}%
\providecommand \url  [0]{\begingroup\@sanitize@url \@url }%
\providecommand \@url [1]{\endgroup\@href {#1}{\urlprefix }}%
\providecommand \urlprefix  [0]{URL }%
\providecommand \Eprint [0]{\href }%
\providecommand \doibase [0]{http://dx.doi.org/}%
\providecommand \selectlanguage [0]{\@gobble}%
\providecommand \bibinfo  [0]{\@secondoftwo}%
\providecommand \bibfield  [0]{\@secondoftwo}%
\providecommand \translation [1]{[#1]}%
\providecommand \BibitemOpen [0]{}%
\providecommand \bibitemStop [0]{}%
\providecommand \bibitemNoStop [0]{.\EOS\space}%
\providecommand \EOS [0]{\spacefactor3000\relax}%
\providecommand \BibitemShut  [1]{\csname bibitem#1\endcsname}%
\let\auto@bib@innerbib\@empty
\bibitem [{\citenamefont {Courant}\ and\ \citenamefont
  {Friedrichs}(1999)}]{courant1999supersonic}%
  \BibitemOpen
  \bibfield  {author} {\bibinfo {author} {\bibfnamefont {R.}~\bibnamefont
  {Courant}}\ and\ \bibinfo {author} {\bibfnamefont {K.~O.}\ \bibnamefont
  {Friedrichs}},\ }\href@noop {} {\emph {\bibinfo {title} {{Supersonic Flow and
  Shock Waves}}}}\ (\bibinfo  {publisher} {Springer Science \& Business
  Media},\ \bibinfo {year} {1999})\BibitemShut {NoStop}%
\bibitem [{\citenamefont {Bertin}\ and\ \citenamefont
  {Cummings}(2006)}]{bertin2006critical}%
  \BibitemOpen
  \bibfield  {author} {\bibinfo {author} {\bibfnamefont {J.~J.}\ \bibnamefont
  {Bertin}}\ and\ \bibinfo {author} {\bibfnamefont {R.~M.}\ \bibnamefont
  {Cummings}},\ }\bibfield  {title} {\enquote {\bibinfo {title} {Critical
  hypersonic aerothermodynamic phenomena},}\ }\href@noop {} {\bibfield
  {journal} {\bibinfo  {journal} {Annu. Rev. Fluid Mech.}\ }\textbf {\bibinfo
  {volume} {38}},\ \bibinfo {pages} {129--157} (\bibinfo {year}
  {2006})}\BibitemShut {NoStop}%
\bibitem [{\citenamefont {Candler}(2019)}]{candler2019rate}%
  \BibitemOpen
  \bibfield  {author} {\bibinfo {author} {\bibfnamefont {G.~V.}\ \bibnamefont
  {Candler}},\ }\bibfield  {title} {\enquote {\bibinfo {title} {Rate effects in
  hypersonic flows},}\ }\href@noop {} {\bibfield  {journal} {\bibinfo
  {journal} {Annu. Rev. Fluid Mech.}\ }\textbf {\bibinfo {volume} {51}},\
  \bibinfo {pages} {379--402} (\bibinfo {year} {2019})}\BibitemShut {NoStop}%
\bibitem [{\citenamefont {Gaitonde}(2015)}]{gaitonde2015progress}%
  \BibitemOpen
  \bibfield  {author} {\bibinfo {author} {\bibfnamefont {D.~V.}\ \bibnamefont
  {Gaitonde}},\ }\bibfield  {title} {\enquote {\bibinfo {title} {Progress in
  shock wave/boundary layer interactions},}\ }\href@noop {} {\bibfield
  {journal} {\bibinfo  {journal} {Prog. Aerosp. Sci.}\ }\textbf {\bibinfo
  {volume} {72}},\ \bibinfo {pages} {80--99} (\bibinfo {year}
  {2015})}\BibitemShut {NoStop}%
\bibitem [{\citenamefont {Huh}\ and\ \citenamefont
  {Lee}(2018)}]{huh2018numerical}%
  \BibitemOpen
  \bibfield  {author} {\bibinfo {author} {\bibfnamefont {J.}~\bibnamefont
  {Huh}}\ and\ \bibinfo {author} {\bibfnamefont {S.}~\bibnamefont {Lee}},\
  }\bibfield  {title} {\enquote {\bibinfo {title} {Numerical study on lateral
  jet interaction in supersonic crossflows},}\ }\href@noop {} {\bibfield
  {journal} {\bibinfo  {journal} {Aerosp. Sci. Technol.}\ }\textbf {\bibinfo
  {volume} {80}},\ \bibinfo {pages} {315--328} (\bibinfo {year}
  {2018})}\BibitemShut {NoStop}%
\bibitem [{\citenamefont {Pan}\ and\ \citenamefont
  {Scannapieco}(2010)}]{pan2010mixing}%
  \BibitemOpen
  \bibfield  {author} {\bibinfo {author} {\bibfnamefont {L.}~\bibnamefont
  {Pan}}\ and\ \bibinfo {author} {\bibfnamefont {E.}~\bibnamefont
  {Scannapieco}},\ }\bibfield  {title} {\enquote {\bibinfo {title} {Mixing in
  supersonic turbulence},}\ }\href@noop {} {\bibfield  {journal} {\bibinfo
  {journal} {Astrophys. J.}\ }\textbf {\bibinfo {volume} {721}},\ \bibinfo
  {pages} {1765} (\bibinfo {year} {2010})}\BibitemShut {NoStop}%
\bibitem [{\citenamefont {Cao}\ \emph {et~al.}(2024)\citenamefont {Cao},
  \citenamefont {Tang}, \citenamefont {Zhu},\ and\ \citenamefont
  {He}}]{cao2024data}%
  \BibitemOpen
  \bibfield  {author} {\bibinfo {author} {\bibfnamefont {F.}~\bibnamefont
  {Cao}}, \bibinfo {author} {\bibfnamefont {Z.}~\bibnamefont {Tang}}, \bibinfo
  {author} {\bibfnamefont {C.}~\bibnamefont {Zhu}}, \ and\ \bibinfo {author}
  {\bibfnamefont {X.}~\bibnamefont {He}},\ }\bibfield  {title} {\enquote
  {\bibinfo {title} {Data-driven hierarchical collaborative optimization method
  with multi-fidelity modeling for aerodynamic optimization},}\ }\href@noop {}
  {\bibfield  {journal} {\bibinfo  {journal} {Aerosp. Sci. Technol.}\ }\textbf
  {\bibinfo {volume} {150}},\ \bibinfo {pages} {109206} (\bibinfo {year}
  {2024})}\BibitemShut {NoStop}%
\bibitem [{\citenamefont {Gallais}(2007)}]{gallais2007atmospheric}%
  \BibitemOpen
  \bibfield  {author} {\bibinfo {author} {\bibfnamefont {P.}~\bibnamefont
  {Gallais}},\ }\href@noop {} {\emph {\bibinfo {title} {{Atmospheric Re-Entry
  Vehicle Mechanics}}}}\ (\bibinfo  {publisher} {Springer Science \& Business
  Media},\ \bibinfo {year} {2007})\BibitemShut {NoStop}%
\bibitem [{\citenamefont {Poloni}\ \emph {et~al.}(2022)\citenamefont {Poloni},
  \citenamefont {Bouville}, \citenamefont {Schmid}, \citenamefont {Pelissari},
  \citenamefont {Pandolfelli}, \citenamefont {Sousa}, \citenamefont {Tervoort},
  \citenamefont {Christidis}, \citenamefont {Shklover}, \citenamefont
  {Leuthold} \emph {et~al.}}]{poloni2022carbon}%
  \BibitemOpen
  \bibfield  {author} {\bibinfo {author} {\bibfnamefont {E.}~\bibnamefont
  {Poloni}}, \bibinfo {author} {\bibfnamefont {F.}~\bibnamefont {Bouville}},
  \bibinfo {author} {\bibfnamefont {A.~L.}\ \bibnamefont {Schmid}}, \bibinfo
  {author} {\bibfnamefont {P.~I.}\ \bibnamefont {Pelissari}}, \bibinfo {author}
  {\bibfnamefont {V.~C.}\ \bibnamefont {Pandolfelli}}, \bibinfo {author}
  {\bibfnamefont {M.~L.}\ \bibnamefont {Sousa}}, \bibinfo {author}
  {\bibfnamefont {E.}~\bibnamefont {Tervoort}}, \bibinfo {author}
  {\bibfnamefont {G.}~\bibnamefont {Christidis}}, \bibinfo {author}
  {\bibfnamefont {V.}~\bibnamefont {Shklover}}, \bibinfo {author}
  {\bibfnamefont {J.}~\bibnamefont {Leuthold}},  \emph {et~al.},\ }\bibfield
  {title} {\enquote {\bibinfo {title} {Carbon ablators with porosity tailored
  for aerospace thermal protection during atmospheric re-entry},}\ }\href@noop
  {} {\bibfield  {journal} {\bibinfo  {journal} {Carbon}\ }\textbf {\bibinfo
  {volume} {195}},\ \bibinfo {pages} {80--91} (\bibinfo {year}
  {2022})}\BibitemShut {NoStop}%
\bibitem [{\citenamefont {Du}\ \emph {et~al.}(2023)\citenamefont {Du},
  \citenamefont {Sun}, \citenamefont {Tan}, \citenamefont {Huang},
  \citenamefont {Yan}, \citenamefont {Meng}, \citenamefont {Chen},\ and\
  \citenamefont {Wang}}]{du2023numerical}%
  \BibitemOpen
  \bibfield  {author} {\bibinfo {author} {\bibfnamefont {Y.}~\bibnamefont
  {Du}}, \bibinfo {author} {\bibfnamefont {S.}~\bibnamefont {Sun}}, \bibinfo
  {author} {\bibfnamefont {M.}~\bibnamefont {Tan}}, \bibinfo {author}
  {\bibfnamefont {H.}~\bibnamefont {Huang}}, \bibinfo {author} {\bibfnamefont
  {C.}~\bibnamefont {Yan}}, \bibinfo {author} {\bibfnamefont {X.}~\bibnamefont
  {Meng}}, \bibinfo {author} {\bibfnamefont {X.}~\bibnamefont {Chen}}, \ and\
  \bibinfo {author} {\bibfnamefont {H.}~\bibnamefont {Wang}},\ }\bibfield
  {title} {\enquote {\bibinfo {title} {Numerical study on the non-equilibrium
  characteristics of high-speed atmospheric re-entry flow and radiation of
  aircraft based on fully coupled model},}\ }\href@noop {} {\bibfield
  {journal} {\bibinfo  {journal} {J. Fluid Mech.}\ }\textbf {\bibinfo {volume}
  {977}},\ \bibinfo {pages} {A39} (\bibinfo {year} {2023})}\BibitemShut
  {NoStop}%
\bibitem [{\citenamefont {Jiang}\ \emph {et~al.}(2021)\citenamefont {Jiang},
  \citenamefont {Hajivand}, \citenamefont {Sadeghi}, \citenamefont
  {Gerdroodbary},\ and\ \citenamefont {Li}}]{jiang2021influence}%
  \BibitemOpen
  \bibfield  {author} {\bibinfo {author} {\bibfnamefont {Y.}~\bibnamefont
  {Jiang}}, \bibinfo {author} {\bibfnamefont {M.}~\bibnamefont {Hajivand}},
  \bibinfo {author} {\bibfnamefont {H.}~\bibnamefont {Sadeghi}}, \bibinfo
  {author} {\bibfnamefont {M.~B.}\ \bibnamefont {Gerdroodbary}}, \ and\
  \bibinfo {author} {\bibfnamefont {Z.}~\bibnamefont {Li}},\ }\bibfield
  {title} {\enquote {\bibinfo {title} {Influence of trapezoidal lobe strut on
  fuel mixing and combustion in supersonic combustion chamber},}\ }\href@noop
  {} {\bibfield  {journal} {\bibinfo  {journal} {Aerosp. Sci. Technol.}\
  }\textbf {\bibinfo {volume} {116}},\ \bibinfo {pages} {106841} (\bibinfo
  {year} {2021})}\BibitemShut {NoStop}%
\bibitem [{\citenamefont {Emelyanov}, \citenamefont {Pustovalov},\ and\
  \citenamefont {Volkov}(2019)}]{emelyanov2019supersonic}%
  \BibitemOpen
  \bibfield  {author} {\bibinfo {author} {\bibfnamefont {V.~N.}\ \bibnamefont
  {Emelyanov}}, \bibinfo {author} {\bibfnamefont {A.~V.}\ \bibnamefont
  {Pustovalov}}, \ and\ \bibinfo {author} {\bibfnamefont {K.~N.}\ \bibnamefont
  {Volkov}},\ }\bibfield  {title} {\enquote {\bibinfo {title} {Supersonic jet
  and nozzle flows in uniform-flow and free-vortex aerodynamic windows of gas
  lasers},}\ }\href@noop {} {\bibfield  {journal} {\bibinfo  {journal} {Acta
  Astronaut.}\ }\textbf {\bibinfo {volume} {163}},\ \bibinfo {pages} {232--243}
  (\bibinfo {year} {2019})}\BibitemShut {NoStop}%
\bibitem [{\citenamefont {Tahsini}\ and\ \citenamefont
  {Mousavi}(2015)}]{tahsini2015investigating}%
  \BibitemOpen
  \bibfield  {author} {\bibinfo {author} {\bibfnamefont {A.~M.}\ \bibnamefont
  {Tahsini}}\ and\ \bibinfo {author} {\bibfnamefont {S.~T.}\ \bibnamefont
  {Mousavi}},\ }\bibfield  {title} {\enquote {\bibinfo {title} {Investigating
  the supersonic combustion efficiency for the jet-in-cross-flow},}\
  }\href@noop {} {\bibfield  {journal} {\bibinfo  {journal} {Int. J. Hydrogen
  Energy}\ }\textbf {\bibinfo {volume} {40}},\ \bibinfo {pages} {3091--3097}
  (\bibinfo {year} {2015})}\BibitemShut {NoStop}%
\bibitem [{\citenamefont {Huang}\ \emph {et~al.}(2013)\citenamefont {Huang},
  \citenamefont {Liu}, \citenamefont {Yan},\ and\ \citenamefont
  {Jin}}]{huang2013multiobjective}%
  \BibitemOpen
  \bibfield  {author} {\bibinfo {author} {\bibfnamefont {W.}~\bibnamefont
  {Huang}}, \bibinfo {author} {\bibfnamefont {J.}~\bibnamefont {Liu}}, \bibinfo
  {author} {\bibfnamefont {L.}~\bibnamefont {Yan}}, \ and\ \bibinfo {author}
  {\bibfnamefont {L.}~\bibnamefont {Jin}},\ }\bibfield  {title} {\enquote
  {\bibinfo {title} {Multiobjective design optimization of the performance for
  the cavity flameholder in supersonic flows},}\ }\href@noop {} {\bibfield
  {journal} {\bibinfo  {journal} {Aerosp. Sci. Technol.}\ }\textbf {\bibinfo
  {volume} {30}},\ \bibinfo {pages} {246--254} (\bibinfo {year}
  {2013})}\BibitemShut {NoStop}%
\bibitem [{\citenamefont {Gan}\ \emph {et~al.}(2025)\citenamefont {Gan},
  \citenamefont {Zhuang}, \citenamefont {Yang}, \citenamefont {Xu},
  \citenamefont {Zhang}, \citenamefont {Chen}, \citenamefont {Song},\ and\
  \citenamefont {Wu}}]{gan2025supersonic}%
  \BibitemOpen
  \bibfield  {author} {\bibinfo {author} {\bibfnamefont {Y.}~\bibnamefont
  {Gan}}, \bibinfo {author} {\bibfnamefont {Z.}~\bibnamefont {Zhuang}},
  \bibinfo {author} {\bibfnamefont {B.}~\bibnamefont {Yang}}, \bibinfo {author}
  {\bibfnamefont {A.}~\bibnamefont {Xu}}, \bibinfo {author} {\bibfnamefont
  {D.}~\bibnamefont {Zhang}}, \bibinfo {author} {\bibfnamefont
  {F.}~\bibnamefont {Chen}}, \bibinfo {author} {\bibfnamefont {J.}~\bibnamefont
  {Song}}, \ and\ \bibinfo {author} {\bibfnamefont {Y.}~\bibnamefont {Wu}},\
  }\bibfield  {title} {\enquote {\bibinfo {title} {{Supersonic flow kinetics:
  Mesoscale structures, thermodynamic nonequilibrium effects and entropy
  production mechanisms}},}\ }\href@noop {} {\bibfield  {journal} {\bibinfo
  {journal} {arXiv:2502.10832}\ } (\bibinfo {year} {2025})}\BibitemShut
  {NoStop}%
\bibitem [{\citenamefont {Xu}\ \emph {et~al.}(2012)\citenamefont {Xu},
  \citenamefont {Zhang}, \citenamefont {Gan}, \citenamefont {Chen},\ and\
  \citenamefont {Yu}}]{xu2012lattice}%
  \BibitemOpen
  \bibfield  {author} {\bibinfo {author} {\bibfnamefont {A.}~\bibnamefont
  {Xu}}, \bibinfo {author} {\bibfnamefont {G.}~\bibnamefont {Zhang}}, \bibinfo
  {author} {\bibfnamefont {Y.}~\bibnamefont {Gan}}, \bibinfo {author}
  {\bibfnamefont {F.}~\bibnamefont {Chen}}, \ and\ \bibinfo {author}
  {\bibfnamefont {X.}~\bibnamefont {Yu}},\ }\bibfield  {title} {\enquote
  {\bibinfo {title} {Lattice {B}oltzmann modeling and simulation of
  compressible flows},}\ }\href@noop {} {\bibfield  {journal} {\bibinfo
  {journal} {Front. Phys.}\ }\textbf {\bibinfo {volume} {7}},\ \bibinfo {pages}
  {582--600} (\bibinfo {year} {2012})}\BibitemShut {NoStop}%
\bibitem [{\citenamefont {Xu}\ and\ \citenamefont {Zhang}(2019)}]{xu2022BSTP}%
  \BibitemOpen
  \bibfield  {author} {\bibinfo {author} {\bibfnamefont {A.}~\bibnamefont
  {Xu}}\ and\ \bibinfo {author} {\bibfnamefont {Y.}~\bibnamefont {Zhang}},\
  }\href@noop {} {\emph {\bibinfo {title} {{Complex Media Kinetics}}}}\
  (\bibinfo  {publisher} {Beijing Sci. Tech. Press},\ \bibinfo {year}
  {2019})\BibitemShut {NoStop}%
\bibitem [{\citenamefont {Xu}, \citenamefont {Zhang},\ and\ \citenamefont
  {Gan}(2024)}]{xu2024advances}%
  \BibitemOpen
  \bibfield  {author} {\bibinfo {author} {\bibfnamefont {A.}~\bibnamefont
  {Xu}}, \bibinfo {author} {\bibfnamefont {D.}~\bibnamefont {Zhang}}, \ and\
  \bibinfo {author} {\bibfnamefont {Y.}~\bibnamefont {Gan}},\ }\bibfield
  {title} {\enquote {\bibinfo {title} {Advances in the kinetics of heat and
  mass transfer in near-continuous complex flows},}\ }\href@noop {} {\bibfield
  {journal} {\bibinfo  {journal} {Front. Phys.}\ }\textbf {\bibinfo {volume}
  {19}},\ \bibinfo {pages} {42500} (\bibinfo {year} {2024})}\BibitemShut
  {NoStop}%
\bibitem [{\citenamefont {Zapryagaev}, \citenamefont {Kiselev},\ and\
  \citenamefont {Gubanov}(2018)}]{zapryagaev2018shock}%
  \BibitemOpen
  \bibfield  {author} {\bibinfo {author} {\bibfnamefont {V.}~\bibnamefont
  {Zapryagaev}}, \bibinfo {author} {\bibfnamefont {N.}~\bibnamefont {Kiselev}},
  \ and\ \bibinfo {author} {\bibfnamefont {D.}~\bibnamefont {Gubanov}},\
  }\bibfield  {title} {\enquote {\bibinfo {title} {Shock-wave structure of
  supersonic jet flows},}\ }\href@noop {} {\bibfield  {journal} {\bibinfo
  {journal} {Aerospace}\ }\textbf {\bibinfo {volume} {5}},\ \bibinfo {pages}
  {60} (\bibinfo {year} {2018})}\BibitemShut {NoStop}%
\bibitem [{\citenamefont {Gaitonde}\ and\ \citenamefont
  {Adler}(2023)}]{gaitonde2023dynamics}%
  \BibitemOpen
  \bibfield  {author} {\bibinfo {author} {\bibfnamefont {D.~V.}\ \bibnamefont
  {Gaitonde}}\ and\ \bibinfo {author} {\bibfnamefont {M.~C.}\ \bibnamefont
  {Adler}},\ }\bibfield  {title} {\enquote {\bibinfo {title} {Dynamics of
  three-dimensional shock-wave/boundary-layer interactions},}\ }\href@noop {}
  {\bibfield  {journal} {\bibinfo  {journal} {Annu. Rev. Fluid Mech.}\ }\textbf
  {\bibinfo {volume} {55}},\ \bibinfo {pages} {291--321} (\bibinfo {year}
  {2023})}\BibitemShut {NoStop}%
\bibitem [{\citenamefont {Andreopoulos}, \citenamefont {Agui},\ and\
  \citenamefont {Briassulis}(2000)}]{andreopoulos2000shock}%
  \BibitemOpen
  \bibfield  {author} {\bibinfo {author} {\bibfnamefont {Y.}~\bibnamefont
  {Andreopoulos}}, \bibinfo {author} {\bibfnamefont {J.~H.}\ \bibnamefont
  {Agui}}, \ and\ \bibinfo {author} {\bibfnamefont {G.}~\bibnamefont
  {Briassulis}},\ }\bibfield  {title} {\enquote {\bibinfo {title} {Shock
  wave—turbulence interactions},}\ }\href@noop {} {\bibfield  {journal}
  {\bibinfo  {journal} {Annu. Rev. Fluid Mech.}\ }\textbf {\bibinfo {volume}
  {32}},\ \bibinfo {pages} {309--345} (\bibinfo {year} {2000})}\BibitemShut
  {NoStop}%
\bibitem [{\citenamefont {Sun}\ \emph {et~al.}(2019)\citenamefont {Sun},
  \citenamefont {Huang}, \citenamefont {Ou}, \citenamefont {Zhang},\ and\
  \citenamefont {Li}}]{xiwan2019survey}%
  \BibitemOpen
  \bibfield  {author} {\bibinfo {author} {\bibfnamefont {X.}~\bibnamefont
  {Sun}}, \bibinfo {author} {\bibfnamefont {W.}~\bibnamefont {Huang}}, \bibinfo
  {author} {\bibfnamefont {M.}~\bibnamefont {Ou}}, \bibinfo {author}
  {\bibfnamefont {R.}~\bibnamefont {Zhang}}, \ and\ \bibinfo {author}
  {\bibfnamefont {S.}~\bibnamefont {Li}},\ }\bibfield  {title} {\enquote
  {\bibinfo {title} {A survey on numerical simulations of drag and heat
  reduction mechanism in supersonic/hypersonic flows},}\ }\href@noop {}
  {\bibfield  {journal} {\bibinfo  {journal} {Chin. J. Aeronaut.}\ }\textbf
  {\bibinfo {volume} {32}},\ \bibinfo {pages} {771--784} (\bibinfo {year}
  {2019})}\BibitemShut {NoStop}%
\bibitem [{\citenamefont {Boccelli}\ \emph {et~al.}(2024)\citenamefont
  {Boccelli}, \citenamefont {Kaufmann}, \citenamefont {Magin},\ and\
  \citenamefont {McDonald}}]{boccelli2024numerical}%
  \BibitemOpen
  \bibfield  {author} {\bibinfo {author} {\bibfnamefont {S.}~\bibnamefont
  {Boccelli}}, \bibinfo {author} {\bibfnamefont {W.}~\bibnamefont {Kaufmann}},
  \bibinfo {author} {\bibfnamefont {T.~E.}\ \bibnamefont {Magin}}, \ and\
  \bibinfo {author} {\bibfnamefont {J.~G.}\ \bibnamefont {McDonald}},\
  }\bibfield  {title} {\enquote {\bibinfo {title} {Numerical simulation of
  rarefied supersonic flows using a fourth-order maximum-entropy moment method
  with interpolative closure},}\ }\href@noop {} {\bibfield  {journal} {\bibinfo
   {journal} {J. Comput. Phys.}\ }\textbf {\bibinfo {volume} {497}},\ \bibinfo
  {pages} {112631} (\bibinfo {year} {2024})}\BibitemShut {NoStop}%
\bibitem [{\citenamefont {Ding}\ \emph {et~al.}(2022)\citenamefont {Ding},
  \citenamefont {Zhang}, \citenamefont {Sun}, \citenamefont {Yang},\ and\
  \citenamefont {Wen}}]{ding2022numerical}%
  \BibitemOpen
  \bibfield  {author} {\bibinfo {author} {\bibfnamefont {H.}~\bibnamefont
  {Ding}}, \bibinfo {author} {\bibfnamefont {Y.}~\bibnamefont {Zhang}},
  \bibinfo {author} {\bibfnamefont {C.}~\bibnamefont {Sun}}, \bibinfo {author}
  {\bibfnamefont {Y.}~\bibnamefont {Yang}}, \ and\ \bibinfo {author}
  {\bibfnamefont {C.}~\bibnamefont {Wen}},\ }\bibfield  {title} {\enquote
  {\bibinfo {title} {{Numerical simulation of supersonic condensation flows
  using Eulerian-Lagrangian and Eulerian wall film models}},}\ }\href@noop {}
  {\bibfield  {journal} {\bibinfo  {journal} {Energy}\ }\textbf {\bibinfo
  {volume} {258}},\ \bibinfo {pages} {124833} (\bibinfo {year}
  {2022})}\BibitemShut {NoStop}%
\bibitem [{\citenamefont {Bernardini}\ \emph {et~al.}(2023)\citenamefont
  {Bernardini}, \citenamefont {Modesti}, \citenamefont {Salvadore},
  \citenamefont {Sathyanarayana}, \citenamefont {Della~Posta},\ and\
  \citenamefont {Pirozzoli}}]{bernardini2023streams}%
  \BibitemOpen
  \bibfield  {author} {\bibinfo {author} {\bibfnamefont {M.}~\bibnamefont
  {Bernardini}}, \bibinfo {author} {\bibfnamefont {D.}~\bibnamefont {Modesti}},
  \bibinfo {author} {\bibfnamefont {F.}~\bibnamefont {Salvadore}}, \bibinfo
  {author} {\bibfnamefont {S.}~\bibnamefont {Sathyanarayana}}, \bibinfo
  {author} {\bibfnamefont {G.}~\bibnamefont {Della~Posta}}, \ and\ \bibinfo
  {author} {\bibfnamefont {S.}~\bibnamefont {Pirozzoli}},\ }\bibfield  {title}
  {\enquote {\bibinfo {title} {{STREAmS-2.0: Supersonic turbulent accelerated
  Navier-Stokes solver version 2.0}},}\ }\href@noop {} {\bibfield  {journal}
  {\bibinfo  {journal} {Comput. Phys. Commun.}\ }\textbf {\bibinfo {volume}
  {285}},\ \bibinfo {pages} {108644} (\bibinfo {year} {2023})}\BibitemShut
  {NoStop}%
\bibitem [{\citenamefont {Kl{\'\i}ma}\ and\ \citenamefont
  {Kolafa}(2018)}]{klima2018direct}%
  \BibitemOpen
  \bibfield  {author} {\bibinfo {author} {\bibfnamefont {M.}~\bibnamefont
  {Kl{\'\i}ma}}\ and\ \bibinfo {author} {\bibfnamefont {J.}~\bibnamefont
  {Kolafa}},\ }\bibfield  {title} {\enquote {\bibinfo {title} {Direct molecular
  dynamics simulation of nucleation during supersonic expansion of gas to a
  vacuum},}\ }\href@noop {} {\bibfield  {journal} {\bibinfo  {journal} {J.
  Chem. Theory Comput.}\ }\textbf {\bibinfo {volume} {14}},\ \bibinfo {pages}
  {2332--2340} (\bibinfo {year} {2018})}\BibitemShut {NoStop}%
\bibitem [{\citenamefont {Li}\ \emph {et~al.}(2021)\citenamefont {Li},
  \citenamefont {Hekmatifar}, \citenamefont {Sun}, \citenamefont {Alizadeh},
  \citenamefont {Aly}, \citenamefont {Toghraie}, \citenamefont {Baleanu},\ and\
  \citenamefont {Sabetvand}}]{li2021evaluation}%
  \BibitemOpen
  \bibfield  {author} {\bibinfo {author} {\bibfnamefont {Y.}~\bibnamefont
  {Li}}, \bibinfo {author} {\bibfnamefont {M.}~\bibnamefont {Hekmatifar}},
  \bibinfo {author} {\bibfnamefont {Y.}~\bibnamefont {Sun}}, \bibinfo {author}
  {\bibfnamefont {A.}~\bibnamefont {Alizadeh}}, \bibinfo {author}
  {\bibfnamefont {A.~A.}\ \bibnamefont {Aly}}, \bibinfo {author} {\bibfnamefont
  {D.}~\bibnamefont {Toghraie}}, \bibinfo {author} {\bibfnamefont
  {D.}~\bibnamefont {Baleanu}}, \ and\ \bibinfo {author} {\bibfnamefont
  {R.}~\bibnamefont {Sabetvand}},\ }\bibfield  {title} {\enquote {\bibinfo
  {title} {Evaluation the vibrational behavior of carbon nanotubes in different
  sizes and chiralities and argon flows at supersonic velocity using molecular
  dynamics simulation},}\ }\href@noop {} {\bibfield  {journal} {\bibinfo
  {journal} {J. Mol. Liq.}\ }\textbf {\bibinfo {volume} {339}},\ \bibinfo
  {pages} {116796} (\bibinfo {year} {2021})}\BibitemShut {NoStop}%
\bibitem [{\citenamefont {Tong}, \citenamefont {He},\ and\ \citenamefont
  {Tao}(2019)}]{tong2019review}%
  \BibitemOpen
  \bibfield  {author} {\bibinfo {author} {\bibfnamefont {Z.}~\bibnamefont
  {Tong}}, \bibinfo {author} {\bibfnamefont {Y.}~\bibnamefont {He}}, \ and\
  \bibinfo {author} {\bibfnamefont {W.}~\bibnamefont {Tao}},\ }\bibfield
  {title} {\enquote {\bibinfo {title} {{A review of current progress in
  multiscale simulations for fluid flow and heat transfer problems: The
  frameworks, coupling techniques and future perspectives}},}\ }\href@noop {}
  {\bibfield  {journal} {\bibinfo  {journal} {Int. J. Heat Mass Transf.}\
  }\textbf {\bibinfo {volume} {137}},\ \bibinfo {pages} {1263--1289} (\bibinfo
  {year} {2019})}\BibitemShut {NoStop}%
\bibitem [{\citenamefont {Qiu}\ \emph {et~al.}(2020{\natexlab{a}})\citenamefont
  {Qiu}, \citenamefont {Bao}, \citenamefont {Zhou}, \citenamefont {Che},
  \citenamefont {Chen},\ and\ \citenamefont {You}}]{qiu2020study}%
  \BibitemOpen
  \bibfield  {author} {\bibinfo {author} {\bibfnamefont {R.}~\bibnamefont
  {Qiu}}, \bibinfo {author} {\bibfnamefont {Y.}~\bibnamefont {Bao}}, \bibinfo
  {author} {\bibfnamefont {T.}~\bibnamefont {Zhou}}, \bibinfo {author}
  {\bibfnamefont {H.}~\bibnamefont {Che}}, \bibinfo {author} {\bibfnamefont
  {R.}~\bibnamefont {Chen}}, \ and\ \bibinfo {author} {\bibfnamefont
  {Y.}~\bibnamefont {You}},\ }\bibfield  {title} {\enquote {\bibinfo {title}
  {{Study of regular reflection shock waves using a mesoscopic kinetic
  approach: Curvature pattern and effects of viscosity}},}\ }\href@noop {}
  {\bibfield  {journal} {\bibinfo  {journal} {Phys. Fluids}\ }\textbf {\bibinfo
  {volume} {32}} (\bibinfo {year} {2020}{\natexlab{a}})}\BibitemShut {NoStop}%
\bibitem [{\citenamefont {Xu}, \citenamefont {Mao},\ and\ \citenamefont
  {Tang}(2005)}]{xu2005multidimensional}%
  \BibitemOpen
  \bibfield  {author} {\bibinfo {author} {\bibfnamefont {K.}~\bibnamefont
  {Xu}}, \bibinfo {author} {\bibfnamefont {M.}~\bibnamefont {Mao}}, \ and\
  \bibinfo {author} {\bibfnamefont {L.}~\bibnamefont {Tang}},\ }\bibfield
  {title} {\enquote {\bibinfo {title} {{A multidimensional gas-kinetic BGK
  scheme for hypersonic viscous flow}},}\ }\href@noop {} {\bibfield  {journal}
  {\bibinfo  {journal} {J. Comput. Phys.}\ }\textbf {\bibinfo {volume} {203}},\
  \bibinfo {pages} {405--421} (\bibinfo {year} {2005})}\BibitemShut {NoStop}%
\bibitem [{\citenamefont {Sun}\ \emph {et~al.}(2021)\citenamefont {Sun},
  \citenamefont {Qu}, \citenamefont {Liu}, \citenamefont {Yao},\ and\
  \citenamefont {Bai}}]{sun2021numerical}%
  \BibitemOpen
  \bibfield  {author} {\bibinfo {author} {\bibfnamefont {D.}~\bibnamefont
  {Sun}}, \bibinfo {author} {\bibfnamefont {F.}~\bibnamefont {Qu}}, \bibinfo
  {author} {\bibfnamefont {C.}~\bibnamefont {Liu}}, \bibinfo {author}
  {\bibfnamefont {F.}~\bibnamefont {Yao}}, \ and\ \bibinfo {author}
  {\bibfnamefont {J.}~\bibnamefont {Bai}},\ }\bibfield  {title} {\enquote
  {\bibinfo {title} {Numerical study of the suction flow control of the
  supersonic boundary layer transition in a framework of gas-kinetic scheme},}\
  }\href@noop {} {\bibfield  {journal} {\bibinfo  {journal} {Aerosp. Sci.
  Technol.}\ }\textbf {\bibinfo {volume} {109}},\ \bibinfo {pages} {106397}
  (\bibinfo {year} {2021})}\BibitemShut {NoStop}%
\bibitem [{\citenamefont {Zhu}, \citenamefont {Zhong},\ and\ \citenamefont
  {Xu}(2016)}]{zhu2016implicit}%
  \BibitemOpen
  \bibfield  {author} {\bibinfo {author} {\bibfnamefont {Y.}~\bibnamefont
  {Zhu}}, \bibinfo {author} {\bibfnamefont {C.}~\bibnamefont {Zhong}}, \ and\
  \bibinfo {author} {\bibfnamefont {K.}~\bibnamefont {Xu}},\ }\bibfield
  {title} {\enquote {\bibinfo {title} {Implicit unified gas-kinetic scheme for
  steady state solutions in all flow regimes},}\ }\href@noop {} {\bibfield
  {journal} {\bibinfo  {journal} {J. Comput. Phys.}\ }\textbf {\bibinfo
  {volume} {315}},\ \bibinfo {pages} {16--38} (\bibinfo {year}
  {2016})}\BibitemShut {NoStop}%
\bibitem [{\citenamefont {Zhu}, \citenamefont {Zhong},\ and\ \citenamefont
  {Xu}(2017)}]{zhu2017unified}%
  \BibitemOpen
  \bibfield  {author} {\bibinfo {author} {\bibfnamefont {Y.}~\bibnamefont
  {Zhu}}, \bibinfo {author} {\bibfnamefont {C.}~\bibnamefont {Zhong}}, \ and\
  \bibinfo {author} {\bibfnamefont {K.}~\bibnamefont {Xu}},\ }\bibfield
  {title} {\enquote {\bibinfo {title} {Unified gas-kinetic scheme with
  multigrid convergence for rarefied flow study},}\ }\href@noop {} {\bibfield
  {journal} {\bibinfo  {journal} {Phys. Fluids}\ }\textbf {\bibinfo {volume}
  {29}} (\bibinfo {year} {2017})}\BibitemShut {NoStop}%
\bibitem [{\citenamefont {Guo}, \citenamefont {Xu},\ and\ \citenamefont
  {Wang}(2013)}]{guo2013discrete}%
  \BibitemOpen
  \bibfield  {author} {\bibinfo {author} {\bibfnamefont {Z.}~\bibnamefont
  {Guo}}, \bibinfo {author} {\bibfnamefont {K.}~\bibnamefont {Xu}}, \ and\
  \bibinfo {author} {\bibfnamefont {R.}~\bibnamefont {Wang}},\ }\bibfield
  {title} {\enquote {\bibinfo {title} {{Discrete unified gas kinetic scheme for
  all Knudsen number flows: Low-speed isothermal case}},}\ }\href@noop {}
  {\bibfield  {journal} {\bibinfo  {journal} {Phys. Rev. E}\ }\textbf {\bibinfo
  {volume} {88}},\ \bibinfo {pages} {033305} (\bibinfo {year}
  {2013})}\BibitemShut {NoStop}%
\bibitem [{\citenamefont {Guo}\ and\ \citenamefont
  {Xu}(2021)}]{guo2021progress}%
  \BibitemOpen
  \bibfield  {author} {\bibinfo {author} {\bibfnamefont {Z.}~\bibnamefont
  {Guo}}\ and\ \bibinfo {author} {\bibfnamefont {K.}~\bibnamefont {Xu}},\
  }\bibfield  {title} {\enquote {\bibinfo {title} {Progress of discrete unified
  gas-kinetic scheme for multiscale flows},}\ }\href@noop {} {\bibfield
  {journal} {\bibinfo  {journal} {Adv. Aerodyn.}\ }\textbf {\bibinfo {volume}
  {3}},\ \bibinfo {pages} {1--42} (\bibinfo {year} {2021})}\BibitemShut
  {NoStop}%
\bibitem [{\citenamefont {Yang}\ \emph
  {et~al.}(2019{\natexlab{a}})\citenamefont {Yang}, \citenamefont {Shu},
  \citenamefont {Yang},\ and\ \citenamefont {Wu}}]{yang2019improved}%
  \BibitemOpen
  \bibfield  {author} {\bibinfo {author} {\bibfnamefont {L.}~\bibnamefont
  {Yang}}, \bibinfo {author} {\bibfnamefont {C.}~\bibnamefont {Shu}}, \bibinfo
  {author} {\bibfnamefont {W.}~\bibnamefont {Yang}}, \ and\ \bibinfo {author}
  {\bibfnamefont {J.}~\bibnamefont {Wu}},\ }\bibfield  {title} {\enquote
  {\bibinfo {title} {{An improved three-dimensional implicit discrete velocity
  method on unstructured meshes for all Knudsen number flows}},}\ }\href@noop
  {} {\bibfield  {journal} {\bibinfo  {journal} {J. Comput. Phys.}\ }\textbf
  {\bibinfo {volume} {396}},\ \bibinfo {pages} {738--760} (\bibinfo {year}
  {2019}{\natexlab{a}})}\BibitemShut {NoStop}%
\bibitem [{\citenamefont {Yang}\ \emph {et~al.}(2022)\citenamefont {Yang},
  \citenamefont {Shu}, \citenamefont {Wu}, \citenamefont {Liu},\ and\
  \citenamefont {Shen}}]{yang2022efficient}%
  \BibitemOpen
  \bibfield  {author} {\bibinfo {author} {\bibfnamefont {L.}~\bibnamefont
  {Yang}}, \bibinfo {author} {\bibfnamefont {C.}~\bibnamefont {Shu}}, \bibinfo
  {author} {\bibfnamefont {J.}~\bibnamefont {Wu}}, \bibinfo {author}
  {\bibfnamefont {Y.}~\bibnamefont {Liu}}, \ and\ \bibinfo {author}
  {\bibfnamefont {X.}~\bibnamefont {Shen}},\ }\bibfield  {title} {\enquote
  {\bibinfo {title} {An efficient discrete velocity method with inner iteration
  for steady flows in all flow regimes},}\ }\href@noop {} {\bibfield  {journal}
  {\bibinfo  {journal} {Phys. Fluids}\ }\textbf {\bibinfo {volume} {34}}
  (\bibinfo {year} {2022})}\BibitemShut {NoStop}%
\bibitem [{\citenamefont {Succi}(2018)}]{succi2018lattice}%
  \BibitemOpen
  \bibfield  {author} {\bibinfo {author} {\bibfnamefont {S.}~\bibnamefont
  {Succi}},\ }\href@noop {} {\emph {\bibinfo {title} {{The Lattice Boltzmann
  Equation: For Complex States of Flowing Matter}}}}\ (\bibinfo  {publisher}
  {Oxford university press},\ \bibinfo {year} {2018})\BibitemShut {NoStop}%
\bibitem [{\citenamefont {Watari}\ and\ \citenamefont
  {Tsutahara}(2006)}]{watari2006supersonic}%
  \BibitemOpen
  \bibfield  {author} {\bibinfo {author} {\bibfnamefont {M.}~\bibnamefont
  {Watari}}\ and\ \bibinfo {author} {\bibfnamefont {M.}~\bibnamefont
  {Tsutahara}},\ }\bibfield  {title} {\enquote {\bibinfo {title} {Supersonic
  flow simulations by a three-dimensional multispeed thermal model of the
  finite difference lattice {B}oltzmann method},}\ }\href@noop {} {\bibfield
  {journal} {\bibinfo  {journal} {Physica A}\ }\textbf {\bibinfo {volume}
  {364}},\ \bibinfo {pages} {129--144} (\bibinfo {year} {2006})}\BibitemShut
  {NoStop}%
\bibitem [{\citenamefont {Tran}\ \emph {et~al.}(2022)\citenamefont {Tran},
  \citenamefont {Leong}, \citenamefont {Le},\ and\ \citenamefont
  {Le}}]{tran2022lattice}%
  \BibitemOpen
  \bibfield  {author} {\bibinfo {author} {\bibfnamefont {S.~B.~Q.}\
  \bibnamefont {Tran}}, \bibinfo {author} {\bibfnamefont {F.~Y.}\ \bibnamefont
  {Leong}}, \bibinfo {author} {\bibfnamefont {Q.~T.}\ \bibnamefont {Le}}, \
  and\ \bibinfo {author} {\bibfnamefont {D.~V.}\ \bibnamefont {Le}},\
  }\bibfield  {title} {\enquote {\bibinfo {title} {{Lattice Boltzmann method
  for high Reynolds number compressible flow}},}\ }\href@noop {} {\bibfield
  {journal} {\bibinfo  {journal} {Comput. Fluids}\ }\textbf {\bibinfo {volume}
  {249}},\ \bibinfo {pages} {105701} (\bibinfo {year} {2022})}\BibitemShut
  {NoStop}%
\bibitem [{\citenamefont {Latt}\ \emph {et~al.}(2020)\citenamefont {Latt},
  \citenamefont {Coreixas}, \citenamefont {Beny},\ and\ \citenamefont
  {Parmigiani}}]{latt2020efficient}%
  \BibitemOpen
  \bibfield  {author} {\bibinfo {author} {\bibfnamefont {J.}~\bibnamefont
  {Latt}}, \bibinfo {author} {\bibfnamefont {C.}~\bibnamefont {Coreixas}},
  \bibinfo {author} {\bibfnamefont {J.}~\bibnamefont {Beny}}, \ and\ \bibinfo
  {author} {\bibfnamefont {A.}~\bibnamefont {Parmigiani}},\ }\bibfield  {title}
  {\enquote {\bibinfo {title} {Efficient supersonic flow simulations using
  lattice {B}oltzmann methods based on numerical equilibria},}\ }\href@noop {}
  {\bibfield  {journal} {\bibinfo  {journal} {Philos. Trans. R. Soc. A}\
  }\textbf {\bibinfo {volume} {378}},\ \bibinfo {pages} {20190559} (\bibinfo
  {year} {2020})}\BibitemShut {NoStop}%
\bibitem [{\citenamefont {Zhan}\ \emph {et~al.}(2025)\citenamefont {Zhan},
  \citenamefont {Liu}, \citenamefont {Chai},\ and\ \citenamefont
  {Shi}}]{zhan2025thermodynamically}%
  \BibitemOpen
  \bibfield  {author} {\bibinfo {author} {\bibfnamefont {C.}~\bibnamefont
  {Zhan}}, \bibinfo {author} {\bibfnamefont {X.}~\bibnamefont {Liu}}, \bibinfo
  {author} {\bibfnamefont {Z.}~\bibnamefont {Chai}}, \ and\ \bibinfo {author}
  {\bibfnamefont {B.}~\bibnamefont {Shi}},\ }\bibfield  {title} {\enquote
  {\bibinfo {title} {A thermodynamically consistent and conservative
  diffuse-interface model for gas/liquid-liquid-solid flows},}\ }\href@noop {}
  {\bibfield  {journal} {\bibinfo  {journal} {J. Comput. Phys.}\ ,\ \bibinfo
  {pages} {113949}} (\bibinfo {year} {2025})}\BibitemShut {NoStop}%
\bibitem [{\citenamefont {Wang}, \citenamefont {Wei},\ and\ \citenamefont
  {Qian}(2022)}]{wang2022novel}%
  \BibitemOpen
  \bibfield  {author} {\bibinfo {author} {\bibfnamefont {Z.}~\bibnamefont
  {Wang}}, \bibinfo {author} {\bibfnamefont {Y.}~\bibnamefont {Wei}}, \ and\
  \bibinfo {author} {\bibfnamefont {Y.}~\bibnamefont {Qian}},\ }\bibfield
  {title} {\enquote {\bibinfo {title} {A novel thermal lattice {B}oltzmann
  model with heat source and its application in incompressible flow},}\
  }\href@noop {} {\bibfield  {journal} {\bibinfo  {journal} {Appl. Math.
  Comput.}\ }\textbf {\bibinfo {volume} {427}},\ \bibinfo {pages} {127167}
  (\bibinfo {year} {2022})}\BibitemShut {NoStop}%
\bibitem [{\citenamefont {Fei}, \citenamefont {Derome},\ and\ \citenamefont
  {Carmeliet}(2024)}]{fei2024pore}%
  \BibitemOpen
  \bibfield  {author} {\bibinfo {author} {\bibfnamefont {L.}~\bibnamefont
  {Fei}}, \bibinfo {author} {\bibfnamefont {D.}~\bibnamefont {Derome}}, \ and\
  \bibinfo {author} {\bibfnamefont {J.}~\bibnamefont {Carmeliet}},\ }\bibfield
  {title} {\enquote {\bibinfo {title} {Pore-scale study on the effect of
  heterogeneity on evaporation in porous media},}\ }\href@noop {} {\bibfield
  {journal} {\bibinfo  {journal} {J. Fluid Mech.}\ }\textbf {\bibinfo {volume}
  {983}},\ \bibinfo {pages} {A6} (\bibinfo {year} {2024})}\BibitemShut
  {NoStop}%
\bibitem [{\citenamefont {Yang}\ \emph {et~al.}(2025)\citenamefont {Yang},
  \citenamefont {Tu}, \citenamefont {Shan}, \citenamefont {Zhang},
  \citenamefont {Chen},\ and\ \citenamefont {Li}}]{yang2025acoustic}%
  \BibitemOpen
  \bibfield  {author} {\bibinfo {author} {\bibfnamefont {Y.}~\bibnamefont
  {Yang}}, \bibinfo {author} {\bibfnamefont {J.}~\bibnamefont {Tu}}, \bibinfo
  {author} {\bibfnamefont {M.}~\bibnamefont {Shan}}, \bibinfo {author}
  {\bibfnamefont {Z.}~\bibnamefont {Zhang}}, \bibinfo {author} {\bibfnamefont
  {C.}~\bibnamefont {Chen}}, \ and\ \bibinfo {author} {\bibfnamefont
  {H.}~\bibnamefont {Li}},\ }\bibfield  {title} {\enquote {\bibinfo {title}
  {{Acoustic cavitation dynamics of bubble clusters near solid wall: A
  multiphase lattice Boltzmann approach}},}\ }\href@noop {} {\bibfield
  {journal} {\bibinfo  {journal} {Ultrason. Sonochem.}\ }\textbf {\bibinfo
  {volume} {114}},\ \bibinfo {pages} {107261} (\bibinfo {year}
  {2025})}\BibitemShut {NoStop}%
\bibitem [{\citenamefont {Hou}\ \emph {et~al.}(2025)\citenamefont {Hou},
  \citenamefont {Fei}, \citenamefont {Lin},\ and\ \citenamefont
  {Yan}}]{hou2025lattice}%
  \BibitemOpen
  \bibfield  {author} {\bibinfo {author} {\bibfnamefont {G.}~\bibnamefont
  {Hou}}, \bibinfo {author} {\bibfnamefont {L.}~\bibnamefont {Fei}}, \bibinfo
  {author} {\bibfnamefont {C.}~\bibnamefont {Lin}}, \ and\ \bibinfo {author}
  {\bibfnamefont {W.}~\bibnamefont {Yan}},\ }\bibfield  {title} {\enquote
  {\bibinfo {title} {Lattice {B}oltzmann modeling of droplet impact on moving
  superhydrophobic wall},}\ }\href@noop {} {\bibfield  {journal} {\bibinfo
  {journal} {Phys. Fluids.}\ }\textbf {\bibinfo {volume} {37}} (\bibinfo {year}
  {2025})}\BibitemShut {NoStop}%
\bibitem [{\citenamefont {Zhang}\ \emph {et~al.}(2016)\citenamefont {Zhang},
  \citenamefont {Xu}, \citenamefont {Zhang}, \citenamefont {Zhu},\ and\
  \citenamefont {Lin}}]{zhang2016kinetic}%
  \BibitemOpen
  \bibfield  {author} {\bibinfo {author} {\bibfnamefont {Y.}~\bibnamefont
  {Zhang}}, \bibinfo {author} {\bibfnamefont {A.}~\bibnamefont {Xu}}, \bibinfo
  {author} {\bibfnamefont {G.}~\bibnamefont {Zhang}}, \bibinfo {author}
  {\bibfnamefont {C.}~\bibnamefont {Zhu}}, \ and\ \bibinfo {author}
  {\bibfnamefont {C.}~\bibnamefont {Lin}},\ }\bibfield  {title} {\enquote
  {\bibinfo {title} {Kinetic modeling of detonation and effects of negative
  temperature coefficient},}\ }\href@noop {} {\bibfield  {journal} {\bibinfo
  {journal} {Combust. Flame}\ }\textbf {\bibinfo {volume} {173}},\ \bibinfo
  {pages} {483--492} (\bibinfo {year} {2016})}\BibitemShut {NoStop}%
\bibitem [{\citenamefont {Lin}\ and\ \citenamefont
  {Luo}(2019)}]{lin2019discrete}%
  \BibitemOpen
  \bibfield  {author} {\bibinfo {author} {\bibfnamefont {C.}~\bibnamefont
  {Lin}}\ and\ \bibinfo {author} {\bibfnamefont {K.}~\bibnamefont {Luo}},\
  }\bibfield  {title} {\enquote {\bibinfo {title} {Discrete {B}oltzmann
  modeling of unsteady reactive flows with nonequilibrium effects},}\
  }\href@noop {} {\bibfield  {journal} {\bibinfo  {journal} {Phys. Rev. E}\
  }\textbf {\bibinfo {volume} {99}},\ \bibinfo {pages} {012142} (\bibinfo
  {year} {2019})}\BibitemShut {NoStop}%
\bibitem [{\citenamefont {Gan}\ \emph {et~al.}(2018)\citenamefont {Gan},
  \citenamefont {Xu}, \citenamefont {Zhang}, \citenamefont {Zhang},\ and\
  \citenamefont {Succi}}]{gan2018discrete}%
  \BibitemOpen
  \bibfield  {author} {\bibinfo {author} {\bibfnamefont {Y.}~\bibnamefont
  {Gan}}, \bibinfo {author} {\bibfnamefont {A.}~\bibnamefont {Xu}}, \bibinfo
  {author} {\bibfnamefont {G.}~\bibnamefont {Zhang}}, \bibinfo {author}
  {\bibfnamefont {Y.}~\bibnamefont {Zhang}}, \ and\ \bibinfo {author}
  {\bibfnamefont {S.}~\bibnamefont {Succi}},\ }\bibfield  {title} {\enquote
  {\bibinfo {title} {Discrete {B}oltzmann trans-scale modeling of high-speed
  compressible flows},}\ }\href@noop {} {\bibfield  {journal} {\bibinfo
  {journal} {Phys. Rev. E}\ }\textbf {\bibinfo {volume} {97}},\ \bibinfo
  {pages} {053312} (\bibinfo {year} {2018})}\BibitemShut {NoStop}%
\bibitem [{\citenamefont {Gan}\ \emph {et~al.}(2022)\citenamefont {Gan},
  \citenamefont {Xu}, \citenamefont {Lai}, \citenamefont {Li}, \citenamefont
  {Sun},\ and\ \citenamefont {Succi}}]{gan2022discrete}%
  \BibitemOpen
  \bibfield  {author} {\bibinfo {author} {\bibfnamefont {Y.}~\bibnamefont
  {Gan}}, \bibinfo {author} {\bibfnamefont {A.}~\bibnamefont {Xu}}, \bibinfo
  {author} {\bibfnamefont {H.}~\bibnamefont {Lai}}, \bibinfo {author}
  {\bibfnamefont {W.}~\bibnamefont {Li}}, \bibinfo {author} {\bibfnamefont
  {G.}~\bibnamefont {Sun}}, \ and\ \bibinfo {author} {\bibfnamefont
  {S.}~\bibnamefont {Succi}},\ }\bibfield  {title} {\enquote {\bibinfo {title}
  {Discrete {B}oltzmann multi-scale modelling of non-equilibrium multiphase
  flows},}\ }\href@noop {} {\bibfield  {journal} {\bibinfo  {journal} {J. Fluid
  Mech.}\ }\textbf {\bibinfo {volume} {951}},\ \bibinfo {pages} {A8} (\bibinfo
  {year} {2022})}\BibitemShut {NoStop}%
\bibitem [{\citenamefont {Zhang}\ \emph {et~al.}(2022)\citenamefont {Zhang},
  \citenamefont {Xu}, \citenamefont {Zhang}, \citenamefont {Gan},\ and\
  \citenamefont {Li}}]{zhang2022discrete}%
  \BibitemOpen
  \bibfield  {author} {\bibinfo {author} {\bibfnamefont {D.}~\bibnamefont
  {Zhang}}, \bibinfo {author} {\bibfnamefont {A.}~\bibnamefont {Xu}}, \bibinfo
  {author} {\bibfnamefont {Y.}~\bibnamefont {Zhang}}, \bibinfo {author}
  {\bibfnamefont {Y.}~\bibnamefont {Gan}}, \ and\ \bibinfo {author}
  {\bibfnamefont {Y.}~\bibnamefont {Li}},\ }\bibfield  {title} {\enquote
  {\bibinfo {title} {Discrete {B}oltzmann modeling of high-speed compressible
  flows with various depths of non-equilibrium},}\ }\href@noop {} {\bibfield
  {journal} {\bibinfo  {journal} {Phys. Fluids}\ }\textbf {\bibinfo {volume}
  {34}} (\bibinfo {year} {2022})}\BibitemShut {NoStop}%
\bibitem [{\citenamefont {Guo}\ \emph {et~al.}(2025)\citenamefont {Guo},
  \citenamefont {Gan}, \citenamefont {Yang}, \citenamefont {Wu}, \citenamefont
  {Lai},\ and\ \citenamefont {Xu}}]{guo2025POF}%
  \BibitemOpen
  \bibfield  {author} {\bibinfo {author} {\bibfnamefont {Q.}~\bibnamefont
  {Guo}}, \bibinfo {author} {\bibfnamefont {Y.}~\bibnamefont {Gan}}, \bibinfo
  {author} {\bibfnamefont {B.}~\bibnamefont {Yang}}, \bibinfo {author}
  {\bibfnamefont {Y.}~\bibnamefont {Wu}}, \bibinfo {author} {\bibfnamefont
  {H.}~\bibnamefont {Lai}}, \ and\ \bibinfo {author} {\bibfnamefont
  {A.}~\bibnamefont {Xu}},\ }\bibfield  {title} {\enquote {\bibinfo {title}
  {{Thermodynamic nonequilibrium effects in three-dimensional high-speed
  compressible flows: Multiscale modeling and simulation via the discrete
  Boltzmann method}},}\ }\href@noop {} {\bibfield  {journal} {\bibinfo
  {journal} {Phys. Fluids}\ }\textbf {\bibinfo {volume} {37}} (\bibinfo {year}
  {2025})}\BibitemShut {NoStop}%
\bibitem [{\citenamefont {Sun}\ \emph {et~al.}(2024)\citenamefont {Sun},
  \citenamefont {Gan}, \citenamefont {Xu},\ and\ \citenamefont
  {Shi}}]{sun2024droplet}%
  \BibitemOpen
  \bibfield  {author} {\bibinfo {author} {\bibfnamefont {G.}~\bibnamefont
  {Sun}}, \bibinfo {author} {\bibfnamefont {Y.}~\bibnamefont {Gan}}, \bibinfo
  {author} {\bibfnamefont {A.}~\bibnamefont {Xu}}, \ and\ \bibinfo {author}
  {\bibfnamefont {Q.}~\bibnamefont {Shi}},\ }\bibfield  {title} {\enquote
  {\bibinfo {title} {{Droplet coalescence kinetics: Thermodynamic
  non-equilibrium effects and entropy production mechanism}},}\ }\href@noop {}
  {\bibfield  {journal} {\bibinfo  {journal} {Phys. Fluids}\ }\textbf {\bibinfo
  {volume} {36}} (\bibinfo {year} {2024})}\BibitemShut {NoStop}%
\bibitem [{\citenamefont {Song}\ \emph
  {et~al.}(2024{\natexlab{a}})\citenamefont {Song}, \citenamefont {Xu},
  \citenamefont {Miao}, \citenamefont {Chen}, \citenamefont {Liu},
  \citenamefont {Wang}, \citenamefont {Wang},\ and\ \citenamefont
  {Hou}}]{song2024plasma}%
  \BibitemOpen
  \bibfield  {author} {\bibinfo {author} {\bibfnamefont {J.}~\bibnamefont
  {Song}}, \bibinfo {author} {\bibfnamefont {A.}~\bibnamefont {Xu}}, \bibinfo
  {author} {\bibfnamefont {L.}~\bibnamefont {Miao}}, \bibinfo {author}
  {\bibfnamefont {F.}~\bibnamefont {Chen}}, \bibinfo {author} {\bibfnamefont
  {Z.}~\bibnamefont {Liu}}, \bibinfo {author} {\bibfnamefont {L.}~\bibnamefont
  {Wang}}, \bibinfo {author} {\bibfnamefont {N.}~\bibnamefont {Wang}}, \ and\
  \bibinfo {author} {\bibfnamefont {X.}~\bibnamefont {Hou}},\ }\bibfield
  {title} {\enquote {\bibinfo {title} {{Plasma kinetics: Discrete Boltzmann
  modeling and Richtmyer--Meshkov instability}},}\ }\href@noop {} {\bibfield
  {journal} {\bibinfo  {journal} {Phys. Fluids}\ }\textbf {\bibinfo {volume}
  {36}} (\bibinfo {year} {2024}{\natexlab{a}})}\BibitemShut {NoStop}%
\bibitem [{\citenamefont {Song}\ \emph {et~al.}(2025)\citenamefont {Song},
  \citenamefont {Miao}, \citenamefont {Chen}, \citenamefont {Gan},
  \citenamefont {Xu},\ and\ \citenamefont {Li}}]{song2025discrete}%
  \BibitemOpen
  \bibfield  {author} {\bibinfo {author} {\bibfnamefont {J.}~\bibnamefont
  {Song}}, \bibinfo {author} {\bibfnamefont {L.}~\bibnamefont {Miao}}, \bibinfo
  {author} {\bibfnamefont {F.}~\bibnamefont {Chen}}, \bibinfo {author}
  {\bibfnamefont {Y.}~\bibnamefont {Gan}}, \bibinfo {author} {\bibfnamefont
  {A.}~\bibnamefont {Xu}}, \ and\ \bibinfo {author} {\bibfnamefont
  {L.}~\bibnamefont {Li}},\ }\bibfield  {title} {\enquote {\bibinfo {title}
  {{Discrete Boltzmann modeling of Kelvin--Helmholtz instability in plasma}},}\
  }\href@noop {} {\bibfield  {journal} {\bibinfo  {journal} {Phys. Fluids}\
  }\textbf {\bibinfo {volume} {37}} (\bibinfo {year} {2025})}\BibitemShut
  {NoStop}%
\bibitem [{\citenamefont {Li}\ and\ \citenamefont {Lin}(2024)}]{li2024kinetic}%
  \BibitemOpen
  \bibfield  {author} {\bibinfo {author} {\bibfnamefont {Y.}~\bibnamefont
  {Li}}\ and\ \bibinfo {author} {\bibfnamefont {C.}~\bibnamefont {Lin}},\
  }\bibfield  {title} {\enquote {\bibinfo {title} {{Kinetic investigation of
  Kelvin--Helmholtz instability with nonequilibrium effects in a force
  field}},}\ }\href@noop {} {\bibfield  {journal} {\bibinfo  {journal} {Phys.
  Fluids}\ }\textbf {\bibinfo {volume} {36}} (\bibinfo {year}
  {2024})}\BibitemShut {NoStop}%
\bibitem [{\citenamefont {Xu}, \citenamefont {Lin},\ and\ \citenamefont
  {Lai}(2025)}]{xu2025influence}%
  \BibitemOpen
  \bibfield  {author} {\bibinfo {author} {\bibfnamefont {H.}~\bibnamefont
  {Xu}}, \bibinfo {author} {\bibfnamefont {C.}~\bibnamefont {Lin}}, \ and\
  \bibinfo {author} {\bibfnamefont {H.}~\bibnamefont {Lai}},\ }\bibfield
  {title} {\enquote {\bibinfo {title} {{Influence of phase difference and
  amplitude ratio on Kelvin--Helmholtz instability with dual-mode interface
  perturbations}},}\ }\href@noop {} {\bibfield  {journal} {\bibinfo  {journal}
  {Phys. Fluids}\ }\textbf {\bibinfo {volume} {37}} (\bibinfo {year}
  {2025})}\BibitemShut {NoStop}%
\bibitem [{\citenamefont {He}\ \emph {et~al.}(2025)\citenamefont {He},
  \citenamefont {Gan}, \citenamefont {Yang}, \citenamefont {Li}, \citenamefont
  {Lai},\ and\ \citenamefont {Xu}}]{he2025POF}%
  \BibitemOpen
  \bibfield  {author} {\bibinfo {author} {\bibfnamefont {Z.}~\bibnamefont
  {He}}, \bibinfo {author} {\bibfnamefont {Y.}~\bibnamefont {Gan}}, \bibinfo
  {author} {\bibfnamefont {B.}~\bibnamefont {Yang}}, \bibinfo {author}
  {\bibfnamefont {D.}~\bibnamefont {Li}}, \bibinfo {author} {\bibfnamefont
  {H.}~\bibnamefont {Lai}}, \ and\ \bibinfo {author} {\bibfnamefont
  {A.}~\bibnamefont {Xu}},\ }\bibfield  {title} {\enquote {\bibinfo {title}
  {{Multiscale thermodynamic nonequilibrium effects in Kelvin–Helmholtz
  instability and their relative importance}},}\ }\href@noop {} {\bibfield
  {journal} {\bibinfo  {journal} {Phys. Fluids}\ }\textbf {\bibinfo {volume}
  {37}},\ \bibinfo {pages} {034131} (\bibinfo {year} {2025})}\BibitemShut
  {NoStop}%
\bibitem [{\citenamefont {Watari}\ and\ \citenamefont
  {Tsutahara}(2003)}]{watari2003two}%
  \BibitemOpen
  \bibfield  {author} {\bibinfo {author} {\bibfnamefont {M.}~\bibnamefont
  {Watari}}\ and\ \bibinfo {author} {\bibfnamefont {M.}~\bibnamefont
  {Tsutahara}},\ }\bibfield  {title} {\enquote {\bibinfo {title}
  {Two-dimensional thermal model of the finite-difference lattice {B}oltzmann
  method with high spatial isotropy},}\ }\href@noop {} {\bibfield  {journal}
  {\bibinfo  {journal} {Phys. Rev. E}\ }\textbf {\bibinfo {volume} {67}},\
  \bibinfo {pages} {036306} (\bibinfo {year} {2003})}\BibitemShut {NoStop}%
\bibitem [{\citenamefont {Kataoka}\ and\ \citenamefont
  {Tsutahara}(2004{\natexlab{a}})}]{kataoka2004lattice}%
  \BibitemOpen
  \bibfield  {author} {\bibinfo {author} {\bibfnamefont {T.}~\bibnamefont
  {Kataoka}}\ and\ \bibinfo {author} {\bibfnamefont {M.}~\bibnamefont
  {Tsutahara}},\ }\bibfield  {title} {\enquote {\bibinfo {title} {{Lattice
  Boltzmann model for the compressible Navier-Stokes equations with flexible
  specific-heat ratio}},}\ }\href@noop {} {\bibfield  {journal} {\bibinfo
  {journal} {Phys. Rev. E}\ }\textbf {\bibinfo {volume} {69}},\ \bibinfo
  {pages} {035701} (\bibinfo {year} {2004}{\natexlab{a}})}\BibitemShut
  {NoStop}%
\bibitem [{\citenamefont {Gan}\ \emph {et~al.}(2008)\citenamefont {Gan},
  \citenamefont {Xu}, \citenamefont {Zhang}, \citenamefont {Yu},\ and\
  \citenamefont {Li}}]{gan2008two}%
  \BibitemOpen
  \bibfield  {author} {\bibinfo {author} {\bibfnamefont {Y.}~\bibnamefont
  {Gan}}, \bibinfo {author} {\bibfnamefont {A.}~\bibnamefont {Xu}}, \bibinfo
  {author} {\bibfnamefont {G.}~\bibnamefont {Zhang}}, \bibinfo {author}
  {\bibfnamefont {X.}~\bibnamefont {Yu}}, \ and\ \bibinfo {author}
  {\bibfnamefont {Y.}~\bibnamefont {Li}},\ }\bibfield  {title} {\enquote
  {\bibinfo {title} {{Two-dimensional lattice Boltzmann model for compressible
  flows with high Mach number}},}\ }\href@noop {} {\bibfield  {journal}
  {\bibinfo  {journal} {Physica A}\ }\textbf {\bibinfo {volume} {387}},\
  \bibinfo {pages} {1721--1732} (\bibinfo {year} {2008})}\BibitemShut {NoStop}%
\bibitem [{\citenamefont {Qiu}\ \emph {et~al.}(2017)\citenamefont {Qiu},
  \citenamefont {You}, \citenamefont {Zhu},\ and\ \citenamefont
  {Chen}}]{qiu2017lattice}%
  \BibitemOpen
  \bibfield  {author} {\bibinfo {author} {\bibfnamefont {R.}~\bibnamefont
  {Qiu}}, \bibinfo {author} {\bibfnamefont {Y.}~\bibnamefont {You}}, \bibinfo
  {author} {\bibfnamefont {C.}~\bibnamefont {Zhu}}, \ and\ \bibinfo {author}
  {\bibfnamefont {R.}~\bibnamefont {Chen}},\ }\bibfield  {title} {\enquote
  {\bibinfo {title} {Lattice {B}oltzmann simulation for high-speed compressible
  viscous flows with a boundary layer},}\ }\href@noop {} {\bibfield  {journal}
  {\bibinfo  {journal} {Appl. Math. Model.}\ }\textbf {\bibinfo {volume}
  {48}},\ \bibinfo {pages} {567--583} (\bibinfo {year} {2017})}\BibitemShut
  {NoStop}%
\bibitem [{\citenamefont {Qiu}\ \emph {et~al.}(2021)\citenamefont {Qiu},
  \citenamefont {Zhou}, \citenamefont {Bao}, \citenamefont {Zhou},
  \citenamefont {Che},\ and\ \citenamefont {You}}]{qiu2021mesoscopic}%
  \BibitemOpen
  \bibfield  {author} {\bibinfo {author} {\bibfnamefont {R.}~\bibnamefont
  {Qiu}}, \bibinfo {author} {\bibfnamefont {T.}~\bibnamefont {Zhou}}, \bibinfo
  {author} {\bibfnamefont {Y.}~\bibnamefont {Bao}}, \bibinfo {author}
  {\bibfnamefont {K.}~\bibnamefont {Zhou}}, \bibinfo {author} {\bibfnamefont
  {H.}~\bibnamefont {Che}}, \ and\ \bibinfo {author} {\bibfnamefont
  {Y.}~\bibnamefont {You}},\ }\bibfield  {title} {\enquote {\bibinfo {title}
  {Mesoscopic kinetic approach for studying nonequilibrium hydrodynamic and
  thermodynamic effects of shock wave, contact discontinuity, and rarefaction
  wave in the unsteady shock tube},}\ }\href@noop {} {\bibfield  {journal}
  {\bibinfo  {journal} {Phys. Rev. E}\ }\textbf {\bibinfo {volume} {103}},\
  \bibinfo {pages} {053113} (\bibinfo {year} {2021})}\BibitemShut {NoStop}%
\bibitem [{\citenamefont {Gan}\ \emph {et~al.}(2011{\natexlab{a}})\citenamefont
  {Gan}, \citenamefont {Xu}, \citenamefont {Zhang},\ and\ \citenamefont
  {Li}}]{gan2011lattice}%
  \BibitemOpen
  \bibfield  {author} {\bibinfo {author} {\bibfnamefont {Y.}~\bibnamefont
  {Gan}}, \bibinfo {author} {\bibfnamefont {A.}~\bibnamefont {Xu}}, \bibinfo
  {author} {\bibfnamefont {G.}~\bibnamefont {Zhang}}, \ and\ \bibinfo {author}
  {\bibfnamefont {Y.}~\bibnamefont {Li}},\ }\bibfield  {title} {\enquote
  {\bibinfo {title} {{Lattice Boltzmann study on Kelvin-Helmholtz instability:
  Roles of velocity and density gradients}},}\ }\href@noop {} {\bibfield
  {journal} {\bibinfo  {journal} {Phys. Rev. E}\ }\textbf {\bibinfo {volume}
  {83}},\ \bibinfo {pages} {056704} (\bibinfo {year}
  {2011}{\natexlab{a}})}\BibitemShut {NoStop}%
\bibitem [{\citenamefont {Hejranfar}\ and\ \citenamefont
  {Abotalebi}(2024)}]{hejranfar2024computation}%
  \BibitemOpen
  \bibfield  {author} {\bibinfo {author} {\bibfnamefont {K.}~\bibnamefont
  {Hejranfar}}\ and\ \bibinfo {author} {\bibfnamefont {M.}~\bibnamefont
  {Abotalebi}},\ }\bibfield  {title} {\enquote {\bibinfo {title} {Computation
  of three-dimensional incompressible flows using high-order weighted
  essentially non-oscillatory finite-difference lattice {B}oltzmann method},}\
  }\href@noop {} {\bibfield  {journal} {\bibinfo  {journal} {Phys. Fluids}\
  }\textbf {\bibinfo {volume} {36}} (\bibinfo {year} {2024})}\BibitemShut
  {NoStop}%
\bibitem [{\citenamefont {Sofonea}\ \emph {et~al.}(2004)\citenamefont
  {Sofonea}, \citenamefont {Lamura}, \citenamefont {Gonnella},\ and\
  \citenamefont {Cristea}}]{sofonea2004finite}%
  \BibitemOpen
  \bibfield  {author} {\bibinfo {author} {\bibfnamefont {V.}~\bibnamefont
  {Sofonea}}, \bibinfo {author} {\bibfnamefont {A.}~\bibnamefont {Lamura}},
  \bibinfo {author} {\bibfnamefont {G.}~\bibnamefont {Gonnella}}, \ and\
  \bibinfo {author} {\bibfnamefont {A.}~\bibnamefont {Cristea}},\ }\bibfield
  {title} {\enquote {\bibinfo {title} {Finite-difference lattice {B}oltzmann
  model with flux limiters for liquid-vapor systems},}\ }\href@noop {}
  {\bibfield  {journal} {\bibinfo  {journal} {Phys. Rev. E}\ }\textbf {\bibinfo
  {volume} {70}},\ \bibinfo {pages} {046702} (\bibinfo {year}
  {2004})}\BibitemShut {NoStop}%
\bibitem [{\citenamefont {Gan}\ \emph {et~al.}(2011{\natexlab{b}})\citenamefont
  {Gan}, \citenamefont {Xu}, \citenamefont {Zhang},\ and\ \citenamefont
  {Li}}]{gan2011flux}%
  \BibitemOpen
  \bibfield  {author} {\bibinfo {author} {\bibfnamefont {Y.}~\bibnamefont
  {Gan}}, \bibinfo {author} {\bibfnamefont {A.}~\bibnamefont {Xu}}, \bibinfo
  {author} {\bibfnamefont {G.}~\bibnamefont {Zhang}}, \ and\ \bibinfo {author}
  {\bibfnamefont {Y.}~\bibnamefont {Li}},\ }\bibfield  {title} {\enquote
  {\bibinfo {title} {{Flux limiter lattice Boltzmann scheme approach to
  compressible flows with flexible specific-heat ratio and Prandtl number}},}\
  }\href@noop {} {\bibfield  {journal} {\bibinfo  {journal} {Commun. Theor.
  Phys.}\ }\textbf {\bibinfo {volume} {56}},\ \bibinfo {pages} {490} (\bibinfo
  {year} {2011}{\natexlab{b}})}\BibitemShut {NoStop}%
\bibitem [{\citenamefont {Huang}\ \emph {et~al.}(2024)\citenamefont {Huang},
  \citenamefont {Jin}, \citenamefont {Li}, \citenamefont {Li},\ and\
  \citenamefont {Zheng}}]{huang2024implicit}%
  \BibitemOpen
  \bibfield  {author} {\bibinfo {author} {\bibfnamefont {H.}~\bibnamefont
  {Huang}}, \bibinfo {author} {\bibfnamefont {K.}~\bibnamefont {Jin}}, \bibinfo
  {author} {\bibfnamefont {K.}~\bibnamefont {Li}}, \bibinfo {author}
  {\bibfnamefont {H.}~\bibnamefont {Li}}, \ and\ \bibinfo {author}
  {\bibfnamefont {X.}~\bibnamefont {Zheng}},\ }\bibfield  {title} {\enquote
  {\bibinfo {title} {An implicit lattice {B}oltzmann method for simulations of
  compressible plasma kinetics},}\ }\href@noop {} {\bibfield  {journal}
  {\bibinfo  {journal} {Phys. Fluids}\ }\textbf {\bibinfo {volume} {36}}
  (\bibinfo {year} {2024})}\BibitemShut {NoStop}%
\bibitem [{\citenamefont {Li}\ \emph {et~al.}(2007)\citenamefont {Li},
  \citenamefont {He}, \citenamefont {Wang},\ and\ \citenamefont
  {Tao}}]{li2007coupled}%
  \BibitemOpen
  \bibfield  {author} {\bibinfo {author} {\bibfnamefont {Q.}~\bibnamefont
  {Li}}, \bibinfo {author} {\bibfnamefont {Y.}~\bibnamefont {He}}, \bibinfo
  {author} {\bibfnamefont {Y.}~\bibnamefont {Wang}}, \ and\ \bibinfo {author}
  {\bibfnamefont {W.}~\bibnamefont {Tao}},\ }\bibfield  {title} {\enquote
  {\bibinfo {title} {{Coupled double-distribution-function lattice Boltzmann
  method for the compressible Navier-Stokes equations}},}\ }\href@noop {}
  {\bibfield  {journal} {\bibinfo  {journal} {Phys. Rev. E}\ }\textbf {\bibinfo
  {volume} {76}},\ \bibinfo {pages} {056705} (\bibinfo {year}
  {2007})}\BibitemShut {NoStop}%
\bibitem [{\citenamefont {Gan}\ \emph {et~al.}(2013)\citenamefont {Gan},
  \citenamefont {Xu}, \citenamefont {Zhang},\ and\ \citenamefont
  {Yang}}]{gan2013lattice}%
  \BibitemOpen
  \bibfield  {author} {\bibinfo {author} {\bibfnamefont {Y.}~\bibnamefont
  {Gan}}, \bibinfo {author} {\bibfnamefont {A.}~\bibnamefont {Xu}}, \bibinfo
  {author} {\bibfnamefont {G.}~\bibnamefont {Zhang}}, \ and\ \bibinfo {author}
  {\bibfnamefont {Y.}~\bibnamefont {Yang}},\ }\bibfield  {title} {\enquote
  {\bibinfo {title} {{Lattice BGK kinetic model for high-speed compressible
  flows: Hydrodynamic and nonequilibrium behaviors}},}\ }\href@noop {}
  {\bibfield  {journal} {\bibinfo  {journal} {Europhys. Lett.}\ }\textbf
  {\bibinfo {volume} {103}},\ \bibinfo {pages} {24003} (\bibinfo {year}
  {2013})}\BibitemShut {NoStop}%
\bibitem [{\citenamefont {Frapolli}, \citenamefont {Chikatamarla},\ and\
  \citenamefont {Karlin}(2015)}]{frapolli2015entropic}%
  \BibitemOpen
  \bibfield  {author} {\bibinfo {author} {\bibfnamefont {N.}~\bibnamefont
  {Frapolli}}, \bibinfo {author} {\bibfnamefont {S.~S.}\ \bibnamefont
  {Chikatamarla}}, \ and\ \bibinfo {author} {\bibfnamefont {I.~V.}\
  \bibnamefont {Karlin}},\ }\bibfield  {title} {\enquote {\bibinfo {title}
  {Entropic lattice {B}oltzmann model for compressible flows},}\ }\href@noop {}
  {\bibfield  {journal} {\bibinfo  {journal} {Phys. Rev. E}\ }\textbf {\bibinfo
  {volume} {92}},\ \bibinfo {pages} {061301} (\bibinfo {year}
  {2015})}\BibitemShut {NoStop}%
\bibitem [{\citenamefont {Frapolli}, \citenamefont {Chikatamarla},\ and\
  \citenamefont {Karlin}(2020)}]{frapolli2020theory}%
  \BibitemOpen
  \bibfield  {author} {\bibinfo {author} {\bibfnamefont {N.}~\bibnamefont
  {Frapolli}}, \bibinfo {author} {\bibfnamefont {S.}~\bibnamefont
  {Chikatamarla}}, \ and\ \bibinfo {author} {\bibfnamefont {I.}~\bibnamefont
  {Karlin}},\ }\bibfield  {title} {\enquote {\bibinfo {title} {Theory,
  analysis, and applications of the entropic lattice {B}oltzmann model for
  compressible flows},}\ }\href@noop {} {\bibfield  {journal} {\bibinfo
  {journal} {Entropy}\ }\textbf {\bibinfo {volume} {22}},\ \bibinfo {pages}
  {370} (\bibinfo {year} {2020})}\BibitemShut {NoStop}%
\bibitem [{\citenamefont {Yeoh}, \citenamefont {Ooi},\ and\ \citenamefont
  {Foo}(2020)}]{yeoh2020lattice}%
  \BibitemOpen
  \bibfield  {author} {\bibinfo {author} {\bibfnamefont {C.~V.}\ \bibnamefont
  {Yeoh}}, \bibinfo {author} {\bibfnamefont {E.~H.}\ \bibnamefont {Ooi}}, \
  and\ \bibinfo {author} {\bibfnamefont {J.~J.}\ \bibnamefont {Foo}},\
  }\bibfield  {title} {\enquote {\bibinfo {title} {{Lattice-Boltzmann
  hydrodynamics of single-square-grid generated turbulence-a partial entropic
  stabilisation approach}},}\ }\href@noop {} {\bibfield  {journal} {\bibinfo
  {journal} {Comput. Math. Appl.}\ }\textbf {\bibinfo {volume} {80}},\ \bibinfo
  {pages} {1301--1326} (\bibinfo {year} {2020})}\BibitemShut {NoStop}%
\bibitem [{\citenamefont {Hosseini}\ \emph {et~al.}(2023)\citenamefont
  {Hosseini}, \citenamefont {Atif}, \citenamefont {Ansumali},\ and\
  \citenamefont {Karlin}}]{hosseini2023entropic}%
  \BibitemOpen
  \bibfield  {author} {\bibinfo {author} {\bibfnamefont {S.~A.}\ \bibnamefont
  {Hosseini}}, \bibinfo {author} {\bibfnamefont {M.}~\bibnamefont {Atif}},
  \bibinfo {author} {\bibfnamefont {S.}~\bibnamefont {Ansumali}}, \ and\
  \bibinfo {author} {\bibfnamefont {I.~V.}\ \bibnamefont {Karlin}},\ }\bibfield
   {title} {\enquote {\bibinfo {title} {{Entropic lattice Boltzmann methods: A
  review}},}\ }\href@noop {} {\bibfield  {journal} {\bibinfo  {journal}
  {Comput. Fluids}\ }\textbf {\bibinfo {volume} {259}},\ \bibinfo {pages}
  {105884} (\bibinfo {year} {2023})}\BibitemShut {NoStop}%
\bibitem [{\citenamefont {Dorschner}, \citenamefont {B{\"o}sch},\ and\
  \citenamefont {Karlin}(2018)}]{dorschner2018particles}%
  \BibitemOpen
  \bibfield  {author} {\bibinfo {author} {\bibfnamefont {B.}~\bibnamefont
  {Dorschner}}, \bibinfo {author} {\bibfnamefont {F.}~\bibnamefont
  {B{\"o}sch}}, \ and\ \bibinfo {author} {\bibfnamefont {I.~V.}\ \bibnamefont
  {Karlin}},\ }\bibfield  {title} {\enquote {\bibinfo {title} {Particles on
  demand for kinetic theory},}\ }\href@noop {} {\bibfield  {journal} {\bibinfo
  {journal} {Phys. Rev. Lett.}\ }\textbf {\bibinfo {volume} {121}},\ \bibinfo
  {pages} {130602} (\bibinfo {year} {2018})}\BibitemShut {NoStop}%
\bibitem [{\citenamefont {Kallikounis}, \citenamefont {Dorschner},\ and\
  \citenamefont {Karlin}(2022)}]{kallikounis2022particles}%
  \BibitemOpen
  \bibfield  {author} {\bibinfo {author} {\bibfnamefont {N.~G.}\ \bibnamefont
  {Kallikounis}}, \bibinfo {author} {\bibfnamefont {B.}~\bibnamefont
  {Dorschner}}, \ and\ \bibinfo {author} {\bibfnamefont {I.~V.}\ \bibnamefont
  {Karlin}},\ }\bibfield  {title} {\enquote {\bibinfo {title} {Particles on
  demand for flows with strong discontinuities},}\ }\href@noop {} {\bibfield
  {journal} {\bibinfo  {journal} {Phys. Rev. E}\ }\textbf {\bibinfo {volume}
  {106}},\ \bibinfo {pages} {015301} (\bibinfo {year} {2022})}\BibitemShut
  {NoStop}%
\bibitem [{\citenamefont {Yang}\ \emph
  {et~al.}(2019{\natexlab{b}})\citenamefont {Yang}, \citenamefont {Wang},
  \citenamefont {Yang},\ and\ \citenamefont {Shu}}]{yang2019development}%
  \BibitemOpen
  \bibfield  {author} {\bibinfo {author} {\bibfnamefont {T.}~\bibnamefont
  {Yang}}, \bibinfo {author} {\bibfnamefont {J.}~\bibnamefont {Wang}}, \bibinfo
  {author} {\bibfnamefont {L.}~\bibnamefont {Yang}}, \ and\ \bibinfo {author}
  {\bibfnamefont {C.}~\bibnamefont {Shu}},\ }\bibfield  {title} {\enquote
  {\bibinfo {title} {Development of multicomponent lattice {B}oltzmann flux
  solver for simulation of compressible viscous reacting flows},}\ }\href@noop
  {} {\bibfield  {journal} {\bibinfo  {journal} {Phys. Rev. E}\ }\textbf
  {\bibinfo {volume} {100}},\ \bibinfo {pages} {033315} (\bibinfo {year}
  {2019}{\natexlab{b}})}\BibitemShut {NoStop}%
\bibitem [{\citenamefont {Song}\ \emph
  {et~al.}(2024{\natexlab{b}})\citenamefont {Song}, \citenamefont {Yang},
  \citenamefont {Du}, \citenamefont {Xiao},\ and\ \citenamefont
  {Shu}}]{song2024double}%
  \BibitemOpen
  \bibfield  {author} {\bibinfo {author} {\bibfnamefont {Y.}~\bibnamefont
  {Song}}, \bibinfo {author} {\bibfnamefont {L.}~\bibnamefont {Yang}}, \bibinfo
  {author} {\bibfnamefont {Y.}~\bibnamefont {Du}}, \bibinfo {author}
  {\bibfnamefont {Y.}~\bibnamefont {Xiao}}, \ and\ \bibinfo {author}
  {\bibfnamefont {C.}~\bibnamefont {Shu}},\ }\bibfield  {title} {\enquote
  {\bibinfo {title} {Double distribution function-based lattice {B}oltzmann
  flux solver for simulation of compressible viscous flows},}\ }\href@noop {}
  {\bibfield  {journal} {\bibinfo  {journal} {Phys. Fluids}\ }\textbf {\bibinfo
  {volume} {36}} (\bibinfo {year} {2024}{\natexlab{b}})}\BibitemShut {NoStop}%
\bibitem [{\citenamefont {Yeoh}, \citenamefont {Ooi},\ and\ \citenamefont
  {Foo}(2021)}]{yeoh2021utilization}%
  \BibitemOpen
  \bibfield  {author} {\bibinfo {author} {\bibfnamefont {C.~V.}\ \bibnamefont
  {Yeoh}}, \bibinfo {author} {\bibfnamefont {E.~H.}\ \bibnamefont {Ooi}}, \
  and\ \bibinfo {author} {\bibfnamefont {J.~J.}\ \bibnamefont {Foo}},\
  }\bibfield  {title} {\enquote {\bibinfo {title} {{Utilization of pressure
  wave-dynamics in accelerating convergence of the lattice-Boltzmann method for
  steady and unsteady flows}},}\ }\href@noop {} {\bibfield  {journal} {\bibinfo
   {journal} {Appl. Math. Comput.}\ }\textbf {\bibinfo {volume} {411}},\
  \bibinfo {pages} {126498} (\bibinfo {year} {2021})}\BibitemShut {NoStop}%
\bibitem [{\citenamefont {Chen}\ \emph {et~al.}(2010)\citenamefont {Chen},
  \citenamefont {Xu}, \citenamefont {Zhang}, \citenamefont {Li},\ and\
  \citenamefont {Succi}}]{chen2010multiple}%
  \BibitemOpen
  \bibfield  {author} {\bibinfo {author} {\bibfnamefont {F.}~\bibnamefont
  {Chen}}, \bibinfo {author} {\bibfnamefont {A.}~\bibnamefont {Xu}}, \bibinfo
  {author} {\bibfnamefont {G.}~\bibnamefont {Zhang}}, \bibinfo {author}
  {\bibfnamefont {Y.}~\bibnamefont {Li}}, \ and\ \bibinfo {author}
  {\bibfnamefont {S.}~\bibnamefont {Succi}},\ }\bibfield  {title} {\enquote
  {\bibinfo {title} {{Multiple-relaxation-time lattice Boltzmann approach to
  compressible flows with flexible specific-heat ratio and Prandtl number}},}\
  }\href@noop {} {\bibfield  {journal} {\bibinfo  {journal} {Europhys. Lett.}\
  }\textbf {\bibinfo {volume} {90}},\ \bibinfo {pages} {54003} (\bibinfo {year}
  {2010})}\BibitemShut {NoStop}%
\bibitem [{\citenamefont {Li}\ and\ \citenamefont
  {Zhong}(2015)}]{li2015multiple}%
  \BibitemOpen
  \bibfield  {author} {\bibinfo {author} {\bibfnamefont {K.}~\bibnamefont
  {Li}}\ and\ \bibinfo {author} {\bibfnamefont {C.}~\bibnamefont {Zhong}},\
  }\bibfield  {title} {\enquote {\bibinfo {title} {A multiple-relaxation-time
  lattice {B}oltzmann method for high-speed compressible flows},}\ }\href@noop
  {} {\bibfield  {journal} {\bibinfo  {journal} {Chin. Phys. B}\ }\textbf
  {\bibinfo {volume} {24}},\ \bibinfo {pages} {050501} (\bibinfo {year}
  {2015})}\BibitemShut {NoStop}%
\bibitem [{\citenamefont {Ghadyani}, \citenamefont {Esfahanian},\ and\
  \citenamefont {Taeibi-Rahni}(2015)}]{ghadyani2015use}%
  \BibitemOpen
  \bibfield  {author} {\bibinfo {author} {\bibfnamefont {M.}~\bibnamefont
  {Ghadyani}}, \bibinfo {author} {\bibfnamefont {V.}~\bibnamefont
  {Esfahanian}}, \ and\ \bibinfo {author} {\bibfnamefont {M.}~\bibnamefont
  {Taeibi-Rahni}},\ }\bibfield  {title} {\enquote {\bibinfo {title} {{The use
  of shock-detecting sensor to improve the stability of lattice Boltzmann model
  for high Mach number flows}},}\ }\href@noop {} {\bibfield  {journal}
  {\bibinfo  {journal} {Int. J. Mod. Phys. C}\ }\textbf {\bibinfo {volume}
  {26}},\ \bibinfo {pages} {1550006} (\bibinfo {year} {2015})}\BibitemShut
  {NoStop}%
\bibitem [{\citenamefont {Esfahanian}\ and\ \citenamefont
  {Ghadyani}(2015)}]{esfahanian2015improvement}%
  \BibitemOpen
  \bibfield  {author} {\bibinfo {author} {\bibfnamefont {V.}~\bibnamefont
  {Esfahanian}}\ and\ \bibinfo {author} {\bibfnamefont {M.}~\bibnamefont
  {Ghadyani}},\ }\bibfield  {title} {\enquote {\bibinfo {title} {Improvement of
  the instability of compressible lattice {B}oltzmann model by shock-detecting
  sensor},}\ }\href@noop {} {\bibfield  {journal} {\bibinfo  {journal} {J.
  Mech. Sci. Technol.}\ }\textbf {\bibinfo {volume} {29}},\ \bibinfo {pages}
  {1981--1991} (\bibinfo {year} {2015})}\BibitemShut {NoStop}%
\bibitem [{\citenamefont {Li}\ \emph {et~al.}(2023)\citenamefont {Li},
  \citenamefont {Wang}, \citenamefont {Gopalakrishnan}, \citenamefont {Li},
  \citenamefont {Zhang},\ and\ \citenamefont {Chen}}]{li2023hybrid}%
  \BibitemOpen
  \bibfield  {author} {\bibinfo {author} {\bibfnamefont {W.}~\bibnamefont
  {Li}}, \bibinfo {author} {\bibfnamefont {J.}~\bibnamefont {Wang}}, \bibinfo
  {author} {\bibfnamefont {P.}~\bibnamefont {Gopalakrishnan}}, \bibinfo
  {author} {\bibfnamefont {Y.}~\bibnamefont {Li}}, \bibinfo {author}
  {\bibfnamefont {R.}~\bibnamefont {Zhang}}, \ and\ \bibinfo {author}
  {\bibfnamefont {H.}~\bibnamefont {Chen}},\ }\bibfield  {title} {\enquote
  {\bibinfo {title} {Hybrid lattice {B}oltzmann approach for simulation of
  high-speed flows},}\ }\href@noop {} {\bibfield  {journal} {\bibinfo
  {journal} {AIAA J.}\ }\textbf {\bibinfo {volume} {61}},\ \bibinfo {pages}
  {5548--5557} (\bibinfo {year} {2023})}\BibitemShut {NoStop}%
\bibitem [{\citenamefont {Zhao}\ \emph {et~al.}(2020)\citenamefont {Zhao},
  \citenamefont {Farag}, \citenamefont {Boivin},\ and\ \citenamefont
  {Sagaut}}]{zhao2020toward}%
  \BibitemOpen
  \bibfield  {author} {\bibinfo {author} {\bibfnamefont {S.}~\bibnamefont
  {Zhao}}, \bibinfo {author} {\bibfnamefont {G.}~\bibnamefont {Farag}},
  \bibinfo {author} {\bibfnamefont {P.}~\bibnamefont {Boivin}}, \ and\ \bibinfo
  {author} {\bibfnamefont {P.}~\bibnamefont {Sagaut}},\ }\bibfield  {title}
  {\enquote {\bibinfo {title} {Toward fully conservative hybrid lattice
  {B}oltzmann methods for compressible flows},}\ }\href@noop {} {\bibfield
  {journal} {\bibinfo  {journal} {Phys. Fluids}\ }\textbf {\bibinfo {volume}
  {32}} (\bibinfo {year} {2020})}\BibitemShut {NoStop}%
\bibitem [{\citenamefont {Guo}\ and\ \citenamefont
  {Feng}(2024)}]{guo2024hybrid}%
  \BibitemOpen
  \bibfield  {author} {\bibinfo {author} {\bibfnamefont {S.}~\bibnamefont
  {Guo}}\ and\ \bibinfo {author} {\bibfnamefont {Y.}~\bibnamefont {Feng}},\
  }\bibfield  {title} {\enquote {\bibinfo {title} {Hybrid compressible lattice
  {B}oltzmann method for supersonic flows with strong discontinuities},}\
  }\href@noop {} {\bibfield  {journal} {\bibinfo  {journal} {Phys. Fluids}\
  }\textbf {\bibinfo {volume} {36}} (\bibinfo {year} {2024})}\BibitemShut
  {NoStop}%
\bibitem [{\citenamefont {Feng}\ \emph {et~al.}(2019)\citenamefont {Feng},
  \citenamefont {Boivin}, \citenamefont {Jacob},\ and\ \citenamefont
  {Sagaut}}]{feng2019hybrid}%
  \BibitemOpen
  \bibfield  {author} {\bibinfo {author} {\bibfnamefont {Y.}~\bibnamefont
  {Feng}}, \bibinfo {author} {\bibfnamefont {P.}~\bibnamefont {Boivin}},
  \bibinfo {author} {\bibfnamefont {J.}~\bibnamefont {Jacob}}, \ and\ \bibinfo
  {author} {\bibfnamefont {P.}~\bibnamefont {Sagaut}},\ }\bibfield  {title}
  {\enquote {\bibinfo {title} {Hybrid recursive regularized thermal lattice
  {B}oltzmann model for high subsonic compressible flows},}\ }\href@noop {}
  {\bibfield  {journal} {\bibinfo  {journal} {J. Comput. Phys.}\ }\textbf
  {\bibinfo {volume} {394}},\ \bibinfo {pages} {82--99} (\bibinfo {year}
  {2019})}\BibitemShut {NoStop}%
\bibitem [{\citenamefont {Feng}\ \emph {et~al.}(2020)\citenamefont {Feng},
  \citenamefont {Guo}, \citenamefont {Jacob},\ and\ \citenamefont
  {Sagaut}}]{feng2020grid}%
  \BibitemOpen
  \bibfield  {author} {\bibinfo {author} {\bibfnamefont {Y.}~\bibnamefont
  {Feng}}, \bibinfo {author} {\bibfnamefont {S.}~\bibnamefont {Guo}}, \bibinfo
  {author} {\bibfnamefont {J.}~\bibnamefont {Jacob}}, \ and\ \bibinfo {author}
  {\bibfnamefont {P.}~\bibnamefont {Sagaut}},\ }\bibfield  {title} {\enquote
  {\bibinfo {title} {Grid refinement in the three-dimensional hybrid recursive
  regularized lattice {B}oltzmann method for compressible aerodynamics},}\
  }\href@noop {} {\bibfield  {journal} {\bibinfo  {journal} {Phys. Rev. E}\
  }\textbf {\bibinfo {volume} {101}},\ \bibinfo {pages} {063302} (\bibinfo
  {year} {2020})}\BibitemShut {NoStop}%
\bibitem [{\citenamefont {Qiu}\ \emph {et~al.}(2020{\natexlab{b}})\citenamefont
  {Qiu}, \citenamefont {Che}, \citenamefont {Zhou}, \citenamefont {Zhu},\ and\
  \citenamefont {You}}]{qiu2020lattice}%
  \BibitemOpen
  \bibfield  {author} {\bibinfo {author} {\bibfnamefont {R.}~\bibnamefont
  {Qiu}}, \bibinfo {author} {\bibfnamefont {H.}~\bibnamefont {Che}}, \bibinfo
  {author} {\bibfnamefont {T.}~\bibnamefont {Zhou}}, \bibinfo {author}
  {\bibfnamefont {J.}~\bibnamefont {Zhu}}, \ and\ \bibinfo {author}
  {\bibfnamefont {Y.}~\bibnamefont {You}},\ }\bibfield  {title} {\enquote
  {\bibinfo {title} {Lattice {B}oltzmann simulation for unsteady shock
  wave/boundary layer interaction in a shock tube},}\ }\href@noop {} {\bibfield
   {journal} {\bibinfo  {journal} {Comput. Math. Appl.}\ }\textbf {\bibinfo
  {volume} {80}},\ \bibinfo {pages} {2241--2257} (\bibinfo {year}
  {2020}{\natexlab{b}})}\BibitemShut {NoStop}%
\bibitem [{\citenamefont {Wilde}\ \emph {et~al.}(2020)\citenamefont {Wilde},
  \citenamefont {Kr{\"a}mer}, \citenamefont {Reith},\ and\ \citenamefont
  {Foysi}}]{wilde2020semi}%
  \BibitemOpen
  \bibfield  {author} {\bibinfo {author} {\bibfnamefont {D.}~\bibnamefont
  {Wilde}}, \bibinfo {author} {\bibfnamefont {A.}~\bibnamefont {Kr{\"a}mer}},
  \bibinfo {author} {\bibfnamefont {D.}~\bibnamefont {Reith}}, \ and\ \bibinfo
  {author} {\bibfnamefont {H.}~\bibnamefont {Foysi}},\ }\bibfield  {title}
  {\enquote {\bibinfo {title} {{Semi-Lagrangian lattice Boltzmann method for
  compressible flows}},}\ }\href@noop {} {\bibfield  {journal} {\bibinfo
  {journal} {Phys. Rev. E}\ }\textbf {\bibinfo {volume} {101}},\ \bibinfo
  {pages} {053306} (\bibinfo {year} {2020})}\BibitemShut {NoStop}%
\bibitem [{\citenamefont {Saadat}\ and\ \citenamefont
  {Karlin}(2020)}]{saadat2020arbitrary}%
  \BibitemOpen
  \bibfield  {author} {\bibinfo {author} {\bibfnamefont {M.~H.}\ \bibnamefont
  {Saadat}}\ and\ \bibinfo {author} {\bibfnamefont {I.~V.}\ \bibnamefont
  {Karlin}},\ }\bibfield  {title} {\enquote {\bibinfo {title} {{Arbitrary
  Lagrangian--Eulerian formulation of lattice Boltzmann model for compressible
  flows on unstructured moving meshes}},}\ }\href@noop {} {\bibfield  {journal}
  {\bibinfo  {journal} {Phys. Fluids}\ }\textbf {\bibinfo {volume} {32}}
  (\bibinfo {year} {2020})}\BibitemShut {NoStop}%
\bibitem [{\citenamefont {Lin}\ \emph {et~al.}(2025)\citenamefont {Lin},
  \citenamefont {Su}, \citenamefont {Fei},\ and\ \citenamefont
  {Luo}}]{lin2025central}%
  \BibitemOpen
  \bibfield  {author} {\bibinfo {author} {\bibfnamefont {C.}~\bibnamefont
  {Lin}}, \bibinfo {author} {\bibfnamefont {X.}~\bibnamefont {Su}}, \bibinfo
  {author} {\bibfnamefont {L.}~\bibnamefont {Fei}}, \ and\ \bibinfo {author}
  {\bibfnamefont {K.}~\bibnamefont {Luo}},\ }\bibfield  {title} {\enquote
  {\bibinfo {title} {Central-moment-based discrete {B}oltzmann modeling of
  compressible flows},}\ }\href@noop {} {\bibfield  {journal} {\bibinfo
  {journal} {arXiv:2502.16818}\ } (\bibinfo {year} {2025})}\BibitemShut
  {NoStop}%
\bibitem [{\citenamefont {Kumari}\ and\ \citenamefont
  {Donzis}(2021)}]{kumari2021generalized}%
  \BibitemOpen
  \bibfield  {author} {\bibinfo {author} {\bibfnamefont {K.}~\bibnamefont
  {Kumari}}\ and\ \bibinfo {author} {\bibfnamefont {D.~A.}\ \bibnamefont
  {Donzis}},\ }\bibfield  {title} {\enquote {\bibinfo {title} {{A generalized
  von Neumann analysis for multi-level schemes: Stability and spectral
  accuracy}},}\ }\href@noop {} {\bibfield  {journal} {\bibinfo  {journal} {J.
  Comput. Phys.}\ }\textbf {\bibinfo {volume} {424}},\ \bibinfo {pages}
  {109868} (\bibinfo {year} {2021})}\BibitemShut {NoStop}%
\bibitem [{\citenamefont {Deriaz}\ and\ \citenamefont
  {Haldenwang}(2020)}]{deriaz2020non}%
  \BibitemOpen
  \bibfield  {author} {\bibinfo {author} {\bibfnamefont {E.}~\bibnamefont
  {Deriaz}}\ and\ \bibinfo {author} {\bibfnamefont {P.}~\bibnamefont
  {Haldenwang}},\ }\bibfield  {title} {\enquote {\bibinfo {title} {{Non-linear
  CFL conditions issued from the von Neumann stability analysis for the
  transport equation}},}\ }\href@noop {} {\bibfield  {journal} {\bibinfo
  {journal} {J. Sci. Comput.}\ }\textbf {\bibinfo {volume} {85}},\ \bibinfo
  {pages} {5} (\bibinfo {year} {2020})}\BibitemShut {NoStop}%
\bibitem [{\citenamefont {Sterling}\ and\ \citenamefont
  {Chen}(1996)}]{sterling1996stability}%
  \BibitemOpen
  \bibfield  {author} {\bibinfo {author} {\bibfnamefont {J.~D.}\ \bibnamefont
  {Sterling}}\ and\ \bibinfo {author} {\bibfnamefont {S.}~\bibnamefont
  {Chen}},\ }\bibfield  {title} {\enquote {\bibinfo {title} {Stability analysis
  of lattice {B}oltzmann methods},}\ }\href@noop {} {\bibfield  {journal}
  {\bibinfo  {journal} {J. Comput. Phys.}\ }\textbf {\bibinfo {volume} {123}},\
  \bibinfo {pages} {196--206} (\bibinfo {year} {1996})}\BibitemShut {NoStop}%
\bibitem [{\citenamefont {Seta}\ and\ \citenamefont
  {Takahashi}(2001{\natexlab{a}})}]{seta2001numerical}%
  \BibitemOpen
  \bibfield  {author} {\bibinfo {author} {\bibfnamefont {T.}~\bibnamefont
  {Seta}}\ and\ \bibinfo {author} {\bibfnamefont {R.}~\bibnamefont
  {Takahashi}},\ }\bibfield  {title} {\enquote {\bibinfo {title} {Numerical
  stability analysis for semi-implicit {FDLBM}},}\ }\href@noop {} {\bibfield
  {journal} {\bibinfo  {journal} {Trans. Jpn. Soc. Mech. Eng. B}\ }\textbf
  {\bibinfo {volume} {67}},\ \bibinfo {pages} {1662--1671} (\bibinfo {year}
  {2001}{\natexlab{a}})}\BibitemShut {NoStop}%
\bibitem [{\citenamefont {Seta}\ and\ \citenamefont
  {Takahashi}(2002)}]{seta2002numerical}%
  \BibitemOpen
  \bibfield  {author} {\bibinfo {author} {\bibfnamefont {T.}~\bibnamefont
  {Seta}}\ and\ \bibinfo {author} {\bibfnamefont {R.}~\bibnamefont
  {Takahashi}},\ }\bibfield  {title} {\enquote {\bibinfo {title} {Numerical
  stability analysis of {FDLBM}},}\ }\href@noop {} {\bibfield  {journal}
  {\bibinfo  {journal} {J. Stat. Phys.}\ }\textbf {\bibinfo {volume} {107}},\
  \bibinfo {pages} {557--572} (\bibinfo {year} {2002})}\BibitemShut {NoStop}%
\bibitem [{\citenamefont {Seta}\ and\ \citenamefont
  {Takahashi}(2001{\natexlab{b}})}]{seta2001stability}%
  \BibitemOpen
  \bibfield  {author} {\bibinfo {author} {\bibfnamefont {T.}~\bibnamefont
  {Seta}}\ and\ \bibinfo {author} {\bibfnamefont {R.}~\bibnamefont
  {Takahashi}},\ }\bibfield  {title} {\enquote {\bibinfo {title} {Stability
  analysis of thermal lattice {B}oltzmann models},}\ }\href@noop {} {\bibfield
  {journal} {\bibinfo  {journal} {Trans. Jpn. Soc. Mech. Eng. B}\ }\textbf
  {\bibinfo {volume} {67}},\ \bibinfo {pages} {2927--2936} (\bibinfo {year}
  {2001}{\natexlab{b}})}\BibitemShut {NoStop}%
\bibitem [{\citenamefont {Niu}\ \emph {et~al.}(2004)\citenamefont {Niu},
  \citenamefont {Shu}, \citenamefont {Chew},\ and\ \citenamefont
  {Wang}}]{niu2004investigation}%
  \BibitemOpen
  \bibfield  {author} {\bibinfo {author} {\bibfnamefont {X.}~\bibnamefont
  {Niu}}, \bibinfo {author} {\bibfnamefont {C.}~\bibnamefont {Shu}}, \bibinfo
  {author} {\bibfnamefont {Y.}~\bibnamefont {Chew}}, \ and\ \bibinfo {author}
  {\bibfnamefont {T.}~\bibnamefont {Wang}},\ }\bibfield  {title} {\enquote
  {\bibinfo {title} {Investigation of stability and hydrodynamics of different
  lattice {B}oltzmann models},}\ }\href@noop {} {\bibfield  {journal} {\bibinfo
   {journal} {J. Stat. Phys.}\ }\textbf {\bibinfo {volume} {117}},\ \bibinfo
  {pages} {665--680} (\bibinfo {year} {2004})}\BibitemShut {NoStop}%
\bibitem [{\citenamefont {Chen}\ \emph {et~al.}(2009)\citenamefont {Chen},
  \citenamefont {Xu}, \citenamefont {Zhang}, \citenamefont {Gan}, \citenamefont
  {Cheng},\ and\ \citenamefont {Li}}]{chen2009highly}%
  \BibitemOpen
  \bibfield  {author} {\bibinfo {author} {\bibfnamefont {F.}~\bibnamefont
  {Chen}}, \bibinfo {author} {\bibfnamefont {A.}~\bibnamefont {Xu}}, \bibinfo
  {author} {\bibfnamefont {G.}~\bibnamefont {Zhang}}, \bibinfo {author}
  {\bibfnamefont {Y.}~\bibnamefont {Gan}}, \bibinfo {author} {\bibfnamefont
  {T.}~\bibnamefont {Cheng}}, \ and\ \bibinfo {author} {\bibfnamefont
  {Y.}~\bibnamefont {Li}},\ }\bibfield  {title} {\enquote {\bibinfo {title}
  {{Highly efficient lattice Boltzmann model for compressible fluids:
  Two-dimensional case}},}\ }\href@noop {} {\bibfield  {journal} {\bibinfo
  {journal} {Commun. Theor. Phys.}\ }\textbf {\bibinfo {volume} {52}},\
  \bibinfo {pages} {681} (\bibinfo {year} {2009})}\BibitemShut {NoStop}%
\bibitem [{\citenamefont {Chen}, \citenamefont {Zhang},\ and\ \citenamefont
  {Li}(2010)}]{chen2010three}%
  \BibitemOpen
  \bibfield  {author} {\bibinfo {author} {\bibfnamefont {F.}~\bibnamefont
  {Chen}}, \bibinfo {author} {\bibfnamefont {G.}~\bibnamefont {Zhang}}, \ and\
  \bibinfo {author} {\bibfnamefont {Y.}~\bibnamefont {Li}},\ }\bibfield
  {title} {\enquote {\bibinfo {title} {Three-dimensional lattice {B}oltzmann
  model for high-speed compressible flows},}\ }\href@noop {} {\bibfield
  {journal} {\bibinfo  {journal} {Commun. Theor. Phys.}\ }\textbf {\bibinfo
  {volume} {54}},\ \bibinfo {pages} {1121} (\bibinfo {year}
  {2010})}\BibitemShut {NoStop}%
\bibitem [{\citenamefont {Servan-Camas}\ and\ \citenamefont
  {Tsai}(2009)}]{servan2009non}%
  \BibitemOpen
  \bibfield  {author} {\bibinfo {author} {\bibfnamefont {B.}~\bibnamefont
  {Servan-Camas}}\ and\ \bibinfo {author} {\bibfnamefont {F.~T.~C.}\
  \bibnamefont {Tsai}},\ }\bibfield  {title} {\enquote {\bibinfo {title}
  {Non-negativity and stability analyses of lattice {B}oltzmann method for
  advection--diffusion equation},}\ }\href@noop {} {\bibfield  {journal}
  {\bibinfo  {journal} {J. Comput. Phys.}\ }\textbf {\bibinfo {volume} {228}},\
  \bibinfo {pages} {236--256} (\bibinfo {year} {2009})}\BibitemShut {NoStop}%
\bibitem [{\citenamefont {Kuzmin}, \citenamefont {Ginzburg},\ and\
  \citenamefont {Mohamad}(2011)}]{kuzmin2011role}%
  \BibitemOpen
  \bibfield  {author} {\bibinfo {author} {\bibfnamefont {A.}~\bibnamefont
  {Kuzmin}}, \bibinfo {author} {\bibfnamefont {I.}~\bibnamefont {Ginzburg}}, \
  and\ \bibinfo {author} {\bibfnamefont {A.~A.}\ \bibnamefont {Mohamad}},\
  }\bibfield  {title} {\enquote {\bibinfo {title} {The role of the kinetic
  parameter in the stability of two-relaxation-time advection--diffusion
  lattice {B}oltzmann schemes},}\ }\href@noop {} {\bibfield  {journal}
  {\bibinfo  {journal} {Comput. Math. Appl.}\ }\textbf {\bibinfo {volume}
  {61}},\ \bibinfo {pages} {3417--3442} (\bibinfo {year} {2011})}\BibitemShut
  {NoStop}%
\bibitem [{\citenamefont {Chen}, \citenamefont {Shu},\ and\ \citenamefont
  {Tan}(2017{\natexlab{a}})}]{chen2017truly}%
  \BibitemOpen
  \bibfield  {author} {\bibinfo {author} {\bibfnamefont {Z.}~\bibnamefont
  {Chen}}, \bibinfo {author} {\bibfnamefont {C.}~\bibnamefont {Shu}}, \ and\
  \bibinfo {author} {\bibfnamefont {D.~S.}\ \bibnamefont {Tan}},\ }\bibfield
  {title} {\enquote {\bibinfo {title} {A truly second-order and unconditionally
  stable thermal lattice {B}oltzmann method},}\ }\href@noop {} {\bibfield
  {journal} {\bibinfo  {journal} {Appl. Sci.}\ }\textbf {\bibinfo {volume}
  {7}},\ \bibinfo {pages} {277} (\bibinfo {year}
  {2017}{\natexlab{a}})}\BibitemShut {NoStop}%
\bibitem [{\citenamefont {Chen}, \citenamefont {Shu},\ and\ \citenamefont
  {Tan}(2017{\natexlab{b}})}]{chen2017three}%
  \BibitemOpen
  \bibfield  {author} {\bibinfo {author} {\bibfnamefont {Z.}~\bibnamefont
  {Chen}}, \bibinfo {author} {\bibfnamefont {C.}~\bibnamefont {Shu}}, \ and\
  \bibinfo {author} {\bibfnamefont {D.}~\bibnamefont {Tan}},\ }\bibfield
  {title} {\enquote {\bibinfo {title} {Three-dimensional simplified and
  unconditionally stable lattice {B}oltzmann method for incompressible
  isothermal and thermal flows},}\ }\href@noop {} {\bibfield  {journal}
  {\bibinfo  {journal} {Phys. Fluids}\ }\textbf {\bibinfo {volume} {29}}
  (\bibinfo {year} {2017}{\natexlab{b}})}\BibitemShut {NoStop}%
\bibitem [{\citenamefont {Krivovichev}\ and\ \citenamefont
  {Mikheev}(2019)}]{krivovichev2019stability}%
  \BibitemOpen
  \bibfield  {author} {\bibinfo {author} {\bibfnamefont {G.~V.}\ \bibnamefont
  {Krivovichev}}\ and\ \bibinfo {author} {\bibfnamefont {S.~A.}\ \bibnamefont
  {Mikheev}},\ }\bibfield  {title} {\enquote {\bibinfo {title} {On the
  stability of multi-step finite-difference-based lattice {B}oltzmann
  schemes},}\ }\href@noop {} {\bibfield  {journal} {\bibinfo  {journal} {Int.
  J. Comput. Methods}\ }\textbf {\bibinfo {volume} {16}},\ \bibinfo {pages}
  {1850087} (\bibinfo {year} {2019})}\BibitemShut {NoStop}%
\bibitem [{\citenamefont {Krivovichev}(2020)}]{krivovichev2020parametric}%
  \BibitemOpen
  \bibfield  {author} {\bibinfo {author} {\bibfnamefont {G.~V.}\ \bibnamefont
  {Krivovichev}},\ }\bibfield  {title} {\enquote {\bibinfo {title} {Parametric
  schemes for the simulation of the advection process in
  finite-difference-based single-relaxation-time lattice {B}oltzmann
  methods},}\ }\href@noop {} {\bibfield  {journal} {\bibinfo  {journal} {J.
  Comput. Sci.}\ }\textbf {\bibinfo {volume} {44}},\ \bibinfo {pages} {101151}
  (\bibinfo {year} {2020})}\BibitemShut {NoStop}%
\bibitem [{\citenamefont {Krivovichev}\ and\ \citenamefont
  {Marnopolskaya}(2020)}]{krivovichev2020approach}%
  \BibitemOpen
  \bibfield  {author} {\bibinfo {author} {\bibfnamefont {G.~V.}\ \bibnamefont
  {Krivovichev}}\ and\ \bibinfo {author} {\bibfnamefont {E.~S.}\ \bibnamefont
  {Marnopolskaya}},\ }\bibfield  {title} {\enquote {\bibinfo {title} {The
  approach to optimization of finite-difference schemes for the advective stage
  of finite-difference-based lattice {B}oltzmann method},}\ }\href@noop {}
  {\bibfield  {journal} {\bibinfo  {journal} {Int. J. Model. Simul. Sci.
  Comput.}\ }\textbf {\bibinfo {volume} {11}},\ \bibinfo {pages} {2050002}
  (\bibinfo {year} {2020})}\BibitemShut {NoStop}%
\bibitem [{\citenamefont {Masset}\ and\ \citenamefont
  {Wissocq}(2020)}]{masset2020linear}%
  \BibitemOpen
  \bibfield  {author} {\bibinfo {author} {\bibfnamefont {P.~A.}\ \bibnamefont
  {Masset}}\ and\ \bibinfo {author} {\bibfnamefont {G.}~\bibnamefont
  {Wissocq}},\ }\bibfield  {title} {\enquote {\bibinfo {title} {Linear
  hydrodynamics and stability of the discrete velocity {B}oltzmann
  equations},}\ }\href@noop {} {\bibfield  {journal} {\bibinfo  {journal} {J.
  Fluid Mech.}\ }\textbf {\bibinfo {volume} {897}},\ \bibinfo {pages} {A29}
  (\bibinfo {year} {2020})}\BibitemShut {NoStop}%
\bibitem [{\citenamefont {Wissocq}, \citenamefont {Coreixas},\ and\
  \citenamefont {Boussuge}(2020)}]{wissocq2020linear}%
  \BibitemOpen
  \bibfield  {author} {\bibinfo {author} {\bibfnamefont {G.}~\bibnamefont
  {Wissocq}}, \bibinfo {author} {\bibfnamefont {C.}~\bibnamefont {Coreixas}}, \
  and\ \bibinfo {author} {\bibfnamefont {J.~F.}\ \bibnamefont {Boussuge}},\
  }\bibfield  {title} {\enquote {\bibinfo {title} {Linear stability and
  isotropy properties of athermal regularized lattice {B}oltzmann methods},}\
  }\href@noop {} {\bibfield  {journal} {\bibinfo  {journal} {Phys. Rev. E}\
  }\textbf {\bibinfo {volume} {102}},\ \bibinfo {pages} {053305} (\bibinfo
  {year} {2020})}\BibitemShut {NoStop}%
\bibitem [{\citenamefont {Worthing}, \citenamefont {Mozer},\ and\ \citenamefont
  {Seeley}(1997)}]{worthing1997stability}%
  \BibitemOpen
  \bibfield  {author} {\bibinfo {author} {\bibfnamefont {R.~A.}\ \bibnamefont
  {Worthing}}, \bibinfo {author} {\bibfnamefont {J.}~\bibnamefont {Mozer}}, \
  and\ \bibinfo {author} {\bibfnamefont {G.}~\bibnamefont {Seeley}},\
  }\bibfield  {title} {\enquote {\bibinfo {title} {Stability of lattice
  {B}oltzmann methods in hydrodynamic regimes},}\ }\href@noop {} {\bibfield
  {journal} {\bibinfo  {journal} {Phys. Rev. E}\ }\textbf {\bibinfo {volume}
  {56}},\ \bibinfo {pages} {2243} (\bibinfo {year} {1997})}\BibitemShut
  {NoStop}%
\bibitem [{\citenamefont {Servan-Camas}\ and\ \citenamefont
  {Tsai}(2008)}]{servan2008lattice}%
  \BibitemOpen
  \bibfield  {author} {\bibinfo {author} {\bibfnamefont {B.}~\bibnamefont
  {Servan-Camas}}\ and\ \bibinfo {author} {\bibfnamefont {F.~T.~C.}\
  \bibnamefont {Tsai}},\ }\bibfield  {title} {\enquote {\bibinfo {title}
  {{Lattice Boltzmann method with two relaxation times for advection--diffusion
  equation: Third order analysis and stability analysis}},}\ }\href@noop {}
  {\bibfield  {journal} {\bibinfo  {journal} {Adv. Water Resour.}\ }\textbf
  {\bibinfo {volume} {31}},\ \bibinfo {pages} {1113--1126} (\bibinfo {year}
  {2008})}\BibitemShut {NoStop}%
\bibitem [{\citenamefont {Siebert}, \citenamefont {Hegele~Jr},\ and\
  \citenamefont {Philippi}(2008)}]{siebert2008lattice}%
  \BibitemOpen
  \bibfield  {author} {\bibinfo {author} {\bibfnamefont {D.~N.}\ \bibnamefont
  {Siebert}}, \bibinfo {author} {\bibfnamefont {L.~A.}\ \bibnamefont
  {Hegele~Jr}}, \ and\ \bibinfo {author} {\bibfnamefont {P.~C.}\ \bibnamefont
  {Philippi}},\ }\bibfield  {title} {\enquote {\bibinfo {title} {{Lattice
  Boltzmann equation linear stability analysis: Thermal and athermal
  models}},}\ }\href@noop {} {\bibfield  {journal} {\bibinfo  {journal} {Phys.
  Rev. E}\ }\textbf {\bibinfo {volume} {77}},\ \bibinfo {pages} {026707}
  (\bibinfo {year} {2008})}\BibitemShut {NoStop}%
\bibitem [{\citenamefont {P{\'e}rez}, \citenamefont {Aguilar},\ and\
  \citenamefont {Theofilis}(2017)}]{perez2017lattice}%
  \BibitemOpen
  \bibfield  {author} {\bibinfo {author} {\bibfnamefont {J.~M.}\ \bibnamefont
  {P{\'e}rez}}, \bibinfo {author} {\bibfnamefont {A.}~\bibnamefont {Aguilar}},
  \ and\ \bibinfo {author} {\bibfnamefont {V.}~\bibnamefont {Theofilis}},\
  }\bibfield  {title} {\enquote {\bibinfo {title} {Lattice {B}oltzmann methods
  for global linear instability analysis},}\ }\href@noop {} {\bibfield
  {journal} {\bibinfo  {journal} {Theor. Comput. Fluid Dyn.}\ }\textbf
  {\bibinfo {volume} {31}},\ \bibinfo {pages} {643--664} (\bibinfo {year}
  {2017})}\BibitemShut {NoStop}%
\bibitem [{\citenamefont {Yang}, \citenamefont {Zhao},\ and\ \citenamefont
  {Lin}(2024)}]{yang2024automatic}%
  \BibitemOpen
  \bibfield  {author} {\bibinfo {author} {\bibfnamefont {J.}~\bibnamefont
  {Yang}}, \bibinfo {author} {\bibfnamefont {W.}~\bibnamefont {Zhao}}, \ and\
  \bibinfo {author} {\bibfnamefont {P.}~\bibnamefont {Lin}},\ }\bibfield
  {title} {\enquote {\bibinfo {title} {An automatic approach for the stability
  analysis of multi-relaxation-time lattice {B}oltzmann models},}\ }\href@noop
  {} {\bibfield  {journal} {\bibinfo  {journal} {J. Comput. Phys.}\ }\textbf
  {\bibinfo {volume} {519}},\ \bibinfo {pages} {113432} (\bibinfo {year}
  {2024})}\BibitemShut {NoStop}%
\bibitem [{\citenamefont {Ren}\ \emph {et~al.}(2024)\citenamefont {Ren},
  \citenamefont {Xie}, \citenamefont {Zhang}, \citenamefont {Yu}, \citenamefont
  {Tian},\ and\ \citenamefont {Zhu}}]{ren2024numerical}%
  \BibitemOpen
  \bibfield  {author} {\bibinfo {author} {\bibfnamefont {W.}~\bibnamefont
  {Ren}}, \bibinfo {author} {\bibfnamefont {W.}~\bibnamefont {Xie}}, \bibinfo
  {author} {\bibfnamefont {Y.}~\bibnamefont {Zhang}}, \bibinfo {author}
  {\bibfnamefont {H.}~\bibnamefont {Yu}}, \bibinfo {author} {\bibfnamefont
  {Z.}~\bibnamefont {Tian}}, \ and\ \bibinfo {author} {\bibfnamefont
  {J.}~\bibnamefont {Zhu}},\ }\bibfield  {title} {\enquote {\bibinfo {title}
  {{Numerical stability analysis of Godunov-type schemes for high Mach number
  flow simulations}},}\ }\href@noop {} {\bibfield  {journal} {\bibinfo
  {journal} {Phys. Fluids}\ }\textbf {\bibinfo {volume} {36}} (\bibinfo {year}
  {2024})}\BibitemShut {NoStop}%
\bibitem [{\citenamefont {Ren}\ \emph {et~al.}(2025)\citenamefont {Ren},
  \citenamefont {Xie}, \citenamefont {Zhang}, \citenamefont {Yu}, \citenamefont
  {Tian},\ and\ \citenamefont {Jiajun}}]{ren2025numerical}%
  \BibitemOpen
  \bibfield  {author} {\bibinfo {author} {\bibfnamefont {W.}~\bibnamefont
  {Ren}}, \bibinfo {author} {\bibfnamefont {W.}~\bibnamefont {Xie}}, \bibinfo
  {author} {\bibfnamefont {Y.}~\bibnamefont {Zhang}}, \bibinfo {author}
  {\bibfnamefont {H.}~\bibnamefont {Yu}}, \bibinfo {author} {\bibfnamefont
  {Z.}~\bibnamefont {Tian}}, \ and\ \bibinfo {author} {\bibfnamefont
  {Z.}~\bibnamefont {Jiajun}},\ }\bibfield  {title} {\enquote {\bibinfo {title}
  {{Numerical stability analysis of shock-capturing methods for strong shocks
  II: High-order finite-volume schemes}},}\ }\href@noop {} {\bibfield
  {journal} {\bibinfo  {journal} {J. Comput. Phys.}\ }\textbf {\bibinfo
  {volume} {523}},\ \bibinfo {pages} {113649} (\bibinfo {year}
  {2025})}\BibitemShut {NoStop}%
\bibitem [{\citenamefont {Watari}\ and\ \citenamefont
  {Tsutahara}(2004)}]{watari2004possibility}%
  \BibitemOpen
  \bibfield  {author} {\bibinfo {author} {\bibfnamefont {M.}~\bibnamefont
  {Watari}}\ and\ \bibinfo {author} {\bibfnamefont {M.}~\bibnamefont
  {Tsutahara}},\ }\bibfield  {title} {\enquote {\bibinfo {title} {{Possibility
  of constructing a multispeed Bhatnagar-Gross-Krook thermal model of the
  lattice Boltzmann method}},}\ }\href@noop {} {\bibfield  {journal} {\bibinfo
  {journal} {Phys. Rev. E}\ }\textbf {\bibinfo {volume} {70}},\ \bibinfo
  {pages} {016703} (\bibinfo {year} {2004})}\BibitemShut {NoStop}%
\bibitem [{\citenamefont {Watari}(2007)}]{watari2007finite}%
  \BibitemOpen
  \bibfield  {author} {\bibinfo {author} {\bibfnamefont {M.}~\bibnamefont
  {Watari}},\ }\bibfield  {title} {\enquote {\bibinfo {title} {Finite
  difference lattice {B}oltzmann method with arbitrary specific heat ratio
  applicable to supersonic flow simulations},}\ }\href@noop {} {\bibfield
  {journal} {\bibinfo  {journal} {Physica A}\ }\textbf {\bibinfo {volume}
  {382}},\ \bibinfo {pages} {502--522} (\bibinfo {year} {2007})}\BibitemShut
  {NoStop}%
\bibitem [{\citenamefont {Alexander}, \citenamefont {Chen},\ and\ \citenamefont
  {Sterling}(1993)}]{alexander1993lattice}%
  \BibitemOpen
  \bibfield  {author} {\bibinfo {author} {\bibfnamefont {F.~J.}\ \bibnamefont
  {Alexander}}, \bibinfo {author} {\bibfnamefont {S.}~\bibnamefont {Chen}}, \
  and\ \bibinfo {author} {\bibfnamefont {J.~D.}\ \bibnamefont {Sterling}},\
  }\bibfield  {title} {\enquote {\bibinfo {title} {Lattice {B}oltzmann
  thermohydrodynamics},}\ }\href@noop {} {\bibfield  {journal} {\bibinfo
  {journal} {Phys. Rev. E}\ }\textbf {\bibinfo {volume} {47}},\ \bibinfo
  {pages} {R2249} (\bibinfo {year} {1993})}\BibitemShut {NoStop}%
\bibitem [{\citenamefont {Chen}, \citenamefont {Ohashi},\ and\ \citenamefont
  {Akiyama}(1994)}]{chen1994thermal}%
  \BibitemOpen
  \bibfield  {author} {\bibinfo {author} {\bibfnamefont {Y.}~\bibnamefont
  {Chen}}, \bibinfo {author} {\bibfnamefont {H.}~\bibnamefont {Ohashi}}, \ and\
  \bibinfo {author} {\bibfnamefont {M.}~\bibnamefont {Akiyama}},\ }\bibfield
  {title} {\enquote {\bibinfo {title} {{Thermal lattice Bhatnagar-Gross-Krook
  model without nonlinear deviations in macrodynamic equations}},}\ }\href@noop
  {} {\bibfield  {journal} {\bibinfo  {journal} {Phys. Rev. E}\ }\textbf
  {\bibinfo {volume} {50}},\ \bibinfo {pages} {2776} (\bibinfo {year}
  {1994})}\BibitemShut {NoStop}%
\bibitem [{\citenamefont {Kataoka}\ and\ \citenamefont
  {Tsutahara}(2004{\natexlab{b}})}]{kataoka2004lattice2}%
  \BibitemOpen
  \bibfield  {author} {\bibinfo {author} {\bibfnamefont {T.}~\bibnamefont
  {Kataoka}}\ and\ \bibinfo {author} {\bibfnamefont {M.}~\bibnamefont
  {Tsutahara}},\ }\bibfield  {title} {\enquote {\bibinfo {title} {{Lattice
  Boltzmann method for the compressible Euler equations}},}\ }\href@noop {}
  {\bibfield  {journal} {\bibinfo  {journal} {Phys. Rev. E}\ }\textbf {\bibinfo
  {volume} {69}},\ \bibinfo {pages} {056702} (\bibinfo {year}
  {2004}{\natexlab{b}})}\BibitemShut {NoStop}%
\bibitem [{\citenamefont {Zhang}(1988)}]{zhang1988non}%
  \BibitemOpen
  \bibfield  {author} {\bibinfo {author} {\bibfnamefont {H.}~\bibnamefont
  {Zhang}},\ }\bibfield  {title} {\enquote {\bibinfo {title} {Non-oscillatory
  and non-free-parameter dissipation difference scheme},}\ }\href@noop {}
  {\bibfield  {journal} {\bibinfo  {journal} {Acta Aerodyn. Sin.}\ }\textbf
  {\bibinfo {volume} {6}},\ \bibinfo {pages} {143--165} (\bibinfo {year}
  {1988})}\BibitemShut {NoStop}%
\bibitem [{\citenamefont {Jiang}\ and\ \citenamefont
  {Shu}(1996)}]{jiang1996efficient}%
  \BibitemOpen
  \bibfield  {author} {\bibinfo {author} {\bibfnamefont {G.}~\bibnamefont
  {Jiang}}\ and\ \bibinfo {author} {\bibfnamefont {C.}~\bibnamefont {Shu}},\
  }\bibfield  {title} {\enquote {\bibinfo {title} {Efficient implementation of
  weighted {ENO} schemes},}\ }\href@noop {} {\bibfield  {journal} {\bibinfo
  {journal} {J. Comput. Phys.}\ }\textbf {\bibinfo {volume} {126}},\ \bibinfo
  {pages} {202--228} (\bibinfo {year} {1996})}\BibitemShut {NoStop}%
\bibitem [{\citenamefont {Sofonea}\ and\ \citenamefont
  {Mecke}(1999)}]{sofonea1999morphological}%
  \BibitemOpen
  \bibfield  {author} {\bibinfo {author} {\bibfnamefont {V.}~\bibnamefont
  {Sofonea}}\ and\ \bibinfo {author} {\bibfnamefont {K.~R.}\ \bibnamefont
  {Mecke}},\ }\bibfield  {title} {\enquote {\bibinfo {title} {Morphological
  characterization of spinodal decomposition kinetics},}\ }\href@noop {}
  {\bibfield  {journal} {\bibinfo  {journal} {Eur. Phys. J. B.}\ }\textbf
  {\bibinfo {volume} {8}},\ \bibinfo {pages} {99--112} (\bibinfo {year}
  {1999})}\BibitemShut {NoStop}%
\bibitem [{\citenamefont {Gan}\ \emph {et~al.}(2011{\natexlab{c}})\citenamefont
  {Gan}, \citenamefont {Xu}, \citenamefont {Zhang}, \citenamefont {Li},\ and\
  \citenamefont {Li}}]{gan2011phase}%
  \BibitemOpen
  \bibfield  {author} {\bibinfo {author} {\bibfnamefont {Y.}~\bibnamefont
  {Gan}}, \bibinfo {author} {\bibfnamefont {A.}~\bibnamefont {Xu}}, \bibinfo
  {author} {\bibfnamefont {G.}~\bibnamefont {Zhang}}, \bibinfo {author}
  {\bibfnamefont {Y.}~\bibnamefont {Li}}, \ and\ \bibinfo {author}
  {\bibfnamefont {H.}~\bibnamefont {Li}},\ }\bibfield  {title} {\enquote
  {\bibinfo {title} {{Phase separation in thermal systems: A lattice Boltzmann
  study and morphological characterization}},}\ }\href@noop {} {\bibfield
  {journal} {\bibinfo  {journal} {Phys. Rev. E.}\ }\textbf {\bibinfo {volume}
  {84}},\ \bibinfo {pages} {046715} (\bibinfo {year}
  {2011}{\natexlab{c}})}\BibitemShut {NoStop}%
\bibitem [{\citenamefont {Lakshmikantham}, \citenamefont {Leela},\ and\
  \citenamefont {Martynyuk}(1989)}]{lakshmikantham1989stability}%
  \BibitemOpen
  \bibfield  {author} {\bibinfo {author} {\bibfnamefont {V.}~\bibnamefont
  {Lakshmikantham}}, \bibinfo {author} {\bibfnamefont {S.}~\bibnamefont
  {Leela}}, \ and\ \bibinfo {author} {\bibfnamefont {A.~A.}\ \bibnamefont
  {Martynyuk}},\ }\href@noop {} {\emph {\bibinfo {title} {{Stability Analysis
  of Nonlinear Systems}}}}\ (\bibinfo  {publisher} {Springer},\ \bibinfo {year}
  {1989})\BibitemShut {NoStop}%
\bibitem [{\citenamefont {Sastry}(2013)}]{sastry2013nonlinear}%
  \BibitemOpen
  \bibfield  {author} {\bibinfo {author} {\bibfnamefont {S.}~\bibnamefont
  {Sastry}},\ }\href@noop {} {\emph {\bibinfo {title} {{Nonlinear Systems:
  Analysis, Stability, and Control}}}},\ Vol.~\bibinfo {volume} {10}\ (\bibinfo
   {publisher} {Springer Science \& Business Media},\ \bibinfo {year}
  {2013})\BibitemShut {NoStop}%
\bibitem [{\citenamefont {Tripathi}\ \emph {et~al.}(2024)\citenamefont
  {Tripathi}, \citenamefont {Shankar}, \citenamefont {Mahajan},\ and\
  \citenamefont {Shivakumara}}]{tripathi2024global}%
  \BibitemOpen
  \bibfield  {author} {\bibinfo {author} {\bibfnamefont {V.~K.}\ \bibnamefont
  {Tripathi}}, \bibinfo {author} {\bibfnamefont {B.}~\bibnamefont {Shankar}},
  \bibinfo {author} {\bibfnamefont {A.}~\bibnamefont {Mahajan}}, \ and\
  \bibinfo {author} {\bibfnamefont {I.}~\bibnamefont {Shivakumara}},\
  }\bibfield  {title} {\enquote {\bibinfo {title} {Global nonlinear stability
  of bidispersive porous convection with throughflow and depth-dependent
  viscosity},}\ }\href@noop {} {\bibfield  {journal} {\bibinfo  {journal}
  {Phys. Fluids}\ }\textbf {\bibinfo {volume} {36}} (\bibinfo {year}
  {2024})}\BibitemShut {NoStop}%
\bibitem [{\citenamefont {Dubois}, \citenamefont {Saint-Jean},\ and\
  \citenamefont {Tekitek}(2024)}]{dubois2024beyond}%
  \BibitemOpen
  \bibfield  {author} {\bibinfo {author} {\bibfnamefont {F.}~\bibnamefont
  {Dubois}}, \bibinfo {author} {\bibfnamefont {C.}~\bibnamefont {Saint-Jean}},
  \ and\ \bibinfo {author} {\bibfnamefont {M.~M.}\ \bibnamefont {Tekitek}},\
  }\bibfield  {title} {\enquote {\bibinfo {title} {{Beyond linear analysis:
  Exploring stability of multiple-relaxation-time lattice Boltzmann method for
  nonlinear flows using decision trees and evolutionary algorithms}},}\
  }\href@noop {} {\bibfield  {journal} {\bibinfo  {journal} {Discret. Contin.
  Dyn. Syst. S}\ }\textbf {\bibinfo {volume} {17}},\ \bibinfo {pages}
  {3174--3191} (\bibinfo {year} {2024})}\BibitemShut {NoStop}%
\end{thebibliography}%

\end{document}